\documentclass[11pt]{report}

\usepackage{fullpage}

\usepackage{amssymb, amsmath, epsfig, subfig, feynmp,
  graphics}
\usepackage{amsthm, url, longtable, slashed}

\newtheorem{thm}{Theorem}[chapter]
\newtheorem{lemma}[thm]{Lemma}
\newtheorem{prop}[thm]{Proposition}

\newtheorem{cor}[thm]{Corollary}

\theoremstyle{definition}
\newtheorem{definition}[thm]{Definition}
\newtheorem{example}[thm]{Example}
\newtheorem{assume}[thm]{Assumption}

\newcommand{\Plin}{P_{\mathrm{lin}}}
\newcommand{\Hlin}{\mathcal{H}_{\mathrm{lin}}}
\newcommand{\h}{\mathcal{H}}
\newcommand{\id}{\mathrm{id}}
\newcommand{\sgn}{\mathrm{sign}}

\begin{document}
\allowdisplaybreaks

% This file contains all the necessary setup and commands to create
% the preliminary pages according to the buthesis.sty option.

\title{Growth estimates for Dyson-Schwinger equations}

\author{Karen Amanda Yeats}

\maketitle

% The copyright page is blank except for the notice at the bottom. You
% must provide your name in capitals.

%\copyrightpage

% Now include the approval page based on the readers information
%\approvalpage

% The acknowledgment page should go here. Use something like
% \newpage\section*{Acknowledgments} followed by your text.

\newpage
\chapter*{Acknowledgments}

I would like to thank Dirk Kreimer for his wisdom, insight, and
never-ending store of ideas.  Second, I would like to thank David
Fried for a very detailed reading.  Third, I would like to thank the
remainder of my committee, Maciej Szczesny, David
Rohrlich, and Takashi Kimura.
I would also like to thank everyone who
got excited about the differential equation including Paul Krapivsky,
Cameron Morland, David Uminsky, and Guillaume Van Baalen.  

Cameron has
further been invaluable professionally for his great skill with plots
and personally for hugs, geekiness, and love.  Finally, Russell Morland has
prevented me from getting work done in the best possible way.

The following software was used in the research behind and the
presentation of this work.  This document is typeset in \LaTeX\ using a
thesis style file originally written by Stephen Gildea and modified by
Paolo Gaudiano, Jonathan Polimeni, Janusz Konrad, and Cameron
Morland.  Symbolic and numerical computation was done using GiNaC and
Maple.  Plots were prepared with gnuplot both directly and via octave.

% The abstractpage environment sets up everything on the page except
% the text itself.  The title and other header material are put at the
% top of the page, and the supervisors are listed at the bottom.  A
% new page is begun both before and after.  Of course, an abstract may
% be more than one page itself.  If you need more control over the
% format of the page, you can use the abstract environment, which puts
% the word "Abstract" at the beginning and single spaces its text.

\begin{abstract}
Dyson-Schwinger equations are integral equations in quantum field
theory that describe the Green functions of a theory and mirror the
recursive decomposition of Feynman diagrams 
into subdiagrams.  Taken as
recursive equations, the Dyson-Schwinger equations describe
perturbative quantum field theory.  However, they also contain
non-perturbative information.

Using the Hopf algebra of Feynman graphs we will follow a sequence of
reductions to convert the Dyson-Schwinger equations to the following system of differential equations,
\[
  \gamma_1^r(x) = P_r(x) - \sgn(s_r)\gamma_1^r(x)^2 + \left(\sum_{j \in
  \mathcal{R}}|s_j|\gamma_1^j(x)\right) x \partial_x \gamma_1^r(x) 
\]
where $r \in \mathcal{R}$, $\mathcal{R}$ is the set of amplitudes of the
theory which need renormalization, $\gamma_1^r$ is the anomalous
dimension associated to $r$, $P_r(x)$ is a modified version of the function for the
primitive skeletons contributing to $r$, and $x$ is the coupling constant.

Next, we approach the new system of differential equations as a system
of recursive equations by expanding $\gamma_1^r(x) = \sum_{n \geq
1}\gamma^r_{1,n} x^n$.  We obtain the radius of convergence of $\sum
\gamma^r_{1,n}x^n/n!$ in 
terms of that of $\sum P_r(n)x^n/n!$.  In particular we show that a
Lipatov bound for the growth of the primitives leads to a Lipatov
bound for the whole theory.

Finally, we make a few observations on the new system considered as differential equations.

\end{abstract}

% Now you can include a preface. Again, use something like
% \newpage\section*{Preface} followed by your text

% Table of contents comes after preface
\tableofcontents

% If you have tables, uncomment the following line
\listoftables

% If you have figures, uncomment the following line
\newpage\listoffigures

% List of Abbrevs is NOT optional (Martha Wellman likes all abbrevs listed)
\chapter*{List of Symbols}
  \begin{longtable}{lp{0.75\textwidth}}
    1PI \dotfill & 1-particle irreducible, that is, 2-connected \\
    $A$ \dotfill & the gauge field in QED \\
    $\mathbf{A}^r(x)$ \dotfill & generating function for $a^r_n$ \\
    $\mathbf{A}(x)$ \dotfill & generating function for $a_n$ \\
    $a^r_n$ \dotfill & $\gamma^r_{1,n}/n!$ \\
    $a_n$ \dotfill & $\gamma_{1,n}/n!$ \\
    $a^1_n$, $a^2_n$ \dotfill & coefficients for an example system\\
    $\beta$ \dotfill & the physicists' $\beta$-function describing the
    nonlinearity of a Green function \\
    $B_+$ \dotfill & insertion into a Hopf algebra primitive taken generically
    \\
    $B_+^{\gamma}$ \dotfill & insertion into the primitive $\gamma$ \\
    $B_+^{k,i;r}$ \dotfill & insertion into the $k$-loop primitive
    with residue $r$    indexed by $i$ \\
    $B_+^{k,i}$ \dotfill & insertion into a primitive at $k$ loops, with $i$ an
    index running over primitives; that is, $B_+^{k,i;r}$ in the case
    with only one $r$ \\
    $\mathbf{B}^r(x)$ \dotfill & generating function for $b^r_n$ \\
    $\mathbf{B}(x)$ \dotfill & generating function for $b_n$  \\
    $b^r_n$ \dotfill & a particular lower bound for $a^r_n$ \\
    $b_n$ \dotfill & a particular lower bound for $a_n$ \\
    $\mathbf{bij}(\gamma, X, \Gamma)$ \dotfill & the number of
    bijections of the external edges of $X$ with an insertion place of
    $\gamma$ such that the resulting insertion gives $\Gamma$ \\
    $\mathbf{C}^r(x)$ \dotfill & generating function for $c^r_n$
    implicitly depending 
    on an $\epsilon > 0$ \\
    $\mathbf{C}(x)$ \dotfill & generating function for $c_n$
    implicitly depending     on an  $\epsilon > 0$ \\
    $c^r_n$ \dotfill & a particular upper bound for $a^r_n$ implicitly
    depending on 
    an $\epsilon > 0$ \\
    $c_n$ \dotfill & a particular upper bound for $a_n$ implicitly depending on
    an $\epsilon > 0$ \\
    $\Delta$ \dotfill & the coproduct of $\h$ \\
    $d^4$ \dotfill & integration over $\mathbb{R}^4$ \\
    $D$ \dotfill & dimension of space-time \\
    $\eta$ \dotfill & the counit of $\h$ \\
    $e$ \dotfill & the unit map of $\h$ \\
    $E$ \dotfill & an edge type, viewed as a pair of half edge types \\
    $F_p$ \dotfill & the Mellin transform associated to the Hopf
    algebra primitive $p$ \\
    $F^r_{k,i}$ \dotfill & the Mellin transform associated to the
    $k$-loop primitive with residue $r$ indexed by $i$ \\
    $F_{k,i}(\rho)$ \dotfill & $F^r_{k,i}$ in the case with only one
    $r$ \\
    $f^r(x)$ \dotfill & $\sum_{k \geq 1} x^k p^r(k)/k!$ when $\sum_{k
      \geq 1} x^k 
    p^r(k)$ is Gevrey-1\\
    $f(x)$ \dotfill & $\sum_{k \geq 1} x^k p(k)/k!$ when $\sum_{k \geq 1} x^k
    p(k)$ is Gevrey-1 \\
    $\gamma^r_1$ \dotfill &the anomalous dimension of the Green
    function indexed    by the amplitude $r$ \\
    $\gamma^r_k$ \dotfill & $k$-th leading log term of the Green
    function indexed 
    by the amplitude $r$ \\
    $\gamma_k$ \dotfill & $\gamma_k^r$ in the case
    with only one $r$ \\
    $\gamma^r_{1,n}$ \dotfill & coefficient of $x^n$ in $\gamma^r_1$ \\
    $\gamma_{1,n}$ \dotfill & coefficient of $x^n$ in $\gamma_1$ \\
    $(\gamma|X)$ \dotfill & the number
    of insertion places for $X$ in $\gamma$ \\
    $\gamma \cdot U$ \dotfill & $\sum \gamma_k U^k$ \\
    $G, \Gamma, \gamma$ \dotfill & graphs \\
    $\Gamma(x)$ \dotfill & the $\Gamma$ function extending the
    factorial function to the complex numbers \\
    $G/\gamma$ \dotfill & the graph $G$ with the subgraph $\gamma$ contracted
    \\
    $G^r(x,L)$ \dotfill & Green function indexed by the amplitude $r$ \\
    $\h$ \dotfill & the Hopf algebra of Feynman graphs \\
    $\Hlin$ \dotfill & the linear piece of $\h$ \\
    $H$ \dotfill & set of half edge types \\
    $\mathbb{I}$ \dotfill & the empty graph as the unit element of $\h$ \\
    $\id$ \dotfill & the identity map on $\h$ \\
    $k$ \dotfill & an internal momentum appearing as an integration variable \\
    $\mathcal{L}$ \dotfill & a Lagrangian \\
    $L$ \dotfill & $\log(q^2/\mu^2)$, the second variable on which the Green
    functions depend, where $q^2$ is a kinematical variable
    and $\mu^2$ is a subtraction point \\
    $m$ \dotfill & multiplication on $\h$ or a mass\\
    $\mathrm{maxf}(\Gamma)$ \dotfill & the number of insertion trees
    corresponding to $\Gamma$ \\
    $d\Omega_k$ \dotfill & angular integration over the $D-1$ sphere
    in $\mathbb{R}^D$ where $k \in \mathbb{R}^D$ \\
    $\phi$ \dotfill & a scalar field or the (unrenormalized) Feynman rules\\
    $\phi^3$ \dotfill & scalar field theory with a 3 valent vertex \\
    $\phi^4$ \dotfill & scalar field theory with a 4 valent vertex \\
    $\phi_R$ \dotfill & the renormalized Feynman rules \\
    $\psi$ \dotfill & the fermion field in QED \\
    $P_\epsilon, P_\epsilon^r$ \dotfill & polynomials depending on
    $\epsilon$ \\
    $P_r$ \dotfill & a modified version of the function of the primitive
    skeletons with residue $r$ \\
    $\Plin$ \dotfill & projection onto the linear piece of $\h$ \\
    $p^r_i(k)$ \dotfill & coefficient giving the contribution of
    primitive $i$ at 
    $k$ loops with external leg structure $r$ \\
    $p_i(k)$ \dotfill & $p^r_i(k)$ in the case with only one $r$ \\
    $p^r(k)$ \dotfill & $-\sum_i r_{k,i;r}p^r_i(k)$, the overall
    contribution of all 
    primitives at $k$ loops \\ 
    $p(k)$ \dotfill & $-\sum_i r_{k,i}p_i(k)$, the overall contribution of all
    primitives at $k$ loops in the case with only one $r$ \\ 
    $q$ \dotfill & an external momentum \\
    $Q$ \dotfill & (combinatorial) invariant charge \\
    QCD \dotfill & quantum chromodynamics \\
    QED \dotfill & quantum electrodynamics \\
    $\rho$ \dotfill & the argument of Mellin transforms with 1 insertion place
    or the radius of convergence of $f(x)$ \\
    $\rho_a$ \dotfill & radius of convergence of $\mathbf{A}(x)$ \\
    $\rho_\epsilon$ \dotfill & the radius of convergence of
    $\mathbf{C}(x)$  \\
    $\rho_i$ \dotfill & the argument of the Mellin transform which
    marks the $i$th
    insertion place \\
    $\rho_r$ \dotfill & the radius of convergence of $f^r(x)$ \\
    $\mathcal{R}$ \dotfill & amplitudes which need renormalization, used as an
    index set \\
    $\mathbb{R}$ \dotfill & the real numbers \\
    $R$ \dotfill & the map from Feynman graphs to regularized Feynman
    integrals \\
    $r_{k,i;r}$ \dotfill & residue of $\rho F^r_{k, i}(\rho)$, especially after
    reducing to geometric series \\
    $r_{k,i}$ \dotfill & residue of $\rho F_{k, i}(\rho)$, especially after
    reducing to geometric series \\
    $\star$ \dotfill & the convolution product of functions on $\h$ \\
    $S$ \dotfill & the antipode of $\h$ \\
    $s_r$ \dotfill & the power of $X^r$ in $Q^{-1}$ \\
    $s$ \dotfill & the power of $X$ in $Q^{-1}$ in the case with only one $r$\\
    $\sgn(s)$ \dotfill& the sign of the real number $s$ \\
    $T$ \dotfill & a combinatorial physical theory \\
    $t^r_k$ \dotfill & upper bound for the index over primitives at $k$ loops
    with external leg structure $r$ \\
    $t_k$ \dotfill & upper bound for the index over primitives at $k$
    loops in the case with only one $r$\\
    $V$ \dotfill & a vertex type viewed as a set of half edge types \\
    $\xi$ \dotfill & a gauge variable \\
    $[x^n]$ \dotfill & the coefficient of $x^n$ operator \\
    $x$ \dotfill & the coupling constant used as an indeterminate in series
    with coefficients in $\h$ and used as one of the variables on
    which the Green functions depend \\
    $|X|_\vee$ \dotfill & the number of distinct graphs
    obtainable by permuting the external edges of $X$ \\
    $X^r(x)$ \dotfill & sum of all graphs with external leg structure $r$, as a
    series in the coupling constant $x$ \\
    $X$ \dotfill & $X^r$ in the case with only one
    $r$
 \end{longtable}

% END OF THE PRELIMINARY PAGES
%\newpage
%\input{0_Prelim/quote}

\newpage
%\endofprelim

\chapter{Introduction}

Dyson-Schwinger equations are integral equations in quantum field
theory that describe the Green functions of a theory and mirror the
recursive decomposition of Feynman diagrams 
into subdiagrams.  Taken as
recursive equations, the Dyson-Schwinger equations describe
perturbative quantum field theory, while as integral equations they also contain
non-perturbative information.  

Dyson-Schwinger equations
have a number of
nice features.  Their recursive nature gives them a strong combinatorial
flavor, they tie Feynman diagrams and the rest of perturbation theory
to non-perturbative quantum field theory, and on occasion they can be
solved, for example \cite{bkerfc}.  However, in general they are complicated and 
difficult to extract information from.

The goal of the present work is to show how the Dyson-Schwinger equations
for a physical theory can be transformed into the more manageable
system of equations
\begin{equation}\label{intro system}
  \gamma_1^r(x) = P_r(x) -\sgn(s_r) \gamma_1^r(x)^2 + \left(\sum_{j \in \mathcal{R}}|s_j|\gamma_1^j(x)\right) x \partial_x \gamma_1^r(x)
\end{equation}
where $r$ runs over $\mathcal{R}$, the amplitudes which need
renormalization in the theory, $x$ is the coupling constant,
$\gamma_1^r(x)$ is the anomalous dimension for $r$, and $P_r(x)$ is
a modified version of the function of the primitive
skeletons contributing to $r$, see Chapter \ref{second recursion} for details.

Chapter \ref{background} discusses the general background with a
focus on definitions and examples rather than proofs.  The
approach taken is that Feynman graphs are the primary objects.  In an
attempt to make matters immediately accessible to a wide range of
mathematicians and to accentuate the combinatorial flavor, the
physics itself is mostly glossed over.  Readers with a physics
background may prefer to skip this chapter and refer to existing
surveys, such as \cite{e-fksurvey}, for the Hopf algebra of Feynman graphs.

Chapter \ref{DSE} discusses the more specific background and setup
for Dyson-Schwinger equations and the insertion operators $B_+$ on Feynman
graphs.  Proofs are again primarily left to other sources.
\cite{bergk} covers combinatorially similar material 
for rooted trees.  Some important subtleties concerning $B_+$ for Feynman
diagrams are discussed in more detail in \cite{anatomy} with important
results proved in \cite{vS}.  The approach
to disentangling the analytic and combinatorial information comes from
\cite{etude}.  This chapter leaves us with the following input to the
upcoming analysis: combinatorial Dyson-Schwinger equations and
a Mellin transform for each connected, divergent, primitive graph.  The former consists of
recursive equations at the level of Feynman graphs with the same 
structure as the original analytic Dyson-Schwinger equations.  The
latter contains all the analytic 
information. 

The next four chapters derive \eqref{intro system} expanding upon the
discussion in \cite{radii}.  Chapter \ref{first recursion} derives a
preliminary recursive equation in two different ways, first from the
renormalization group equation, and second from the Connes-Kreimer
scattering-type formula \cite{ckII}.  Chapter \ref{first reduction}
reduces to the case of single variable Mellin transforms and a single
external scale.  The Mellin
transform variables correspond to the different insertion places in
the graph, so we refer to this as the single insertion place case,
though this is only literally true for simple examples.  The cost of
this reduction is that we are forced to consider non-connected
primitive elements in the Hopf algebra.  Chapter \ref{second
reduction} reduces to the case where all Mellin transforms are
geometric series to first order in the scale parameters by exchanging unwanted powers of the Mellin transform
variable for a given primitive with lower powers of the variable for a
primitive with a larger loop number, that is, with a larger number of
independent cycles.  The cost of this reduction is that we lose some
control over the residues of the primitive graphs.  Chapter \ref{second recursion} applies the
previous chapters to derive \eqref{intro system}.

Chapter \ref{radii} considers \eqref{intro system} as a system of
recursions.  It is devoted to the result of \cite{radii} where we
bound the
radii of convergence of the Borel transforms of the $\gamma_1^r$ in terms
of those of $P_r$.  For systems with nonnegative coefficients we
determine the radius exactly as ${\min}\{\rho_r,1/b_1\}$,
where $\rho_r$ is the radius of the Borel transform of $P_r$, the instanton
radius, and $b_1$ the first coefficient of the
$\beta$-function\footnote{This is the physicists' $\beta$-function,
  see Section \ref{rg section}, not the Euler $\beta$ function.}.  In
particular this means that a Lipatov bound\footnote{A Lipatov bound
  for $\sum d_n n^k$ means that $|d_n| \leq c^n n!$ for some $c$.} for the superficially
convergent Green functions leads to a Lipatov bound for the
superficially divergent Green functions.  This generalizes and
mathematizes similar results obtained in particular cases, such as
$\phi^4$, through quite different means
by constructive field theory \cite{Riv2}.  Both approaches require
estimates on the 
convergent Green functions which can also be obtained in some cases from
constructive field theory, for example \cite{Riv}.  

Chapter \ref{dechapter} considers \eqref{intro system} as a system of
differential equations.  We are not able to prove any non-trivial
results, and so simply discuss some tantalizing features of vector field plots
of some important examples.  More substantial results will appear in
\cite{vBKUY}. 

\chapter{Background}\label{background}

\section{Series}

\begin{definition}
If $\{a_n\}_{n \geq 0}$ is a sequence then $\mathbf{A}(x) = \sum_{n
  \geq 0}a_n x^n$ is its (ordinary) generating function and
$\sum_{n\geq 0}a_n x^n/n!$ is its exponential generating function.
\end{definition}  
Bold capital letters are used for the ordinary generating function for the sequence
  denoted by the corresponding lower case letters. 
  $\rho$ will often denote a radius of convergence.

We will make use of the standard combinatorial notation for extracting coefficients.
\begin{definition}\label{coefficient of}
  If $\mathbf{A}(x) = \sum_{n
  \geq 0}a_n x^n$ then $[x^n]\mathbf{A}(x) = a_n$.
\end{definition}

\begin{definition}
Call a power series $\sum_{k \geq 0} a(k) x^k$ \emph{Gevrey-$n$} if
$\sum_{k \geq 0} x^k a(k)/(k!)^n$ has nonzero radius of convergence.
\end{definition}

For example, a convergent power series is Gevrey-$0$ and $\sum_{k \geq
  0} (xk)^k$ is Gevrey-$1$ due to Stirling's formula.
Trivially, a series which is Gevrey-$n$ is also Gevrey-$m$ for all $m \geq n$.

Gevrey-$1$ series are important in perturbative quantum field theory
since being Gevrey-$1$ is necessary (but not sufficient) for Borel
resummation.  Resummation and resurgence are an enormous topic which
will not be touched
further herein; one entry point is
\cite{stingl}.  
Generally very little is known about
the growth rates of 
the series appearing in perturbation theory.  They are usually
thought to be divergent, though this is questioned by some
\cite{cfinite}, and hoped to be Borel resummable.

\section{Feynman graphs as combinatorial objects}
Feynman graphs are graphs, with multiple edges and self loops
permitted, made from a specified set of edge types, which may include both
directed and undirected edges, with a specified set of permissible
edge types which can meet at any given vertex.  Additionally there are
so-called external edges, weights for calculating the degree of
divergence, and there may be additional colorings or orderings as necessary.

There are many possible ways to set up the foundational definitions,
each with sufficient power to fully capture all aspects of the
combinatorial side of Feynman graphs. 
However it is worth picking a setup which is as clean and natural as
possible. 

For the purposes of this thesis graphs are formed out of half edges.
This naturally accounts for external edges and symmetry factors and
permits oriented and unoriented edges to be put on the same footing.

\begin{definition} A \emph{graph} consists of a set $H$ of half edges,
  a set $V$ of vertices, a set of vertex - half edge adjacency
  relations ($\subseteq V \times H$), and a set of half edge - half
  edge adjacency relations ($\subseteq H \times H$), with the
  requirements that each half edge is adjacent to at most one other
  half edge and to exactly one vertex.
\end{definition}

Graphs are considered up to isomorphism.

\begin{definition}
Half edges which are not
adjacent to another half edge are called \emph{external edges}.  Pairs
of adjacent half edges are called \emph{internal edges}.
\end{definition}

\begin{definition} A \emph{half edge labelling} of a graph with half
  edge set $H$ is a bijection $H \rightarrow \{1,2,\ldots, |H|\}$.  A
  graph with a half edge labelling is called a \emph{half edge
    labelled graph}.
\end{definition}

\subsection{Combinatorial physical theories}

Feynman graphs will be graphs with extra information and requirement.  
In order to define this extra structure we
need to isolate the combinatorial information that the
physical theory, such as quantum electrodynamics (QED), scalar
$\phi^4$, or quantum chromodynamics (QCD), requires of the graph.

Each edge in the graph corresponds to a particle and a given physical
theory describes only certain classes of
particles, hence the physical theory determines a finite set of permissible
\emph{edge types}.  For our half edge based setup, an edge type $E$
consists of two, not
necessarily distinct, \emph{half edge types}, with the restriction
that each half edge type appears in exactly one edge type.  An edge composed of two
adjacent half edges, one of each half edge type in $E$, is then an edge of type
$E$.  An edge type made up of the same half edge type twice is called an
\emph{unoriented edge type}.  An edge type made up of two distinct
half edge types is called an \emph{oriented edge type}.
The half edge types themselves contain no further structure and
thus can be identified with $\{1, \ldots, n\}$ for appropriate $n$.

\begin{fmffile}{simpleegs}
\setlength{\unitlength}{0.05mm}
\fmfset{arrow_len}{2mm}
\fmfset{wiggly_len}{2mm}
\fmfset{thin}{0.5pt}
\fmfset{curly_len}{1.5mm}
\fmfset{dot_len}{1mm}
\fmfset{dash_len}{1.5mm}

For example in QED there are two edge
types, an unoriented edge type, 
{\begin{fmfgraph}(100,60)
    \fmfleft{i}
    \fmfright{o}
    \fmf{photon}{i,o}
\end{fmfgraph}}, representing a photon, and an oriented edge type,
{\begin{fmfgraph}(100,60)
    \fmfleft{i}
    \fmfright{o}
    \fmf{fermion}{i,o}
\end{fmfgraph}}, representing an electron or positron\footnote{If we
    chose a way for time to flow through the graph then the edge would
    represent an electron or positron depending on whether it was
    oriented in the direction of time or not.  However part of the
    beauty of Feynman graphs is that both combinatorially and
    analytically they do not depend on a flow of time.}.  At the level
  of half edge types we thus have a half photon, a front half
  electron, and a back half electron.

Each vertex in the graph corresponds to an interaction of particles
and only certain interactions are permitted in a given physical
theory, hence the physical theory also determines a set of permissible
\emph{vertex types}.  A vertex type $V$ consists of a
multiset of half edge types with $3 \leq |V| < \infty$.  A vertex in a graph which is
adjacent to half edges one of each half edge type in $V$ is then a
vertex of type $V$.  For example in QED there is one type of vertex, 
%\[
\raisebox{-0.5ex}{\begin{fmfgraph}(100,100)
    \fmfleft{i}
    \fmfrightn{o}{2}
    \fmf{photon}{i,v}
    \fmf{fermion}{o1,v,o2}
\end{fmfgraph}}.
%\]

The physical theory determines a formal integral expression for
each graph by associating a
factor in the integrand to each edge and vertex according to their
type.  This map is called the Feynman rules, see subsection \ref{feynman rules}.  On the
combinatorial side the only part of the Feynman rules we need is the
net degree of the integration variables appearing in the factor of
the integrand associated to each type.  Traditionally this degree is
taken with a negative sign; specifically for a factor $N/D$ this net
degree is $\deg(D) - \deg(N)$, which we call the  \emph{power counting
  weight} of this vertex type or edge type.

The other thing needed in order to determine the divergence or convergence of these
integrals at large values of the integration variables, which will be discussed further in subsection
\ref{divergence}, is the \emph{dimension of space
  time}.  We 
are not doing anything sophisticated here and this value will be a
nonnegative integer, 4 for most theories of interest.

Thus we define,
\begin{definition}
  A \emph{combinatorial physical theory} $T$ consists of a set of half
  edge types, a set of edge
  types with associated power counting weights, a set of vertex types
  with associated power counting weights, and a nonnegative integer
  dimension of space-time.
\end{definition}

More typically the dimension of space-time is not included in the definition
of the theory, and so one would say a theory $T$ in dimension $D$ to
specify what we have called a physical theory.

Our examples will come from five theories

\begin{example}
  QED describes photons and electrons interacting electromagnetically.
  As a combinatorial physical theory it has 3 half-edge 
  types, a half-photon, a front half-electron, and a back
  half-electron.  This
  leads to two edge types a photon, 
  {\begin{fmfgraph}(100,60)
    \fmfleft{i}
    \fmfright{o}
    \fmf{photon}{i,o}
\end{fmfgraph}}, with weight 2, and an electron, {\begin{fmfgraph}(100,60)
    \fmfleft{i}
    \fmfright{o}
    \fmf{fermion}{i,o}
\end{fmfgraph}}, with weight 1.  There is only one vertex consisting
    of one of each half-edge type and with weight 0.  The dimension of
    space-time is 4.
\end{example}

\begin{example}
Quantum chromodynamics (QCD) is the theory of the interactions of
quarks and gluons.  As a combinatorial physical
theory it has 5 half-edge types, a half-gluon, a front half-fermion, a back
half-fermion, a front half-ghost, and a back half-ghost.  There are 3
edge types and 4 vertex types with weights as described in Table
\ref{qcd weights}. The dimension of space-time is again 4.

\begin{table}
\begin{center}
\begin{tabular}{lll}
name & graph & weight \\
\hline
gluon & 
\raisebox{-1ex}{\begin{fmfgraph}(200,120)
    \fmfleft{i}
    \fmfright{o}
    \fmf{gluon}{i,o}
\end{fmfgraph}}
& 2 \\
fermion &
\raisebox{-1ex}{\begin{fmfgraph}(200,120)
    \fmfleft{i}
    \fmfright{o}
    \fmf{fermion}{i,o}
\end{fmfgraph}}
& 1 \\
ghost &
\raisebox{-1ex}{\begin{fmfgraph}(200,120)
    \fmfleft{i}
    \fmfright{o}
    \fmf{ghost}{i,o}
\end{fmfgraph}}
& 1 \\
& \raisebox{-1ex}{\begin{fmfgraph}(200,200)
    \fmfleft{i}
    \fmfrightn{o}{2}
    \fmf{gluon}{i,v}
    \fmf{fermion}{o1,v,o2}
\end{fmfgraph}}
& 0 \\
& \raisebox{-1ex}{\begin{fmfgraph}(200,200)
    \fmfleft{i}
    \fmfrightn{o}{2}
    \fmf{gluon}{i,v}
    \fmf{ghost}{o1,v,o2}
\end{fmfgraph}}
& 0 \\
& \raisebox{-1ex}{\begin{fmfgraph}(200,200)
    \fmfleft{i}
    \fmfrightn{o}{2}
    \fmf{gluon}{i,v}
    \fmf{gluon}{o1,v,o2}
\end{fmfgraph}}
& -1 \\
& \raisebox{-1ex}{\begin{fmfgraph}(200,200)
    \fmfleftn{i}{2}
    \fmfrightn{o}{2}
    \fmf{gluon}{i1,v,i2}
    \fmf{gluon}{o1,v,o2}
\end{fmfgraph}}
& 0 
\end{tabular}
\end{center}
\caption{Edge and vertex types in QCD with power counting weights}
\label{qcd weights}
\end{table}

\end{example}

\begin{example}
  $\phi^4$, a scalar field theory, is the arguably the simplest renormalizable quantum field
  theory and is often used as an example in quantum field theory
  textbooks.  As a combinatorial theory it consists of one half-edge
  type, one edge type, 
  {\begin{fmfgraph}(100,60)
      \fmfleft{i}
      \fmfright{o}
      \fmf{plain}{i,o}
\end{fmfgraph}}, with weight 2, one vertex type,
  {\begin{fmfgraph}(100,60)
      \fmfleftn{i}{2}
      \fmfrightn{o}{2}
      \fmf{plain}{i1,v,i2}
      \fmf{plain}{o1,v,o2}
    \end{fmfgraph}}, with weight 0, and space-time dimension 4.
\end{example}

\begin{example}
$\phi^3$, also a scalar field theory, is another candidate for the
  simplest renormalizable quantum field theory.  It is not as
  physical since to be renormalizable the dimension of space-time must
  be 6, and hence it is not as pedagogically popular.  However the
  Feynman graphs in $\phi^3$ are a little simpler in some respects and
  so it will be used here in longer examples such as Example \ref{first reduction eg}.  $\phi^3$ consists of half-edges and edges as
  in $\phi^4$ but the single vertex type, which has weight 0, is 3-valent.
\end{example}

\begin{example}
  The final physical theory which we will use for examples is Yukawa
  theory in 4 dimensions, which has 3 half-edge types, a half-meson
  edge, a front half-fermion edge, and a back half-fermion edge.  The
  edge types are a meson edge,   
  {\begin{fmfgraph}(100,60)
      \fmfleft{i}
      \fmfright{o}
      \fmf{dashes}{i,o}
\end{fmfgraph}}, with weight 2 and a fermion edge
  {\begin{fmfgraph}(100,60)
      \fmfleft{i}
      \fmfright{o}
      \fmf{fermion}{i,o}
\end{fmfgraph}}, with weight 1.  There is one vertex type,
  \raisebox{-1ex}{\begin{fmfgraph}(100,100)
      \fmfleft{i}
      \fmfrightn{o}{2}
      \fmf{dashes}{i,v}
      \fmf{fermion}{o1,v,o2}
      \end{fmfgraph}}, with weight 0.
This example arises for us because of \cite{bkerfc}.
\end{example}

\end{fmffile}

\subsection{Feynman graphs}
\begin{fmffile}{basicfeynmanegs}
\setlength{\unitlength}{0.05mm}
\fmfset{arrow_len}{2mm}
\fmfset{wiggly_len}{2mm}
\fmfset{thin}{0.5pt}
\fmfset{curly_len}{1.5mm}
\fmfset{dot_len}{1mm}
\fmfset{dash_len}{1.5mm}

Notice that given a graph $G$, a combinatorial physical theory $T$,
and a map from the half edge of $G$ to the half edge types of $T$,
there is at most one induced map from the internal edges of $G$ to the
edge types of $T$ and at most one induced map from the vertices of $G$
to the vertex types of $T$.  Thus we can make the following definition.

\begin{definition}
  A \emph{Feynman graph} in a combinatorial physical theory $T$ is 
  \begin{itemize}
    \item a graph $G$,
    \item a map from the half edges of $G$ to the
      half edge types of $T$ which is compatible with the edges and
      vertices of $G$ in the sense that it induces a map from the internal
      edges of $G$ to the edge types of $T$ and induces a map from the
      vertices of $G$ to the vertex types of $T$, and
    \item a bijection from the external edges of $G$ to $\{1,\ldots n\}$
      where $n$ is the number of external edges.
  \end{itemize}
\end{definition}

The final point serves to fix the external edges of $G$, which is
traditional among physicists. 
%A homomorphism of Feynman graphs is a homomorphism of the underlying
%graph which is also compatible with the additional structure.

\begin{lemma}\label{aut}
  Let $G$ be a connected Feynman graph with $n$ half edges.  Let $m$
  be the number of half edge labelled 
  Feynman graphs (up to isomorphism as labelled Feynman graphs) giving $G$ upon forgetting the labelling, and let
  $\mathrm{Aut}$ be the automorphism group of $G$.  Then
  \[
    \frac{m}{n!} = \frac{1}{|\mathrm{Aut}|}
  \]
\end{lemma}

\begin{proof}
  $\mathrm{Aut}$ acts freely on the $n!$ half edge labellings of $G$.
  The orbits are the $m$ isomorphism classes of half edge labellings.
  The result follows by elementary group theory.
  %Every automorphism of $G$ permutes the half
  %edges, with distinct automorphisms determining distinct permutations
  %since isolated vertices are not allowed.  Also due to the absence
  %of isolated vertices distinct permutations determine distinct
  %automorphisms of $G$.  The result then follows from Lagrange's
  %theorem of elementary group theory. 
\end{proof}

The primary consequence of Lemma \ref{aut} is that the exponential
generating function for half-edge labelled graphs is identical to the
generating function for Feynman graphs weighted with
$1/|\mathrm{Aut}|$.  $1/|\mathrm{Aut}|$ is known as the symmetry factor
of the graph.  Table \ref{symmetry factors} gives some examples.

\begin{table}
\begin{center}
\begin{tabular}{ll}
graph & symmetry factor \\
\hline
\raisebox{-1ex}{\begin{fmfgraph}(200,120)
    \fmfleft{i}
    \fmfright{o}
    \fmf{plain}{i,v1}
    \fmf{plain}{v2,o}
    \fmf{plain,left,tension=0.25}{v1,v2,v1}
\end{fmfgraph}} & $\frac{1}{2}$ \\
\raisebox{-1ex}{\begin{fmfgraph}(200,120)
    \fmfleft{i}
    \fmfright{o}
    \fmf{photon}{i,v1}
    \fmf{photon}{v2,o}
    \fmf{fermion,left,tension=0.25}{v1,v2,v1}
\end{fmfgraph}} & $1$ \\
\raisebox{-1ex}{\begin{fmfgraph}(200,120)
    \fmfleftn{i}{2}
    \fmfrightn{o}{2}
    \fmf{plain}{i1,v1}
    \fmf{plain}{i2,v1}
    \fmf{plain,left,tension=0.25}{v1,v2,v1}
    \fmf{plain,left,tension=0.25}{v2,v3,v2}
    \fmf{plain}{v3,o1}
    \fmf{plain}{v3,o2}
  \end{fmfgraph}} & $\frac{1}{4}$ \\
\raisebox{-1ex}{\begin{fmfgraph}(200,120)
    \fmfleft{i}
    \fmfright{o}
    \fmf{plain}{i,v,o}
    \fmf{plain,right}{v,v}
\end{fmfgraph}} & $\frac{1}{2}$ \\
\raisebox{-1ex}{\begin{fmfgraph}(200,120)
    \fmfleft{i}
    \fmfright{o}
    \fmf{plain}{i,v1}
    \fmf{plain}{v2,o}
    \fmf{plain,left,tension=0.25}{v1,v2,v1}
    \fmffreeze
    \fmf{plain}{v1,v2}
  \end{fmfgraph}} & $\frac{1}{6}$
\end{tabular}
\end{center}
\caption{Examples of symmetry factors}\label{symmetry factors}
\end{table}

We will be concerned from now on with Feynman graphs which are
connected and which 
remain connected upon removal of any one internal edge, a property which
physicists call \emph{one particle irreducible} (1PI) and which
combinatorialists call \emph{2-edge connected}.  Another way to look
at this definition is that a 1PI graph is a unions of cycles and
external edges.  We'll generally be interested
in Feynman graphs with each connected component 1PI.

\subsection{Operations}\label{operations subsec}

For us subgraphs are always \emph{full} in the
sense that all half edges adjacent to a vertex in a subgraph must
themselves be in the subgraph.  

The most important operations are contraction of subgraphs and
insertion of graphs.  
To set these definitions up cleanly we need a
preliminary definition.

\begin{definition}
  The set of external edges of a connected Feynman graph is called the \emph{external
    leg structure} of the Feynman graph.  The set of half edge types
  associated to the external edges of a Feynman graph can be
  identified with at most one edge or vertex type.  This edge or
  vertex type, if it exists, is also called the \emph{external leg structure}.
\end{definition}

\begin{definition}
  Let $G$ be a Feynman graph in a theory $T$, $\gamma$ a connected
  subgraph with external leg structure a vertex type $V$.  Then the
  \emph{contraction} of $\gamma$, denoted $G/\gamma$ is the Feynman
  graph in $T$ with
  \begin{itemize}
    \item vertex set the vertex set of $G$ with all vertices of
      $\gamma$ removed and a new vertex $v$ of type $V$ added,
    \item half edge set the half edge set of $G$ with all half edges
      corresponding to internal edges of $\gamma$ removed,
  \end{itemize}
  and with adjacencies induced from $G$ along with the adjacency of
  the external edges of $\gamma$ with $v$.
\end{definition}

\begin{definition}
  Let $G$ be a Feynman graph in a theory $T$, $\gamma$ a connected
  subgraph with external leg structure an edge type $E$.  Then the
  \emph{contraction} of $\gamma$, denoted $G/\gamma$ is the Feynman
  graph in $T$ with
  \begin{itemize}
    \item vertex set the vertex set of $G$ with all vertices of
      $\gamma$ removed,
    \item half edge set the half edge set of $G$ with all the half edges
      of $\gamma$ removed,
  \end{itemize}
  and with the induced adjacencies from $G$ along with the adjacency
  of the two half edges adjacent to the external edges of $\gamma$ if
  they exist.
\end{definition}

\begin{definition}
  Let $G$ be a Feynman graph in a theory $T$, $\gamma$ a not
  necessarily connected 
  subgraph with the external leg structure of each connected component
  an edge or vertex type in $T$.  Then the \emph{contraction} of
  $\gamma$, also denoted 
  $G/\gamma$ is the graph resulting from contracting each connected
  component of $\gamma$. 
\end{definition}

For example in QED 
\[
  \raisebox{-2ex}{\begin{fmfgraph}(400,180)
\fmfleftn{i}{1}
\fmfrightn{o}{1}
\fmf{photon}{i1,in1}
\fmf{fermion}{in2,in4,in3,in1}
\fmf{photon}{in2,o1}
\fmf{fermion,left,tension=0.25}{in1,in2}
\fmf{photon,left,tension=0.25}{in3,in4}
\end{fmfgraph}} 
/  \raisebox{-2ex}{\begin{fmfgraph}(300,180)
\fmfleftn{i}{1}
\fmfrightn{o}{1}
\fmf{fermion}{o1,in2,in1,i1}
\fmf{photon,left,tension=0.25}{in1,in2}
\end{fmfgraph}}
%\leadsto
%  \raisebox{-2ex}{\begin{fmfgraph}(300,180)
%\fmfleftn{i}{1}
%\fmfrightn{o}{1}
%\fmf{photon}{i1,in1}
%\fmf{fermion}{in2,in3,in1}
%\fmf{photon}{in2,o1}
%\fmf{fermion,left,tension=0.25}{in1,in2}
%\fmf{photon,left,tension=0.4}{in3,in3}
%\end{fmfgraph}}
%\leadsto
%  \raisebox{-2ex}{\begin{fmfgraph}(300,180)
%\fmfleftn{i}{1}
%\fmfrightn{o}{1}
%\fmf{photon}{i1,in1}
%\fmf{fermion}{in2,in3,in1}
%\fmf{photon}{in2,o1}
%\fmf{fermion,left,tension=0.25}{in1,in2}
%\fmfdot{in3}
%\end{fmfgraph}}
%\leadsto
=   \raisebox{-2ex}{\begin{fmfgraph}(300,180)
\fmfleftn{i}{1}
\fmfrightn{o}{1}
\fmf{photon}{i1,in1}
\fmf{photon}{in2,o1}
\fmf{fermion,left,tension=0.25}{in1,in2,in1}
\end{fmfgraph}}
\]  

%In all of our example theories there are no 2-valent vertices and so
%we need not specify $f$ if we simply follow the rule of popping off
%all newly created 2-valent vertices.   The external edges of a graph,
%particularly when viewed as a vertex or edge type in this way, are
%called the \emph{external leg structure} of the graph.
%We will approach the question of characterizing which graphs allow
%contraction to remain in the theory through the idea of
%divergence in subsection \ref{divergence}.   

Also useful is the operation of inserting a subgraph, which is the
opposite of contracting a subgraph.

\begin{definition}
  Let $G$ and $\gamma$ be Feynman graphs in a theory $T$ with $\gamma$
  connected.  Suppose $\gamma$ has
  external leg structure a vertex type and let $v$ be a vertex of $G$
  of the same type.  Let $f$ be a bijection from the external edges of
  $\gamma$ to 
  the half edges adjacent to $v$ preserving half edge type.  Then $G
  \circ_{v,f} \gamma$ is the 
  graph consisting of
  \begin{itemize}
  \item the vertices of $G$ except for $v$, disjoint union with the
    vertices of $\gamma$,
  \item the half edges of $G$ and those of $\gamma$ with the
    identifications given by $f$,
  \end{itemize}
  with the induced adjacencies from $G$ and $\gamma$.
\end{definition}

\begin{definition}
  Let $G$ and $\gamma$ be Feynman graphs in a theory $T$ with $\gamma$
  connected.  Suppose $\gamma$ has
  external leg structure an edge type and let $e$ be an edge of $G$ of
  the same type.  Let $f$ be
  a bijection from the external edges of $\gamma$ to the half edges
  composing $e$, such that if $a$ is an external edge of $G$ then $(a,
  f(a))$ is a permissible half edge - half edge adjacency.   Then $G
  \circ_{e,f} \gamma$ is the 
  graph consisting of
  \begin{itemize}
  \item the vertices of $G$ disjoint union with the
    vertices of $\gamma$,
  \item the half edges of $G$ disjoint union with those of $\gamma$,
  \end{itemize}
  with the adjacency of $a$ and $f(a)$ for each external edge $a$ of
  $\gamma$ along with the induced adjacencies from $G$ and $\gamma$.
\end{definition}

The vertices and edges of $G$ viewed as above are called \emph{insertion places}.

For example if 
\[
G = \gamma = \raisebox{-3ex}{\begin{fmfgraph}(400,240)
    \fmfleft{i}
    \fmfright{o}
    \fmf{fermion}{i,v1,v2,o}
    \fmf{photon,left,tension=0.25}{v1,v2}
\end{fmfgraph}}
\]
then there is only one possible insertion place for $\gamma$ in $G$,
namely the bottom internal edge $e$ of $G$, and there is only one possible
map $f$.  Thus
\[
G \circ_{e,f} \gamma = \raisebox{-3ex}{\begin{fmfgraph}(400,240)
    \fmfleft{i}
    \fmfright{o}
    \fmf{fermion}{i,v1,v3,v4,v2,o}
    \fmf{photon,left,tension=0.25}{v1,v2}
    \fmf{photon,left,tension=0.25}{v3,v4}
\end{fmfgraph}}.
\]
On the other hand if 
\[
G = \gamma = \raisebox{-1ex}{\begin{fmfgraph}(200,120)
    \fmfleftn{i}{2}
    \fmfrightn{o}{2}
    \fmf{plain}{i1,v1}
    \fmf{plain}{i2,v1}
    \fmf{plain,left,tension=0.25}{v1,v2,v1}
     \fmf{plain}{v2,o1}
    \fmf{plain}{v2,o2}
  \end{fmfgraph}} 
\]
then there are 2 possible insertion places for $\gamma$ in $G$,
namely the right vertex and the left vertex.  Let $e$ be the left
vertex.  Then there are also $4!$ possibilities for $f$, however 8 of
them give
\[
G \circ_{f,g} \gamma = \raisebox{-1ex}{\begin{fmfgraph}(200,120)
    \fmfleftn{i}{2}
    \fmfrightn{o}{2}
    \fmf{plain}{i1,v1}
    \fmf{plain}{i2,v1}
    \fmf{plain,left,tension=0.25}{v1,v2,v1}
    \fmf{plain,left,tension=0.25}{v2,v3,v2}
    \fmf{plain}{v3,o1}
    \fmf{plain}{v3,o2}
  \end{fmfgraph}} 
\]
and 16 of them give
\[
G \circ_{f,g} \gamma = \raisebox{-1ex}{\begin{fmfgraph}(200,120)
    \fmfleftn{i}{2}
    \fmfrightn{o}{2}
    \fmf{plain}{i1,v1}
    \fmf{plain}{i2,v2}
    \fmf{plain,left,tension=0.25}{v1,v2,v1}
    \fmf{plain}{v1,v3}
    \fmf{plain}{v2,v3}
     \fmf{plain}{v3,o1}
    \fmf{plain}{v3,o2}
  \end{fmfgraph}}.
\]

\begin{prop}\label{1PI to 1PI}
  \begin{enumerate}
    \item Contracting any subgraph $\gamma$ of a 1PI graph $G$ results in a 1PI graph.  
    \item Inserting a 1PI graph $\gamma$ into a 1PI graph $G$ results in a 1PI
      graph.
    \end{enumerate}
\end{prop}

\begin{proof}
  \begin{enumerate}
    \item Without loss of generality suppose $\gamma$ is connected.
      Suppose the result does not hold and $e$ is an internal edge
      in $\Gamma = G / \gamma$ which disconnects $\Gamma$ upon
      removal.  Since $G$ is 1PI, $e$ cannot be an internal edge of
      $G$ and hence must be the insertion place for $\gamma$ in
      $\Gamma$.  However then removing either half edge of $e$ from
      $G$ would disconnect $G$ which is also impossible.
    \item Suppose $e$ is an internal edge in
      $\Gamma = G \circ \gamma$.  Removing $e$ removes at least one internal half edge
      of $G$ or of $\gamma$ which cannot disconnect either since both
      are themselves 1PI, and hence cannot disconnect $\Gamma$.
  \end{enumerate}
\end{proof}

\subsection{Divergence}\label{divergence}

For a 1PI Feynman graph $G$ and a physical theory $T$ let $w(a)$ be the power
counting weight of $a$ where $a$ is an edge or a vertex of $G$ and let
$D$ be the dimension of space-time.  Then 
the \emph{superficial degree of divergence} is
\[
  D \ell - \sum_{e} w(e) - \sum_{v} w(v)
\]
where $\ell$ is the loop number of the graph, that is, the number of
independent cycles.  If the superficial degree of divergence of a
graph is nonnegative we say the graph is \emph{divergent}.  It is the
divergent graphs and subgraphs which we are primarily interested in.

The notion of superficial divergence comes from the fact that the Feynman rules
associate to a graph a formal integral, as will be explained in
subsection \ref{feynman rules}; the 
corresponding weights $w(a)$ give the degree in the integration variables of
the inverse of each factor of the integrand, while the loop number
$\ell$ gives the
number of independent 
integration variables, each running over $\mathbb{R}^D$.  Thus the
superficial degree of divergence encodes how badly the integral
associated to the graph diverges for large values of the integration
variables.  The adjective \emph{superficial} refers to the fact that
the integral may have different, potentially worse, behavior when
some subset of the integration variables are large, hence the
importance of divergent subgraphs.

In this context we say a theory $T$ (in a given dimension) is
\emph{renormalizable} if graph 
insertion within $T$ does not change the superficial degree of
divergence of the 
graph.  
%The theory in the given dimension is
%\emph{superrenormalizable} if insertion decreases the superficial
%degree of divergence and \emph{unrenormalizable} if insertion
%increases it.  
%Another way to look at this is that in a
%renormalizable theory the superficial degree of divergence of a graph
%only depends on the external edges of a graph.

A theory being renormalizable means more than that the integrals
associated to the graphs of the theory can be renormalized in the
sense of Subsection \ref{renormalization}.  In fact even if insertion
increases the superficial degree of divergence, and so the theory is
called \emph{unrenormalizable}, the individual graphs can typically
still be renormalized.  Rather, a theory being renormalizable refers
to the fact that the theory as a whole can be renormalized, all of
its graphs at all loop orders, without
introducing more than finitely many new parameters.  Combinatorially
this translates into the fact that there are finitely many families of
divergent graphs, typically indexed by external leg structures.  In
the unrenormalizable case by contrast there are infinitely many
families of divergent graphs and, correspondingly, to renormalize the
whole theory would require infinitely many new parameters.  

The interplay of renormalizability and dimension explains our choices
for the dimension of space-time in our examples.  In particular
$\phi^4$, QED, and QCD are all 
renormalizable in 4 dimensions and $\phi^3$ is renormalizable in 6 dimensions.

By viewing a divergent
graph in terms of its divergent subgraphs we see a structural
self-similarity.  This insight leads to the recursive equations which
are the primary object of interest in this thesis.  

Another useful definition is
\begin{definition}
  Suppose $G$ is a Feynman graph and $\gamma$ and $\tau$ are divergent
  subgraphs.  Then $\gamma$ and $\tau$ are \emph{overlapping} if they
  have internal edges or vertices in common, but neither contains the other. 
\end{definition}

\end{fmffile}
\subsection{The Hopf algebra of Feynman graphs}
The algebra structure on divergent 1PI Feynman graphs in a given theory is reasonably simple.

\begin{fmffile}{hopfegs}
\setlength{\unitlength}{0.1mm}
\fmfset{arrow_len}{2mm}
\fmfset{wiggly_len}{2mm}
\fmfset{thin}{0.5pt}
\fmfset{curly_len}{1.5mm}
\fmfset{dot_len}{1mm}
\fmfset{dash_len}{1.5mm}

\begin{definition}
Let $\h$ be the vector space formed by the
$\mathbb{Q}$ span of disjoint unions of divergent 1PI Feynman graphs
including
the empty graph denoted $\mathbb{I}$.
\end{definition}
 
\begin{prop}
$\h$ has an algebra structure where multiplication $m:\h\otimes \h \rightarrow
\h$ is given by disjoint union and the unit by $\mathbb{I}$.
\end{prop}

\begin{proof}
This multiplication can immediately be checked to be commutative and
associative with unit $\mathbb{I}$, and to be a linear map. 
\end{proof} 
Another way to look at this is that as an algebra $\h$ is the
polynomial algebra over $\mathbb{Q}$ in divergent 1PI Feynman graphs with the multiplication
viewed as disjoint union.  Note that we are only considering one graph
with no cycles (the empty graph $\mathbb{I}$); from the physical
perspective this means we are normalizing all the tree-level graphs to $1$.

We will
use the notation $e:\mathbb{Q} \rightarrow \h$ for the unit map $e(q) = q\mathbb{I}$.  Also useful is the notation $\Hlin \subset
\h$ for the $\mathbb{Q}$ span of connected nonempty Feynman
graphs in $\h$ and $\Plin:\h \rightarrow \Hlin$ for the corresponding
projection.  That is $\Hlin$ is the parts of degree 1.
Note that $\h$ is graded by the number of independent cycles in the
graph, which is known as the loop number of the graph.  This grading,
not the degree as a monomial, is the more relevant in most circumstances.

The coalgebra structure encodes, as is common for combinatorial Hopf
algebras, how the objects decompose into subobjects.
\begin{definition}The coproduct $\Delta:\h \rightarrow \h \otimes
\h$ is defined on a connected Feynman graph $\Gamma$ by
\[
  \Delta(\Gamma) = \sum_{\substack{\gamma \subseteq \Gamma \\ \gamma
  \text{ product of divergent}\\ \text{1PI subgraphs}}}\gamma \otimes \Gamma/\gamma  
\]
and extended to $\h$ as an algebra homomorphism.
\end{definition}

Note that the sum in the definition of $\Delta$ includes the cases
$\gamma = \mathbb{I}$ and $\gamma=\Gamma$, since $\Gamma$
is divergent and 1PI, hence includes
the terms $\mathbb{I} \otimes \Gamma + \Gamma \otimes \mathbb{I}$.
Note also that
$\gamma$ may be a product, that is a disjoint union. This is
typically intended in presentations of this Hopf algebra, but not always clear.

\begin{definition} Let $\eta:\h \rightarrow \mathbb{Q}$ be the algebra
  homomorphism with $\eta(\mathbb{I}) = 1$ and $\eta(G) = 0$
for $G$ a non-empty connected Feynman graph.
\end{definition}

\begin{prop}$\h$ has a coalgebra structure with coproduct $\Delta$ and
  counit $\eta$ as
  above.
\end{prop}

\begin{proof}
We will verify only
coassociativity.  Calculate $(\id \otimes \Delta)\Delta \Gamma =
\sum_{\gamma'} \gamma' \otimes \Delta(\Gamma/\gamma') = \sum_{\gamma'}
\sum_{\gamma} \gamma' \otimes \gamma/\gamma' \otimes \Gamma/\gamma$
where $\gamma' \subseteq \gamma \subseteq \Gamma$ with each connected
component of $\gamma'$ and $\gamma/\gamma'$ 1PI divergent.  This
calculation holds because every subgraph of $\Gamma/\gamma'$ is
uniquely of the
form $\gamma/\gamma'$ for some $\gamma' \subseteq \gamma \subseteq
\Gamma$.  Further by Proposition \ref{1PI to 1PI} and
renormalizability each connected component of $\gamma$ is 1PI
divergent, so we can switch the order of summation to see that the
above sum is simply $(\Delta \otimes \id)\Delta \Gamma$ giving coassociativity.
\end{proof}

From now on we will only be concerned with the sort of Feynman graphs
which appear in $\h$, that is, Feynman graphs with connected components
which are divergent and 1PI.

$\h$ is graded by the loop number, that is the first Betti number.
$\h$ is commutative but not in general cocommutative.  For example in
$\phi^3$ theory 
\[
  \Delta \left( \raisebox{-1ex}{\begin{fmfgraph}(100,60)
        \fmfleftn{i}{1}
        \fmfrightn{o}{1}
        \fmf{plain}{i1,in1,in3}
        \fmf{plain}{in4,in2,o1}
        \fmf{plain}{in5,in6}
        \fmf{plain,left,tension=0.25}{in1,in2}
        \fmf{plain,left,tension=0.25}{in3,in5,in3}
        \fmf{plain,left,tension=0.25}{in6,in4,in6}
      \end{fmfgraph}}
  \right)
  = \raisebox{-1ex}{\begin{fmfgraph}(100,60)
        \fmfleftn{i}{1}
        \fmfrightn{o}{1}
        \fmf{plain}{i1,in1,in3}
        \fmf{plain}{in4,in2,o1}
        \fmf{plain}{in5,in6}
        \fmf{plain,left,tension=0.25}{in1,in2}
        \fmf{plain,left,tension=0.25}{in3,in5,in3}
        \fmf{plain,left,tension=0.25}{in6,in4,in6}
      \end{fmfgraph}} \otimes \mathbb{I}
    + \mathbb{I} \otimes \raisebox{-1ex}{\begin{fmfgraph}(100,60)
        \fmfleftn{i}{1}
        \fmfrightn{o}{1}
        \fmf{plain}{i1,in1,in3}
        \fmf{plain}{in4,in2,o1}
        \fmf{plain}{in5,in6}
        \fmf{plain,left,tension=0.25}{in1,in2}
        \fmf{plain,left,tension=0.25}{in3,in5,in3}
        \fmf{plain,left,tension=0.25}{in6,in4,in6}
      \end{fmfgraph}}
   + 2 \raisebox{-1ex}{\begin{fmfgraph}(100,60)
       \fmfleftn{i}{1}
       \fmfrightn{o}{1}
       \fmf{plain}{i1,in1}
       \fmf{plain}{in2,o1}
       \fmf{plain,left,tension=0.25}{in1,in2,in1}
     \end{fmfgraph}}
     \otimes \raisebox{-1ex}{\begin{fmfgraph}(100,60)
         \fmfkeep{2loops}
         \fmfleftn{i}{1}
         \fmfrightn{o}{1}
         \fmf{plain}{i1,in1,in3}
         \fmf{plain}{in4,in2,o1}
         \fmf{plain,left,tension=0.25}{in1,in2}
         \fmf{plain,left,tension=0.25}{in3,in4,in3}
       \end{fmfgraph}}
     + \raisebox{-1ex}{\begin{fmfgraph}(100,60)
       \fmfleftn{i}{1}
       \fmfrightn{o}{1}
       \fmf{plain}{i1,in1}
       \fmf{plain}{in2,o1}
       \fmf{plain,left,tension=0.25}{in1,in2,in1}
       \end{fmfgraph}}
     \raisebox{-1ex}{\begin{fmfgraph}(100,60)
       \fmfleftn{i}{1}
       \fmfrightn{o}{1}
       \fmf{plain}{i1,in1}
       \fmf{plain}{in2,o1}
       \fmf{plain,left,tension=0.25}{in1,in2,in1}
       \end{fmfgraph}}
     \otimes \raisebox{-1ex}{\begin{fmfgraph}(100,60)
       \fmfleftn{i}{1}
       \fmfrightn{o}{1}
       \fmf{plain}{i1,in1}
       \fmf{plain}{in2,o1}
       \fmf{plain,left,tension=0.25}{in1,in2,in1}
     \end{fmfgraph}}.
\]

\begin{definition}
For $f_1, f_2:\h \rightarrow \h$ define the convolution $f_1 \star f_2 = m (f_1\otimes
f_2)\Delta$
\end{definition}

We will use the notation $\id$ for the identity map $\h \rightarrow \h$.

\begin{prop}
  With antipode $S:\h \rightarrow \h$ defined recursively by
  $S(\mathbb{I}) = \mathbb{I}$ and
  \[
    S(\Gamma) = -\Gamma - \sum_{\substack{\gamma \subseteq \Gamma \\\mathbb{I} \neq \gamma \neq \Gamma \\ \gamma
  \text{ product of divergent}\\ \text{1PI subgraphs}}}S(\gamma)\,\Gamma/\gamma
  \] on connected graphs, and extended to all of $\h$ as an antihomomorphism,
  $\h$ is a Hopf algebra
\end{prop}

\begin{proof}
  The defining property of the antipode is $e\eta = S\star \id = \id
  \star S$.   The first equality gives exactly the proposition in view of the
  definitions of $\Delta$ and $\star$, the second equality is then
  standard since $\h$ is commutative, see for instance
  \cite[Proposition 4.0.1]{sweedler}.
\end{proof}

Note that since $\h$ is commutative $S$ is in fact a homomorphism.  $S$ is not, however, an
interesting antipode from the quantum groups perspective since $\h$ is
commutative and thus $S \circ S = \id$ (see again \cite[Proposition 4.0.1]{sweedler}).

\begin{definition}
  An element $\gamma$ of $\h$ is \emph{primitive} if $\Delta(\gamma) =
  \gamma \otimes \mathbb{I} + \mathbb{I} \otimes \gamma$.
\end{definition}
A single Feynman graph is primitive iff it has no divergent subgraphs. 
However appropriate sums of
nonprimitive graphs are also primitive.  For example\begin{align*}
  \Delta \left( \raisebox{-1.5ex}{\begin{fmfgraph}(100,60)
    \fmfleftn{i}{2}
    \fmfrightn{o}{2}
    \fmf{plain}{i1,v1}
    \fmf{plain}{i2,v1}
    \fmf{plain,left,tension=0.25}{v1,v2,v1}
    \fmf{plain,left,tension=0.25}{v2,v3,v2}
    \fmf{plain}{v3,o1}
    \fmf{plain}{v3,o2}
  \end{fmfgraph}} 
 - 2\raisebox{-1.5ex}{\begin{fmfgraph}(100,60)
    \fmfleftn{i}{2}
    \fmfrightn{o}{2}
    \fmf{plain}{i1,v1}
    \fmf{plain}{i2,v2}
    \fmf{plain,left,tension=0.25}{v1,v2,v1}
    \fmf{plain}{v1,v3}
    \fmf{plain}{v2,v3}
     \fmf{plain}{v3,o1}
    \fmf{plain}{v3,o2}
  \end{fmfgraph}}\right)
 & = \left( \raisebox{-1.5ex}{\begin{fmfgraph}(100,60)
    \fmfleftn{i}{2}
    \fmfrightn{o}{2}
    \fmf{plain}{i1,v1}
    \fmf{plain}{i2,v1}
    \fmf{plain,left,tension=0.25}{v1,v2,v1}
    \fmf{plain,left,tension=0.25}{v2,v3,v2}
    \fmf{plain}{v3,o1}
    \fmf{plain}{v3,o2}
  \end{fmfgraph}} 
 - 2\raisebox{-1.5ex}{\begin{fmfgraph}(100,60)
    \fmfleftn{i}{2}
    \fmfrightn{o}{2}
    \fmf{plain}{i1,v1}
    \fmf{plain}{i2,v2}
    \fmf{plain,left,tension=0.25}{v1,v2,v1}
    \fmf{plain}{v1,v3}
    \fmf{plain}{v2,v3}
     \fmf{plain}{v3,o1}
    \fmf{plain}{v3,o2}
  \end{fmfgraph}}\right) \otimes \mathbb{I} + \mathbb{I} \otimes \left( \raisebox{-1.5ex}{\begin{fmfgraph}(100,60)
    \fmfleftn{i}{2}
    \fmfrightn{o}{2}
    \fmf{plain}{i1,v1}
    \fmf{plain}{i2,v1}
    \fmf{plain,left,tension=0.25}{v1,v2,v1}
    \fmf{plain,left,tension=0.25}{v2,v3,v2}
    \fmf{plain}{v3,o1}
    \fmf{plain}{v3,o2}
  \end{fmfgraph}} 
 - 2\raisebox{-1.5ex}{\begin{fmfgraph}(100,60)
    \fmfleftn{i}{2}
    \fmfrightn{o}{2}
    \fmf{plain}{i1,v1}
    \fmf{plain}{i2,v2}
    \fmf{plain,left,tension=0.25}{v1,v2,v1}
    \fmf{plain}{v1,v3}
    \fmf{plain}{v2,v3}
     \fmf{plain}{v3,o1}
    \fmf{plain}{v3,o2}
  \end{fmfgraph}}\right) \\ & \qquad + 2 \raisebox{-1.5ex}{\begin{fmfgraph}(100,60)
       \fmfleftn{i}{2}
       \fmfrightn{o}{2}
       \fmf{plain}{i1,in1}
       \fmf{plain}{in2,o1}
       \fmf{plain}{i2,in1}
       \fmf{plain}{in2,o2}
       \fmf{plain,left,tension=0.25}{in1,in2,in1}
     \end{fmfgraph}} \otimes \raisebox{-1.5ex}{\begin{fmfgraph}(100,60)
       \fmfleftn{i}{2}
       \fmfrightn{o}{2}
       \fmf{plain}{i1,in1}
       \fmf{plain}{in2,o1}
       \fmf{plain}{i2,in1}
       \fmf{plain}{in2,o2}
       \fmf{plain,left,tension=0.25}{in1,in2,in1}
     \end{fmfgraph}} - 2 \raisebox{-1.5ex}{\begin{fmfgraph}(100,60)
       \fmfleftn{i}{2}
       \fmfrightn{o}{2}
       \fmf{plain}{i1,in1}
       \fmf{plain}{in2,o1}
       \fmf{plain}{i2,in1}
       \fmf{plain}{in2,o2}
       \fmf{plain,left,tension=0.25}{in1,in2,in1}
     \end{fmfgraph}} \otimes \raisebox{-1.5ex}{\begin{fmfgraph}(100,60)
       \fmfleftn{i}{2}
       \fmfrightn{o}{2}
       \fmf{plain}{i1,in1}
      \fmf{plain}{in2,o1}
       \fmf{plain}{i2,in1}
       \fmf{plain}{in2,o2}
       \fmf{plain,left,tension=0.25}{in1,in2,in1}
     \end{fmfgraph}} \\
 & = \left( \raisebox{-1.5ex}{\begin{fmfgraph}(100,60)
    \fmfleftn{i}{2}
    \fmfrightn{o}{2}
    \fmf{plain}{i1,v1}
    \fmf{plain}{i2,v1}
    \fmf{plain,left,tension=0.25}{v1,v2,v1}
    \fmf{plain,left,tension=0.25}{v2,v3,v2}
    \fmf{plain}{v3,o1}
    \fmf{plain}{v3,o2}
  \end{fmfgraph}} 
 - 2\raisebox{-1.5ex}{\begin{fmfgraph}(100,60)
    \fmfleftn{i}{2}
    \fmfrightn{o}{2}
    \fmf{plain}{i1,v1}
    \fmf{plain}{i2,v2}
    \fmf{plain,left,tension=0.25}{v1,v2,v1}
    \fmf{plain}{v1,v3}
    \fmf{plain}{v2,v3}
     \fmf{plain}{v3,o1}
    \fmf{plain}{v3,o2}
  \end{fmfgraph}}\right) \otimes \mathbb{I} + \mathbb{I} \otimes \left( \raisebox{-1.5ex}{\begin{fmfgraph}(100,60)
    \fmfleftn{i}{2}
    \fmfrightn{o}{2}
    \fmf{plain}{i1,v1}
    \fmf{plain}{i2,v1}
    \fmf{plain,left,tension=0.25}{v1,v2,v1}
    \fmf{plain,left,tension=0.25}{v2,v3,v2}
    \fmf{plain}{v3,o1}
    \fmf{plain}{v3,o2}
  \end{fmfgraph}} 
 - 2\raisebox{-1.5ex}{\begin{fmfgraph}(100,60)
    \fmfleftn{i}{2}
    \fmfrightn{o}{2}
    \fmf{plain}{i1,v1}
    \fmf{plain}{i2,v2}
    \fmf{plain,left,tension=0.25}{v1,v2,v1}
    \fmf{plain}{v1,v3}
    \fmf{plain}{v2,v3}
     \fmf{plain}{v3,o1}
    \fmf{plain}{v3,o2}
  \end{fmfgraph}}\right)
\end{align*}
This phenomenon will be important in Chapter \ref{first reduction}.

We will make sparing but important use of the Hochschild cohomology of
$\h$.  To define the Hochschild cohomology we will follow the
presentation of Bergbauer and Kreimer \cite{bergk}.  The $n$-cochains
are linear maps $L:\h \rightarrow \h^{\otimes 
  n}$.  The coboundary operator $b$ is defined by
\[
  bL = (\id \otimes L)\Delta + \sum_{i=1}^{n}(-1)^i \Delta_i L +
  (-1)^{n+1}L \otimes \mathbb{I}
\]
where $\Delta_i = \id \otimes \cdots \otimes \id \otimes \Delta
\otimes \id \otimes \cdots \id$ with the $\Delta$ appearing in the
$i$th slot.  $b^2=0$ since $\Delta$ is coassociative and so we get a
cochain complex and hence cohomology.  The only part of the Hochschild
cohomology which will be needed below are the 1-cocycles $L:\h
\rightarrow \h$, whose defining property $bL=0$ gives
\begin{equation}\label{1 cocycle}
  \Delta L = (\id \otimes L)\Delta + L \otimes \mathbb{I}.
\end{equation}

\end{fmffile}
\section{Feynman graphs as physical objects}

\begin{fmffile}{physicalegs}
\setlength{\unitlength}{0.05mm}
\fmfset{arrow_len}{2mm}
\fmfset{wiggly_len}{2mm}
\fmfset{thin}{0.5pt}
\fmfset{curly_len}{1.5mm}
\fmfset{dot_len}{1mm}
\fmfset{dash_len}{1.5mm}

\subsection{Feynman rules}\label{feynman rules}
The information in the Feynman rules is the additional piece of analytic information
contained in a physical theory, so for us we can define a physical
theory to be a combinatorial physical theory along with \emph{Feynman
  rules}.  In the following definition we will use the term
\emph{tensor expression} for a tensor written in terms of the
standard basis for $\mathbb{R}^D$ where $D$ is the dimension of space-time.  Such expressions will be intended to be multiplied and then
interpreted with Einstein summation.  An example of a tensor
expression in indices $\mu$ and $\nu$ is
\[
\frac{g_{\mu,\nu} - \xi\frac{k_\mu k_\nu}{k^2}}{k^2}
\]
where $g$ is the Euclidean metric, $k \in \mathbb{R}^4$, $k^2$ is the
standard dot product of $k$ with itself, and $\xi$ is a variable
called the \emph{gauge}.
Such a tensor expression is meant to be a factor of a larger expression like
\begin{equation}\label{integrand}
  \gamma_\mu \frac{1}{\slashed{k} +\slashed{p} - m}\gamma_\nu \left(\frac{g_{\mu,\nu} - \xi\frac{k_\mu k_\nu}{k^2}}{k^2}\right)
\end{equation}
where the $\gamma_\mu$ are the Dirac gamma matrices, $\slashed{k}$ is
the Feynman slash notation, namely $\slashed{k} = \gamma^\mu k_\mu$,
and $m$ is a variable for the mass.  In this example \eqref{integrand}
is the integrand for the Feynman integral for the graph
\[
\mbox{\begin{fmfgraph*}(400,400)
      \fmfleft{i}
      \fmfright{o}
      \fmflabel{$p$}{i}
      \fmflabel{$p$}{o}
      \fmf{fermion}{i,v1}
      \fmf{fermion}{v2,o}
      \fmf{fermion,label=$k+p$,label.side=left}{v1,v2}
      \fmf{photon,left,tension=0.25,label=$k$,label.side=left}{v2,v1}
    \end{fmfgraph*}}
\]

\begin{definition}
  Let $T$ be a combinatorial physical theory with dimension of
  space-time $D$.  Let $\xi$ be a real variable.  % and $x$ a
  %real variable called the coupling constant.
  \emph{Feynman rules} consist of 3 maps
  \begin{enumerate}
    \item the first takes a half
      edge type (viewed as an external edge), an $\mathbb{R}^D$
      vector (the momentum), and a tensor index $\mu$ to a tensor
      expression in $\mu$,
    \item the second takes an edge type $e$, an $\mathbb{R}^D$
      vector (the momentum), and tensor indices $\mu$, $\nu$ for each half edge
      type making up $e$ to a tensor expression in $\mu$, $\nu$,
    \item the third takes a vertex type $v$ and one tensor index
      $\mu_1, \mu_2, \ldots$ for each
    half edge type making up $v$ to a tensor expression in $\mu_1,
    \mu_2, \ldots$.
  \end{enumerate}
  In each case the tensor expressions may depend on $\xi$.
\end{definition}
If there is a non-trivial dependence on $\xi$ in the Feynman
  rules then we say we are working in a \emph{gauge theory}.  QED and
  QCD are gauge theories.  If the
  Feynman rules are independent of the tensor indices then we say we
  are working in a \emph{scalar field theory}.  $\phi^4$ and $\phi^3$
  are scalar field theories.  Note that unoriented edges have no way
  to distinguish their two tensor indices and hence must be
  independent of them.  For us the Feynman rules do not include a
  dependence on a coupling constant $x$ since we wish to use $x$ at
  the level of Feynman graphs as an indeterminate in which to write
  power series.  This setup ultimately coincides with the more typical
  situation because there the dependence of the Feynman rules on $x$ is contrived so
  that it ultimately counts the loop number of the graph and so
  functions as a counting variable.

Using the Feynman rules we can associate to each graph $\gamma$ in a
theory $T$ a formal integral, that is, an integrand and a space to integrate over but with no assurances that the resulting integral is convergent.  We will denote the integrand by
$\mathrm{Int}_\gamma$ and take it over a Euclidean space
$\mathbb{R}^{D|v_\gamma|}$ where $D$ is the dimension of space time
and $v_\gamma$ is a finite index set corresponding to the set of integration variables appearing
in $\mathrm{Int}_\gamma$.  Then the formal integral
is given by
\[
  \int_{\mathbb{R}^{D|v_\gamma|}} \mathrm{Int}_\gamma \prod_{k\in v_\gamma}d^Dk
\]
where $D$ is the dimension of space-time in $T$ and 
where $\mathrm{Int}_\gamma$ and $v_\gamma$ are defined below.
 
Associate to each half edge of $\gamma$ a tensor index.
  Associate to each internal and external edge of $\gamma$ a variable
  (the momentum, with
  values in $\mathbb{R}^D$) and an orientation of the edge
  with the restriction that for each vertex $v$ the sum of the momenta
  of edges entering $v$ equals the sum of the momenta of edges exiting
  $v$.  Consequently the $\mathbb{R}$-vector space of the edge variables has
  dimension the loop number of the graph.  Let $v_\gamma$ be a basis
  of this vector space.  Let $\mathrm{Int}_\gamma$ be the product of
  the Feynman rules applied to the type of each external edge, internal edge, and
  vertex of $\gamma$, along with the assigned tensor indices and the
  edge variables as the momenta.  

Note that $\mathrm{Int}_\gamma$ depends on the momenta $q_1, \ldots,
q_n$ for the
external edges and that these variables are not ``integrated out'' in
the formal integral.  Consequently we may also use the notation
$\mathrm{Int}_\gamma(q_1, \ldots, q_n)$ to show this dependence.
The factors associated to internal edges are called
\emph{propagators}.

In practice the integrals we obtain in this way are not arbitrarily bad in their divergence.  In
fact for arbitrary $\Lambda < \infty$ each will converge when
integrated over a box consisting of all 
parameters running from $-\Lambda$ to $\Lambda$.

For example consider $\phi^4$ with Euclidean Feynman rules, see
\cite[p.268]{iz}.  The Feynman rules in this case say that an edge
labelled with momentum $k$ is associated to the factor $1/(k^2 + m^2)$,
where the square of a vector means the usual dot product with itself
and $m$ is the mass of the particle.  The Feynman rules say that the vertex is associated to
$-1$ (if the coupling constant $\lambda$ was included in the
Feynman rules the vertex would be associated with $-\lambda$.)  Consider
\[
\gamma = \mbox{\begin{fmfgraph*}(400,40)
    \fmfleftn{i}{2}
    \fmfrightn{o}{2}
    \fmf{plain}{i1,v1}
    \fmf{plain}{i2,v1}
    \fmf{plain,left,tension=0.25,label=$k+p$}{v1,v2}
    \fmf{plain,left,tension=0.25,label=$k$}{v2,v1}
     \fmf{plain}{v2,o1}
    \fmf{plain}{v2,o2}
  \end{fmfgraph*}} 
\]
oriented from left to right with the momenta associated to the two right hand external edges
summing to $p$ and hence the momenta associated to the two left hand
external edges also summing to $p$.  Then the integral
associated to $\gamma$ is
\[
  \int d^4k \frac{1}{(k^2+m^2)((p+k)^2+m^2)}
\] 
where $d^4k = dk_0dk_1dk_2dk_3$ with $k=(k_0,k_1,k_2,k_3)$ and squares
stand for the standard dot product.

The above discussion of Feynman rules is likely to
appear either unmotivated or glib depending on one's background,
particularly the rather crass gloss of gauge theories, so it
is worth briefly mentioning a few important words of context.  

More typically a physical theory might be defined by its
Lagrangian $\mathcal{L}$.  For example for $\phi^4$
\[
  \mathcal{L} = \frac{1}{2}\partial^\mu \phi \partial_\mu \phi -
  \frac{1}{2}m^2\phi^2 - \frac{\lambda}{4!}\phi^4.
\]
There is one term for each vertex and edge of the theory and for
massive particles an additional term.  In this case
$\frac{1}{2}\partial^\mu \phi \partial_\mu \phi$ is the term for the
edge of $\phi^4$, $- \frac{1}{2}m^2\phi^2$ is the mass term, and $-
\frac{\lambda}{4!}\phi^4$ is the term for the vertex.
One of the many important properties of the Lagrangian is that it is
Lorentz invariant.

The Feynman rules can be derived from the Lagrangian in a variety of
ways to suit different tastes, for instance directly
\cite[p.16]{cfieldth}, or by expanding the path integral in the
coupling constant.

Gauge theories are a bit more complicated since they are defined on
a fibre bundle over space-time rather than directly on space-time.  
The structure group of the
fibre bundle is
called the gauge group.  A \emph{gauge field} (for example the photon in
QED or the gluon in QCD) is a connection.
A \emph{gauge} is a local section.  Choosing a gauge brings us back to
something closer to the above situation.  

There are many ways to choose a gauge each with different advantages
and disadvantages.  For the present purpose we're interested in a
1-parameter family of Lorentz covariant gauges called the $R_\xi$
gauges.  The parameter for the family is denoted $\xi$ and is the
$\xi$ which we have called the gauge in the above.   The $R_\xi$ gauges can be put into the Lagrangian in the
sense that in these gauges we can write a Lagrangian for the theory which
depends on $\xi$.  For example, for
QED in the $R_\xi$ gauges we have (see for example \cite[p.504]{cl})
\[
  \mathcal{L} = -\frac{1}{4}(\partial_\mu A_\nu - \partial_\nu A_\mu)^2
  - \frac{1}{2\xi}(\partial_\mu A^\mu)^2 + \bar\psi(i\gamma^\mu (\partial_\mu - ieA_\mu) - m)\psi
\]
where the $\gamma^\mu$ are the Dirac gamma matrices.  Whence $\xi$
also appears in the Feynman rules, giving the definition of gauge
theory used above.

Another perspective, perhaps clearer to many mathematicians is Polyak \cite{polyak}.

\subsection{Renormalization}\label{renormalization}

\begin{definition}
  Let 
  \[
  I = \int_{\substack{\mathbb{R}^{D|v|}}} \mathrm{Int} \prod_{k\in v}
  d^D k
  \]
  be a formal integral.
  I is \emph{logarithmically divergent} if the
  net degree (that is the degree of the numerator minus the degree of
  the denominator) of the integration variables in $\mathrm{Int}$ is
  $-D|v|$.  I \emph{diverges like an $n$th power} (or, is
  \emph{linearly divergent}, \emph{quadratically divergent}, etc.) if the net degree
  of the integration variables in $\mathrm{Int}$ is $-D|v| + n$.
\end{definition}

Let $\phi$ be the Feynman rules viewed as map which associates
formal integrals to elements of $\h$.  Next we need a method (called
renormalization) which can
convert the formal integrals for primitive graphs into convergent
integrals.  There are many possible choices; commonly first a
regularization scheme is chosen to introduce one or more additional
variables which convert the formal integrals to
meromorphic expressions with a pole at the original point.  For instance
one may raise propagators to non-integer powers (analytic
regularization) or take the dimension of space-time to be complex (dimensional
regularization, see for instance \cite{collins} on setting up the
appropriate definitions).  Then a map such as minimal subtraction is chosen
to remove the pole part.

We will take a slightly different approach.  
First we will set
\begin{equation}\label{powers of int var}
%\int
%d^Dk \frac{k_{\mu_1}\cdots k_{\mu_t}}{(k^2)^r} = 0
\int (k^2)^r = 0
\end{equation}
for all $r$. % and $t\geq 0$.  
This is the result which is obtained, for instance, from
dimensional  
regularization and from analytic regularization, but simply taking it
as true allows us to remain
agnostic about the choice of regularization scheme.  To see the origin of this
peculiar identity consider %first $t=0$.  We have 
the following computation with $q \in
\mathbb{R}^D$ and the square of an element of $\mathbb{R}^D$ denoting
its dot product with itself.
\begin{align*}
      & \int d^D k \frac{1}{(k^2)^r((k+q)^2)^s} \\ & = \int d^D k \frac{\Gamma(r+s)}{\Gamma(r)\Gamma(s)}\int_0^1 dx
    \frac{x^{r-1}(1-x)^{s-1}}{(xk^2 + (1-x)(k+q)^2)^{r+s}} \\
    & =  \frac{\Gamma(r+s)}{\Gamma(r)\Gamma(s)}\int_0^1 dx
    x^{r-1}(1-x)^{s-1} \int d^D k \frac{1}{\big(xk^2 +
      (1-x)(k+q)^2\big)^{r+s}} \\
    & =  \frac{\Gamma(r+s)}{\Gamma(r)\Gamma(s)} \int_0^1 dx
    x^{r-1}(1-x)^{s-1} \int d^D k \frac{1}{\big((k+q(1-x))^2 +
      q^2(x-x^2)\big)^{r+s}} \\
    & = \frac{\Gamma(r+s)}{\Gamma(r)\Gamma(s)}  \int_0^1 dx
    x^{r-1}(1-x)^{s-1} \int d^D k \frac{1}{(k^2 + q^2(x-x^2))^{r+s}} \\
    & = \frac{\Gamma(r+s)}{\Gamma(r)\Gamma(s)} \int_0^1 dx x^{r-1}(1-x)^{s-1} \int_0^\infty d|k| \frac{|k|^{D-1}}{(|k|^2 +
      q^2(x-x^2))^{r+s}} \int d \Omega_k \\
    & = \frac{\Gamma(r+s)}{\Gamma(r)\Gamma(s)} \frac{2\pi^{\frac{D}{2}}}{\Gamma(\frac{D}{2})} \int_0^1 dx x^{r-1}(1-x)^{s-1} \int_0^\infty d|k|\frac{|k|^{D-1}}{(|k|^2 +
      q^2(x-x^2))^{r+s}} \\
    & =
    \frac{\Gamma(r+s)}{\Gamma(r)\Gamma(s)}\frac{2\pi^{\frac{D}{2}}}{\Gamma(\frac{D}{2})} \frac{\Gamma(r+s-\frac{D}{2})\Gamma(\frac{D}{2})}{2\Gamma(r+s)} (q^2)^{\frac{D}{2}-r-s}\int_0^1  dx x^{\frac{D}{2}-1-s}(1-x)^{\frac{D}{2}-1-r} \\
    & =
    \frac{\pi^{\frac{D}{2}}\Gamma(r+s-\frac{D}{2})}{\Gamma(r)\Gamma(s)}(q^2)^{\frac{D}{2}-r-s} \frac{\Gamma(\frac{D}{2} - r)\Gamma(\frac{D}{2}-s)}{\Gamma(D-r-s)}
\end{align*}
when $2r+2s>D>0$, $D>2r>0$, and $D>2s>0$, and
  where the first equality is by Feynman parameters:
  \[
    \frac{1}{a^\alpha b^\beta} =
    \frac{\Gamma(\alpha+\beta)}{\Gamma(\alpha)\Gamma(\beta)} \int_0^1
    dx \frac{x^{\alpha-1}(1-x)^{\beta-1}}{(ax+b(1-x))^{\alpha+\beta}}
    \qquad \text{for $\alpha, \beta > 0$}
  \]
  and where
  $d\Omega_k$ refers to the angular integration over the unit
  $D-1$-sphere in $\mathbb{R}^D$.
Now consider just the final line and suppose $s=0$, then since $\Gamma$ has simple poles precisely at
the nonpositive integers, is never 0, and 
\[
\Gamma(x)\Gamma(-x) = \frac{-\pi}{x\sin(\pi x)}
\]
we see that for $D>0$ the result is $0$ for $s=0$ and $r$ not a half-integer.  If we view the
original integral as a function of complex variables $r$ and $s$ for fixed integer $D$
(analytic regularization), or as a function of complex $D$
(dimensional regularization), then by analytic
continuation the above calculations gives \eqref{powers of int var}.
%for $t=0$ and moreover gives $\int d^Dk (k^2)^{-r}((k+q)^2)^{-s} = 0$ for
%all $r$ and for nonpositive integer $s$.
%
%Inductively then suppose
%\[
%\int
%d^Dk \frac{k_{\mu_1}\cdots k_{\mu_{t-1}}}{(k^2)^r(k+q)^s} = 0
%\]
%for $t>0$ and $s$ a nonpositive integer.  Then at level $t$
%consider 
%\begin{align*}
%    & \int d^D k \frac{(k \cdot q_1) \cdots (k \cdot
%    q_t)}{(k^2)^r((k+q_t)^2)^s}  \\
%    & = \int d^D k \frac{(k \cdot q_1) \cdots (k \cdot q_{t-1})}{2(k^2)^r((k+q_t)^2)^{s-1}} - q^2\int d^D k
%    \frac{(k \cdot q_1) \cdots (k \cdot
%      q_{t-1})}{2(k^2)^r((k+q_t)^2)^s} \\
%    & \qquad - \int d^D k
%    \frac{(k \cdot q_1) \cdots (k \cdot
%      q_{t-1})}{2(k^2)^{r-1}((k+q_t)^2)^s} 
%\end{align*}
%For $s$ (and so also $s-1$) a nonpositive integer all three terms of
%the left hand side are zero 

Returning to the question of renormalization, in view of \eqref{powers of int var} we need only consider
logarithmically divergent integrals since by subtracting off $0$ in the
form of a power of
$k^2$ which is equally divergent to the original integral the whole
expression becomes less divergent.   Logarithmically divergent
integrals with no subdivergences can then be made finite
simply by subtracting the same formal integral evaluated at fixed
external momenta.  

Let $R$ be the map which given a formal integral returns
the formal
integral evaluated at the subtraction point.  In our case then $R$ has as domain and range the algebra of
formal integrals where relations are generated by evaluating
convergent integrals and \eqref{powers of int var}.  Let $\phi$ be the
Feynman rules, the algebra homomorphism which
given a graph $G$ returns the formal integral $\phi(G)$.  We suppose
$\phi(\mathbb{I}) = 1$ and $R(1)=1$.

If instead we had chosen to use a regulator and
corresponding renormalization scheme then $\phi$ would give the
regularized integral of a graph, and $R$ would implement the scheme itself.  
One such example
would be dimensional regularization with the minimal subtraction
scheme.  In that case $\phi$ would take
values in the space of Laurent series in the small
parameter $\epsilon$ and $R$ would take such a Laurent series and
return only the part with negative degree in $\epsilon$.  That is
$R\phi(\Gamma)$ is the singular part of $\phi(\Gamma)$, the part one
wishes to ignore.  Note that in this case $R(1)=0$.  The key
requirement in general is that $R$ be a Rota-Baxter operator see \cite{e-fgkII}, \cite{e-fksurvey}.

To deal with graphs containing subdivergences, define $S_R^\phi$
recursively by $S_R^\phi(\mathbb{I}) = 1$,
\[
  S^\phi_R(\Gamma) = - R(\phi(\Gamma)) - \sum_{\substack{\mathbb{I} \neq \gamma \subsetneq \Gamma \\ \gamma
  \text{ product of divergent}\\ \text{1PI subgraphs}}}
  S^\phi_R(\gamma) R(\phi(\Gamma/\gamma))
\]
for connected Feynman graphs $\Gamma$ extended to all of $\h$ as an algebra
homomorphism.  $S_R^\phi$ can be thought of as a twisted antipode; the
defining recursion says that $S_R^\phi \star R\phi = \eta$.  Use $S_R^\phi$ to define the
\emph{renormalized Feynman rules} by 
\[
  \phi_R = S^\phi_R \star \phi.
\]
When $\Gamma$ contains no subdivergences, $\phi_R(\Gamma) =
\phi(\Gamma) - R\phi(\Gamma)$; in view of Subsection
\ref{renormalization} we may assume that $\phi(\Gamma)$ is log
divergent and so $\phi_R(\Gamma)$ is a convergent integral.
Inductively one can show that $\phi_R$ maps $\h$ to convergent
integrals.
This result is the original
purpose of the Hopf algebraic approach to renormalization.  It gives a
consistent algebraic framework to the long-known but ad-hoc
renormalization procedures of physicists.  For more details and more
history see for instance instance the survey
\cite{e-fksurvey} and the references therein.

These integrals lead to interesting transcendental numbers, but that is
very much another story \cite{bek}, \cite{numbers}, \cite{bkknots}.

\begin{example}
  To illustrate the conversion to log divergence and renormalization
  by subtraction 
  consider the following graph in massless
  $\phi^3$
  \[
  \mbox{\begin{fmfgraph*}(400,400)
      \fmfleft{i}
      \fmfright{o}
      \fmflabel{$q$}{i}
      \fmflabel{$q$}{o}
      \fmf{plain}{i,v1}
      \fmf{plain}{v2,o}
      \fmf{plain,left,tension=0.25,label=$k+q$,label.side=left}{v1,v2}
      \fmf{plain,left,tension=0.25,label=$k$,label.side=left}{v2,v1}
    \end{fmfgraph*}}
  \]
  The Feynman rules associate to it the integral
  \[
  I = \frac{1}{q^2}\int d^6k \frac{1}{k^2(k+q)^2}.
  \]
  The factor of $1/q^2$ is there because our
  conventions have that the graphs with no cycles are all normalized
  to $1$.
  This integral is quadratically divergent and so can not be renormalized by a
  simple subtraction.  
  However we take
  \[
  \int d^6k \frac{1}{(k^2)^2} = 0,
  \]
  so 
  \begin{align*}
    I = & \frac{1}{q^2}\int d^6k \frac{1}{k^2(k+q)^2}
    - \frac{1}{q^2}\int d^6k \frac{1}{(k^2)^2} \\
    = & -\frac{2}{q^2}\int d^6k \frac{k\cdot q}{(k^2)^2(k+q)^2} - \int d^6 k
    \frac{1}{(k^2)^2(k+q)^2} \\
    = & -2I_1 - I_2.
  \end{align*}
  Each of the two resulting terms are now less divergent.

  To illustrate renormalization by subtraction consider the integral
  from the second of
  the above terms.  As formal integrals (or carrying along the
  subtraction which we will add below), using the same tricks as the
  calculation earlier this section,
  \begin{align*}
    I_2 = \int d^6 k \frac{1}{(k^2)^2(k+q)^2} = & \int d^6 k \int_0^1 dx
    \frac{2x}{(xk^2 + (1-x)(k+q)^2)^3} \\
    = & 2 \int_0^1 dx x \int d^6 k \frac{1}{\big(xk^2 + (1-x)(k+q)^2\big)^3} \\
    = & 2 \int_0^1 dx x \int d^6 k \frac{1}{\big((k+q(1-x))^2 +
      q^2(x-x^2)\big)^3} \\
    = & 2 \int_0^1 dx x \int d^6 k \frac{1}{(k^2 + q^2(x-x^2))^3} \\
    = & 2 \int_0^1 dx x \int_0^\infty d|k| \frac{|k|^5}{(|k|^2 +
      q^2(x-x^2))^3} \int d \Omega_k \\
    = & 2\pi^3 \int_0^1 dx x \int_0^\infty d|k|\frac{|k|^5}{(|k|^2 +
      q^2(x-x^2))^3}
  \end{align*}
  %where the first equality is by Feynman parameters:
  %\[
  %  \frac{1}{a^\alpha b^\beta} =
  %  \frac{\Gamma(\alpha+\beta)}{\Gamma(\alpha)\Gamma(\beta)} \int_0^1
  %  dx \frac{x^{\alpha-1}(1-x)^{\beta-1}}{(ax+b(1-x))^{\alpha+\beta}}
  %\]
  %and 
  %$d\Omega_k$ refers to the angular integration over the unit
  %$5$-sphere in $\mathbb{R}^6$.  
  Now consider the result of subtracting
  at $q^2 = \mu^2$.  By Maple
  \begin{align*}
    I_2 - RI_2 & = 2\pi^3 \int_0^1 dx x \int_0^\infty d|k|\frac{|k|^5}{(|k|^2 +
      q^2(x-x^2))^3} - \frac{|k|^5}{(|k|^2 +
      \mu^2(x-x^2))^3} \\& =  2\pi^3 \int_0^1 dx x
      \left(-\frac{1}{2}\log(q^2(x-x^2)) +
        \frac{1}{2}\log(\mu^2(x-x^2))\right) \\
      & = -\frac{\pi^3}{2}\log(q^2/\mu^2)
  \end{align*}
  giving us a finite value.

  To finish the example we need to consider the integral
  \[
   I_1 = \frac{1}{q^2}\int d^6 k \frac{2 k \cdot q}{(k^2)^2(k+q)^2}.
  \]
  This integral is linearly divergent so it needs another subtraction
  of $0$.  
  %Notice that using the same calculations as earlier in this section
  %\begin{align*}
  %  & \int d^D k \frac{2 k \cdot q}{(k^2)^r((k+q)^2)^s} \\
  %  & = \int d^D k \frac{1}{(k^2)^r((k+q)^2)^{s-1}} - q^2\int d^D k
  %  \frac{1}{(k^2)^r((k+q)^2)^s} - \int d^D k
  %  \frac{1}{(k^2)^{r-1}((k+q)^2)^s} \\
  %  & = (q^2)^{\frac{D}{2} - r-
  %    s+1}\pi^{\frac{D}{2}}\Bigg(\frac{\Gamma(r+s-1-\frac{D}{2})\Gamma(\frac{D}{2}-r+1)\Gamma(\frac{D}{2}-s)}{\Gamma(r-1)\Gamma(s)\Gamma(D-r-s+1)} \\
  %  & \qquad \qquad - \frac{\Gamma(r+s-\frac{D}{2})\Gamma(\frac{D}{2}-r)\Gamma(\frac{D}{2}-s)}{\Gamma(r)\Gamma(s)\Gamma(D-r-s)} - \frac{\Gamma(r+s-1-\frac{D}{2})\Gamma(\frac{D}{2}-r)\Gamma(\frac{D}{2}-s+1)}{\Gamma(r)\Gamma(s+1)\Gamma(D-r-s+1)}\Bigg)
  %\end{align*}
  %So, since the $\Gamma$ function has a simple pole at each
  %nonpositive integer, in particular at $-1$ as well as at $0$, in the
  %same way as before we get that
  %\[
  %  \int d^6 k \frac{2 k \cdot q}{(k^2)^r} = 0.
  %\]
  However, this time we only need
  \[
    \int d^6 k \frac{2 k \cdot q}{(k^2)^3} = 0.
  \]
  which we can derive from \eqref{powers of int var}.
  Write  $k =
  k_\perp + k_\parallel$ where $k_\parallel$ is the orthogonal
  projection of $k$ onto $\mathrm{span}(q)$ and $k_\perp$ is the
  orthogonal complement, and notice that
  \begin{align*}
    \int d^6 k \frac{2 k \cdot q}{(k^2)^3} & = \int d^6 k \frac{ 2
      k_\parallel |q|}{(k_\parallel^2 + k_\perp^2)^3}\\
& = \int d^5 k_\perp \int_0^\infty d k_\parallel \frac{ 2
      k_\parallel |q|}{(k_\parallel^2 + k_\perp^2)^3} +
\int d^5 k_\perp \int_{-\infty}^0 d k_\parallel \frac{ 2
      k_\parallel |q|}{(k_\parallel^2 + k_\perp^2)^3}\\
    & = \frac{|q|}{2}\int d^5 k_\perp \frac{1}{(k_\perp^2)^2} -
    \frac{|q|}{2}\int d^5 k_\perp \frac{1}{(k_\perp^2)^2} \\
    & = 0-0 = 0.
  \end{align*}

  So returning to $I_1$, as formal integrals,
  \begin{align*}
    I_1 & = \frac{1}{q^2}\int d^6 k
    \frac{2 k \cdot q}{(k^2)^2(k+q)^2} - \frac{1}{q^2}\int d^6 k \frac{2 k \cdot
      q}{(k^2)^3} \\
    & = - \frac{4}{q^2} \int d^6 k \frac{(k \cdot q)^2}{(k^2)^3(k+q)^2} - \int
    d^6 k \frac{ k \cdot q}{(k^2)^3(k+q)^2}.
  \end{align*}
  The second term is convergent and so needs no further
  consideration.  The first term is now log divergent, call it $-4I_3$.  Writing $k =
  k_\perp + k_\parallel$ as above, we get
  \begin{align*}
    I_3 & = \frac{1}{q^2}\int d^6 k \frac{(k \cdot
      q)^2}{(k^2)^3(k+q)^2} \\ 
    & = \frac{1}{q^2}\int d^5 k_\perp \int_{-\infty}^\infty d k_\parallel \frac{(k_\parallel |q|)^2}{(k_\parallel^2 +
      k_\perp^2)^3(k_\perp^2 + (k_\parallel + q)^2)} \\
    & = \int_0^\infty d |k_\perp| \int_{-\infty}^\infty d
    k_\parallel \frac{k_\parallel^2 |k_\perp|^4}{(k_\parallel^2 +
      |k_\perp|^2)^3(|k_\perp|^2 + (k_\parallel + q)^2)} \int
    d\Omega_{k_\perp} \\
    & = \frac{2 \pi^{5/2}}{\Gamma(5/2)} \int_0^\infty d |k_\perp|
    \int_{-\infty}^\infty d k_\parallel \frac{k_\parallel^2 k_\perp^4}{(k_\parallel^2 +
      |k_\perp|^2)^3(|k_\perp|^2 + (k_\parallel + q)^2)}.
  \end{align*}
  The inner integral Maple can do, and then subtracting at $q^2 =
  \mu^2$ the outer integral is again within Maple's powers and we
  finally get a finite answer
  \[
  I_3 - RI_3= -\frac{1}{16}\pi \log(q^2/\mu^2).
  \]
  Combining these various terms together we have finally computed
  $I-RI$.  This completes this example.
\end{example}

\begin{example}
  Subtracting $0$ in this way also plays nicely with analytic
  regularization, and is less messy on top of it.  Consider the example
  \[
    \int d^4k \frac{k \cdot q}{(k^2)^{1+\rho_1}((k+q)^2)^{1+\rho_2}}
  \]

  Then
  \begin{align*}
    \int d^4k \frac{k \cdot q}{(k^2)^{1+\rho_1}((k+q)^2)^{1+\rho_2}}
    & = 
    \int d^4k
    \frac{k \cdot q}{(k^2)^{1+\rho_1}((k+q)^2)^{1+\rho_2}}-
    \frac{k \cdot q}{(k^2)^{2+\rho_1+\rho_2}}   \\ 
    & =\int d^4k
    \frac{k \cdot
      q\Big((k^2)^{1+\rho_2}-((k+q)^2)^{1+\rho_2}\Big)}{(k^2)^{2+\rho_1+\rho_2}((k+q)^2)^{1+\rho_2}} 
    \end{align*}
    which is merely log divergent and so can be renormalized by
    subtracting the same integrand at $q^2 = \mu^2$.  This sort of
    example will be important later on, as we can simply take this
    integral with 
    $q=1$ as the Mellin transform which we need in Section \ref{dse
      section}.
\end{example}

Subtracting off zero in its various forms and subtracting at
fixed momenta should not be confused.  The former consists just of
adding and subtracting zero and so can be done in whatever way is
convenient.  In the following we will assume that it has been
done, and so that all integrals are log divergent.  The latter,
however, we will always explicitly keep track of.  It is our choice
of renormalization scheme and a different choice would give different
results. 

\subsection{Symmetric insertion}\label{symmetric insertion}

For one of the upcoming reductions we will need to define a symmetric
insertion with a single external momentum $q^2$.  Let $p$ be a primitive of $\h$, not necessarily connected.
For the purposes of symmetric insertion define the
Mellin transform $F_p$ of $p$ (see Section \ref{dse section}) as 
\[
F_p(\rho)=(q^2)^\rho\int
{\rm Int}_p(q^2)\left(\frac{1}{|p|}\sum_{i=1}^{|p|}(k_i^2)^{-\rho}\right)\prod_{i=
1}^{|p|}d^4
k_i,\]
 where ${\rm Int}_p(q^2)$ is the integrand determined by
$p$. We'll renormalize by subtraction at $q^2=\mu^2$ and let \[{\rm Int}_p^-(q^2)={\rm Int}_p(q^2)-{\rm
Int}_p(\mu^2).\]

So define renormalized Feynman rules for this symmetric
scheme with subtractions at $q^2=\mu^2$ by 
\[
  \phi_R(B_+^p(X))(q^2/\mu^2)=\int \textrm{Int}_p^-(q^2)\left(\frac{1}{|p|}\sum_{i=1}^{|p|}\phi_R(X)(-k_i^2/\mu^2)\right)\prod_{i=1}^{|p|}d^4
k_i.
\]

We have 
\[
  \phi_R(B_+^p(X))(q^2/\mu^2)=\lim_{\rho\to
0}\phi_R(X)(\partial_{-\rho})F_p(\rho)\left((q^2/\mu^2)^{-\rho}-1\right),
\]
where $\partial_{-\rho}=-\frac{\partial}{\partial \rho}$.

\end{fmffile}

\chapter{Dyson-Schwinger equations}\label{DSE}

\begin{fmffile}{dseegs}
\setlength{\unitlength}{0.05mm}
\fmfset{arrow_len}{2mm}
\fmfset{wiggly_len}{2mm}
\fmfset{thin}{0.5pt}
\fmfset{curly_len}{1.5mm}
\fmfset{dot_len}{1mm}
\fmfset{dash_len}{1.5mm}

\section{$B_+$}\label{B+}

For $\gamma$ a primitive Feynman graph,
$B_+^\gamma$ denotes the operation of insertion into $\gamma$.  There
are, however, a few subtleties which we need to address.
%, but first define $B_+^{a\gamma + b\tau} = aB_+^\gamma + bB_+^\tau$ for
%$a,b\in\mathbb{Q}$ and $\tau$ also a primitive Feynman graph.

In the closely related Connes-Kreimer Hopf algebra of rooted trees
\cite{ck0}, see Chapter \ref{first reduction}, $B_+(F)$ applied to a forest $F$ denotes the operation of constructing a
new tree by adding a new root with children the roots of each tree
from $F$.  For example 
\[
  B_+\left(\raisebox{-1ex}{\begin{fmfgraph}(60,60)\fmftop{t}\fmfdot{t}\end{fmfgraph}}
    \raisebox{-1ex}{\begin{fmfgraph}(60,60)\fmftop{t}\fmfbottom{b}\fmf{plain}{t,b}\fmfdot{t}\fmfdot{b}\end{fmfgraph}}\right)
  = \raisebox{-2ex}{\begin{fmfgraph}(120,120)\fmftop{t}\fmfbottomn{b}{2}\fmf{plain}{t,v1}\fmf{plain}{t,v2}\fmf{phantom}{v1,b1}\fmf{plain}{v2,b2}\fmfdot{t}\fmfdot{v1}\fmfdot{v2}\fmfdot{b2}\end{fmfgraph}}.
\]
$B_+$ in rooted trees is a Hochschild
1-cocycle \cite[Theorem 2]{ck0},
\[
\Delta B_+ = (\id \otimes B_+)\Delta + B_+ \otimes \mathbb{I}.
\]
This 1-cocycle
property is key to many of the arguments below.  The corresponding
property which is desired of the various $B_+$ appearing in the Hopf
algebras of Feynman graphs is that the sum of all $B_+$ associated to
primitives of the same loop number and the same external leg
structures is a Hochschild 1-cocycle.

In the case where all subdivergences are nested rather than
overlapping, and where there is only one way to make each insertion,
a 1PI Feynman graph $\Gamma$ can be uniquely represented by a rooted tree with
labels on each vertex corresponding to the associated subdivergence.
Call such a tree an \emph{insertion tree}.  For example the insertion
tree for the graph in Yukawa theory
\[
\raisebox{-8ex}{\begin{fmfgraph}(1200,480)
    \fmfleft{i}
    \fmfright{o}
    \fmf{fermion}{i,v1,v2,v3,v4,v5,v6,v7,v8,o}
    \fmf{dashes,left,tension=0.25}{v1,v8}
    \fmf{dashes,left,tension=0.25}{v2,v5}
    \fmf{dashes,left,tension=0.25}{v3,v4}
    \fmf{dashes,left,tension=0.25}{v6,v7}
  \end{fmfgraph}}
\] is
\[
\raisebox{-4ex}{\begin{fmfgraph}(800,800)
    \fmftop{t}
    \fmfbottomn{b}{2}
    \fmf{plain}{t,v1}
    \fmf{plain}{t,v2}
    \fmf{plain}{v1,b1}
    \fmf{phantom}{v2,b2}
    \fmfdot{t}
    \fmfdot{v1}
    \fmfdot{v2}
    \fmfdot{b1}
    \begin{fmfsubgraph}(0,0)(60,60)
    \fmfleft{s1i}
    \fmfright{s1o}
    \fmf{fermion}{s1i,s1v1,s1v2,s1o}
    \fmf{dashes,left,tension=0.25}{s1v1,s1v2}
    \end{fmfsubgraph}
    \begin{fmfsubgraph}(0,40)(60,60)
    \fmfleft{s2i}
    \fmfright{s2o}
    \fmf{fermion}{s2i,s2v1,s2v2,s2o}
    \fmf{dashes,left,tension=0.25}{s2v1,s2v2}
    \end{fmfsubgraph}
    \begin{fmfsubgraph}(10,90)(60,60)
    \fmfleft{s3i}
    \fmfright{s3o}
    \fmf{fermion}{s3i,s3v1,s3v2,s3o}
    \fmf{dashes,left,tension=0.25}{s3v1,s3v2}
  \end{fmfsubgraph}
    \begin{fmfsubgraph}(80,40)(60,60)
    \fmfleft{s4i}
    \fmfright{s4o}
    \fmf{fermion}{s4i,s4v1,s4v2,s4o}
    \fmf{dashes,left,tension=0.25}{s4v1,s4v2}
    \end{fmfsubgraph}
  \end{fmfgraph}}.
\]
In such cases $B_+^\gamma$ is the same operation as
the $B_+$ for rooted trees (with the new root labelled by the new
graph).  So the 1-cocycle identity holds for $B_+^\gamma$ too.

However in general there are many possible ways to insert one graph
into another so the tree must also contain the information of which
insertion place to use.  Also when there are overlapping subdivergences different
tree structures of insertions can give rise to the same graph.
For example in $\phi^3$ the graph
\[
\mbox{\begin{fmfgraph}(400,400)
    \fmfleft{i}
    \fmfright{o}
    \fmftop{t}
    \fmfbottom{b}
    \fmf{plain}{i,v1}
    \fmf{plain}{v2,o}
    \fmf{phantom}{t,v3}
    \fmf{phantom}{b,v4}
    \fmf{plain}{v1,v4,v2,v3,v1}
    \fmffreeze
    \fmf{plain}{v3,v4}
    \end{fmfgraph}
    }
\] can be obtained by inserting 
\[
\mbox{\begin{fmfgraph}(200,200)
    \fmfleft{i}
    \fmfrightn{o}{2}
    \fmf{plain}{i,v1}
    \fmf{plain}{o2,v2,v1,v3,o1}
    \fmffreeze
    \fmf{plain}{v2,v3}
    \end{fmfgraph}}
\qquad \text{into}
\qquad
\mbox{\begin{fmfgraph}(200,120)
    \fmfleft{i}
    \fmfright{o}
    \fmf{plain}{i,v1}
    \fmf{plain}{v2,o}
    \fmf{plain,left,tension=0.25}{v1,v2,v1}
\end{fmfgraph}}
\] either at the right vertex or at the left vertex giving two
different insertion trees.  Provided any overlaps are made by multiple copies of
the same graph, as in the previous example, then, since $\gamma$ is primitive,
the same tensor products of 
graphs appear on both sides of \eqref{1 cocycle} but potentially with
different coefficients.  Note that this only requires $\gamma$ to be
primitive, not necessarily connected.  Fortunately it is possible to make a choice of
coefficients in the definition of $B_+^\gamma$ which fixes this
problem.  This is discussed in the first and second sections of
\cite{anatomy}, and the result is the definition

\begin{definition}\label{B+ def}
  For $\gamma$ a connected Feynman graph define
  \[
  B_+^\gamma(X) = \sum_{\Gamma \in \Hlin} \frac{\mathbf{bij}(\gamma,
  X, \Gamma)}{|X|_\vee} \frac{1}{\mathrm{maxf}(\Gamma)}
  \frac{1}{(\gamma|X)} \Gamma
  \]
where $\mathrm{maxf}(\Gamma)$ is the number of insertion trees
corresponding to $\Gamma$, $|X|_\vee$ is the number of distinct graphs
obtainable by permuting the external edges of $X$,
$\mathbf{bij}(\gamma, X, \Gamma)$ is the number of bijections of the
external edges of $X$ with an insertion place of $\gamma$ such that
the resulting insertion gives $\Gamma$, and $(\gamma|X)$ is the number
of insertion places for $X$ in $\gamma$.

Extend $B^\gamma_+$ linearly to all primitives $\gamma$.
\end{definition}
Note that $B_+^\gamma(\mathbb{I}) = \gamma$.  Also with the above
definition we have $B_+^\gamma$ defined even for nonprimitive graphs,
but this was merely our approach to make the definition for primitives
which are sums; now that the definitions are settled we will only
consider $B_+^\gamma$ for primitives.

The messy coefficient in the definition of $B_+^\gamma$ assures that if we sum all
$B_+^\gamma$ running over $\gamma$ primitive 1PI with a given external leg
structure (that is, over all primitives of the Hopf algebra which are
single graphs and which have the
given external leg structure), inserting into all insertion places of each $\gamma$, then
each 1PI graph with that external leg structure
occurs and is weighted by its symmetry factor.  This property is \cite[Theorem 4]{anatomy} and is illustrated in
Example \ref{qed comb system}.

Gauge theories are more general in one way; there may be
overlapping subdivergences with different external leg structures.
Consequently we may be able to form a graph $G$ by inserting one graph
into another but in the coproduct of $G$ there may be subgraphs and
cographs completely different from those which we used to form $G$ as
in the following example.  
\begin{example}\label{bad overlapping}
In QCD
\[
\mbox{\begin{fmfgraph}(400,400)
    \fmfleft{i}
    \fmfright{o}
    \fmftop{t}
    \fmfbottom{b}
    \fmf{gluon}{i,v1}
    \fmf{gluon}{v2,o}
    \fmf{phantom}{t,v3}
    \fmf{phantom}{b,v4}
    \fmf{fermion}{v3,v1,v4}
    \fmf{gluon}{v4,v2,v3}
    \fmffreeze
    \fmf{fermion}{v4,v3}
    \end{fmfgraph}
    }
\]  
can be obtained by inserting 
\[
\mbox{\begin{fmfgraph}(200,200)
    \fmfleft{i}
    \fmfrightn{o}{2}
    \fmf{gluon}{i,v1}
    \fmf{fermion}{v2,v1,v3}
    \fmf{gluon}{o2,v2}
    \fmf{gluon}{v3,o1}
    \fmffreeze
    \fmf{fermion}{v3,v2}
    \end{fmfgraph}}
\qquad  \text{into} 
\qquad
\mbox{\begin{fmfgraph}(200,120)
    \fmfleft{i}
    \fmfright{o}
    \fmf{gluon}{i,v1}
    \fmf{gluon}{v2,o}
    \fmf{gluon,left,tension=0.25}{v1,v2,v1}
\end{fmfgraph}}
\]
or by inserting
\[
\mbox{\begin{fmfgraph}(200,200)
    \fmfleft{i}
    \fmfrightn{o}{2}
    \fmf{gluon}{i,v1}
    \fmf{gluon}{v2,v1,v3}
    \fmf{fermion}{o2,v2}
    \fmf{fermion}{v3,o1}
    \fmffreeze
    \fmf{fermion}{v2,v3}
    \end{fmfgraph}}
\qquad \text{into} 
\qquad
\mbox{\begin{fmfgraph}(200,120)
    \fmfleft{i}
    \fmfright{o}
    \fmf{gluon}{i,v1}
    \fmf{gluon}{v2,o}
    \fmf{fermion,left,tension=0.25}{v1,v2,v1}
\end{fmfgraph}}.
\]
\end{example}
This makes it impossible for every
$B_+^\gamma$ for $\gamma$ primitive to be a Hochschild 1-cocycle since
there may be graphs appearing on the right hand side of \eqref{1
  cocycle} which do not appear on the left.
In these
cases there are identities between graphs, known as Ward
identities for QED and Slavnov-Taylor identities for QCD,
which guarantee that
$\sum B_+^\gamma$ is a 1-cocycle where the sum is over all $\gamma$
with a given loop number and external leg structure.  This phenomenon is discussed in
\cite{anatomy} and the result is proved for QED and QCD by van
Suijlekom \cite{vS}.

For our
purposes we will consider sets of $B_+$ operators, 
\[
 \{B_+^{k,i;r}\}_{i=0}^{t^r_k}
\]
where $k$ is the loop number, $r$ is an index for the external leg
structure, and $i$ is an additional index running over the primitive
graphs under consideration with $k$ loops and external leg structure
$r$.  In the case where there is only one $r$ under consideration
write $\{B_+^{k,i}\}_{i=0}^{t_k}$.  Now assume that in this more
general case, as in QED and QCD, that the required identities form a
Hopf ideal so that by working in a suitable quotient Hopf algebra we
get
\begin{assume}\label{b+ 1 cocycle}
  $\sum_{i=0}^{t^r_k}B_+^{k,i;r}$ is a Hochschild
  1-cocycle.
\end{assume} 
%assume that the identities to make each $B_+^\gamma$ a
%1-cocycle hold and we will assume that these identities form a coideal
%and hence that we can simply work in the quotient, as van Suijlekom proved
%for QED and QCD.
%
%Thus in all cases we have
%\begin{thm}\label{b+ 1 cocycle}
%For any primitive linear combination of Feynman graphs $\gamma$,
%$B_+^\gamma$ is a Hochschild 1-cocycle, that is,
%\[
%\Delta B_+^\gamma = (\id \otimes B_+^\gamma)\Delta + B_+^\gamma \otimes \mathbb{I}.
%\]
%\end{thm}
\section{Dyson-Schwinger equations}\label{dse section}

Consider power series in the indeterminate $x$ with coefficients in
$\h$ where $x$ counts the loop number, that is the coefficient of
$x^k$ lives in the $k$th graded piece of $\h$.
By \emph{combinatorial Dyson-Schwinger equations} we will mean a
recursive equation, or system of recursive equations, in such power
series written in terms of insertion operations $B_+$.  The
particular form of combinatorial Dyson-Schwinger equation which we
will be able to analyze in detail will be discussed further in section
\ref{our setup}.

One of the most important examples is the case where the system of equations expresses the series of
graphs with a given external leg structure in terms of insertion into
all connected primitive graphs with that external leg structure. 
More specifically for a given primitive we insert into each of its
vertices the series for that vertex and for each edge all possible
powers of the series for that edge, that is, a geometric series in the
series for that edge.
The
system of such equations generates all 1PI graphs of the theory.  

\begin{example}\label{qed comb system}
For QED the system to generate all divergent 1PI graphs in the theory is
\begin{align*}
 X^{{\raisebox{-1ex}{\begin{fmfgraph}(100,100)
    \fmfleft{i}
    \fmfrightn{o}{2}
    \fmf{photon}{i,v}
    \fmf{fermion}{o1,v,o2}
\end{fmfgraph}}}} & = \mathbb{I} + \sum_{\substack{\gamma \text{
primitive with}\\\text{external leg structure {\raisebox{-1ex}{\begin{fmfgraph}(100,100)
    \fmfleft{i}
    \fmfrightn{o}{2}
    \fmf{photon}{i,v}
    \fmf{fermion}{o1,v,o2}
\end{fmfgraph}}}}}} x^{|\gamma|} B_+^\gamma\left(\frac{\left(X^{{\raisebox{-1ex}{\begin{fmfgraph}(100,100)
    \fmfleft{i}
    \fmfrightn{o}{2}
    \fmf{photon}{i,v}
    \fmf{fermion}{o1,v,o2}
\end{fmfgraph}}}}\right)^{1+2k}}{\left(X^{{\raisebox{-1ex}{\begin{fmfgraph}(100,60)
    \fmfleft{i}
    \fmfright{o}
    \fmf{photon}{i,o}
\end{fmfgraph}}}}\right)^k \left(X^{{\raisebox{-1ex}{\begin{fmfgraph}(100,60)
    \fmfleft{i}
    \fmfright{o}
    \fmf{fermion}{i,o}
\end{fmfgraph}}}}\right)^{2k}}\right) \\
X^{{\raisebox{-1ex}{\begin{fmfgraph}(100,60)
    \fmfleft{i}
    \fmfright{o}
    \fmf{photon}{i,o}
\end{fmfgraph}}}} 
& = \mathbb{I} - x B_+^{{\raisebox{-2ex}{\begin{fmfgraph}(200,120)
    \fmfleft{i}
    \fmfright{o}
    \fmf{photon}{i,v1}
    \fmf{photon}{v2,o}
    \fmf{fermion,left,tension=0.25}{v1,v2,v1}
\end{fmfgraph}}}}\left(\frac{\left(X^{{\raisebox{-1ex}{\begin{fmfgraph}(100,100)
    \fmfleft{i}
    \fmfrightn{o}{2}
    \fmf{photon}{i,v}
    \fmf{fermion}{o1,v,o2}
\end{fmfgraph}}}}\right)^{2}}{ \left(X^{{\raisebox{-1ex}{\begin{fmfgraph}(100,60)
    \fmfleft{i}
    \fmfright{o}
    \fmf{fermion}{i,o}
\end{fmfgraph}}}}\right)^{2}}\right) \\
X^{{\raisebox{-1ex}{\begin{fmfgraph}(100,60)
    \fmfleft{i}
    \fmfright{o}
    \fmf{fermion}{i,o}
\end{fmfgraph}}}} & = \mathbb{I} - x B_+^{{\raisebox{-2ex}{\begin{fmfgraph}(200,120)
    \fmfleft{i}
    \fmfright{o}
    \fmf{fermion}{i,v1,v2,o}
    \fmf{photon,left,tension=0.25}{v1,v2}
\end{fmfgraph}}}}\left(\frac{\left(X^{{\raisebox{-1ex}{\begin{fmfgraph}(100,100)
    \fmfleft{i}
    \fmfrightn{o}{2}
    \fmf{photon}{i,v}
    \fmf{fermion}{o1,v,o2}
\end{fmfgraph}}}}\right)^{2}}{X^{{\raisebox{-1ex}{\begin{fmfgraph}(100,60)
    \fmfleft{i}
    \fmfright{o}
    \fmf{photon}{i,o}
\end{fmfgraph}}}} X^{{\raisebox{-1ex}{\begin{fmfgraph}(100,60)
    \fmfleft{i}
    \fmfright{o}
    \fmf{fermion}{i,o}
\end{fmfgraph}}}}}\right).
\end{align*}
where $|\gamma|$ is the loop number of $\gamma$.  

$X^{{\raisebox{-1ex}{\begin{fmfgraph}(100,100)
    \fmfleft{i}
    \fmfrightn{o}{2}
    \fmf{photon}{i,v}
    \fmf{fermion}{o1,v,o2}
\end{fmfgraph}}}}$ is the vertex series.  The coefficient of $x^n$ in $X^{{\raisebox{-1ex}{\begin{fmfgraph}(100,100)
    \fmfleft{i}
    \fmfrightn{o}{2}
    \fmf{photon}{i,v}
    \fmf{fermion}{o1,v,o2}
\end{fmfgraph}}}}$ is the sum of all 1PI QED Feynman graphs with external
leg structure \raisebox{-1ex}{\begin{fmfgraph}(100,100)
    \fmfleft{i}
    \fmfrightn{o}{2}
    \fmf{photon}{i,v}
    \fmf{fermion}{o1,v,o2}
\end{fmfgraph}} and $n$ independent cycles.  In QED all graphs have symmetry factor $1$ so this
example hides the fact that in general each graph will appear weighted
with its symmetry factor.
$X^{{\raisebox{-1ex}{\begin{fmfgraph}(100,60)
    \fmfleft{i}
    \fmfright{o}
    \fmf{photon}{i,o}
\end{fmfgraph}}}}$ and $X^{{\raisebox{-1ex}{\begin{fmfgraph}(100,60)
    \fmfleft{i}
    \fmfright{o}
    \fmf{fermion}{i,o}
\end{fmfgraph}}}}$ are the two edge series.  The coefficient of $x^n$
for $n>0$ in $X^{{\raisebox{-1ex}{\begin{fmfgraph}(100,60)
    \fmfleft{i}
    \fmfright{o}
    \fmf{photon}{i,o}
\end{fmfgraph}}}}$ is minus the sum of all 1PI QED Feynman graphs
with external leg structure {\begin{fmfgraph}(100,60)
    \fmfleft{i}
    \fmfright{o}
    \fmf{photon}{i,o}
\end{fmfgraph}} and $n$ independent cycles.  The negative sign appears
in the edge series because when we use these series
we want their inverses; that is, we are interested in the
series where the coefficient of $x^n$ consists of products of graphs
each with a given edge as external leg structure and with total loop
number $n$.  The arguments to each $B_+^\gamma$ consist of a factor of
the vertex series in the numerator for each vertex of $\gamma$, a
factor of the photon edge series in the denominator for each photon
edge of $\gamma$, and a factor of the electron edge series in the
denominator for each electron edge of $\gamma$.

To illustrate these features lets work out the first few coefficients
of each series.  First work out the coefficient of $x$.
\begin{align*}
X^{{\raisebox{-1ex}{\begin{fmfgraph}(100,100)
    \fmfleft{i}
    \fmfrightn{o}{2}
    \fmf{photon}{i,v}
    \fmf{fermion}{o1,v,o2}
\end{fmfgraph}}}} 
& = \mathbb{I} +  
x B_+^{{\raisebox{-1ex}{\begin{fmfgraph}(150,150)
    \fmfleft{i}
    \fmfrightn{o}{2}
    \fmf{photon}{i,v0}
    \fmf{fermion}{o1,v1,v0,v2,o2}
    \fmffreeze
    \fmf{photon}{v1,v2}
\end{fmfgraph}}}}
\left(\frac{\left(X^{{\raisebox{-1ex}{\begin{fmfgraph}(100,100)
    \fmfleft{i}
    \fmfrightn{o}{2}
    \fmf{photon}{i,v}
    \fmf{fermion}{o1,v,o2}
\end{fmfgraph}}}}\right)^{3}}{X^{{\raisebox{-1ex}{\begin{fmfgraph}(100,60)
    \fmfleft{i}
    \fmfright{o}
    \fmf{photon}{i,o}
\end{fmfgraph}}}} \left(X^{{\raisebox{-1ex}{\begin{fmfgraph}(100,60)
    \fmfleft{i}
    \fmfright{o}
    \fmf{fermion}{i,o}
\end{fmfgraph}}}}\right)^2}\right) + O(x^2)\\
& = \mathbb{I} + x B_+^{{\raisebox{-1ex}{\begin{fmfgraph}(150,150)
    \fmfleft{i}
    \fmfrightn{o}{2}
    \fmf{photon}{i,v0}
    \fmf{fermion}{o1,v1,v0,v2,o2}
    \fmffreeze
    \fmf{photon}{v1,v2}
\end{fmfgraph}}}}(\mathbb{I}) + O(x^2) \\
& = \mathbb{I} + x\raisebox{-2.5ex}{\begin{fmfgraph}(200,200)
    \fmfleft{i}
    \fmfrightn{o}{2}
    \fmf{photon}{i,v0}
    \fmf{fermion}{o1,v1,v0,v2,o2}
    \fmffreeze
    \fmf{photon}{v1,v2}
\end{fmfgraph}} + O(x^2)
\end{align*}

\begin{align*}
X^{{\raisebox{-1ex}{\begin{fmfgraph}(100,60)
    \fmfleft{i}
    \fmfright{o}
    \fmf{photon}{i,o}
\end{fmfgraph}}}} 
& = \mathbb{I} - x B_+^{{\raisebox{-2ex}{\begin{fmfgraph}(200,120)
    \fmfleft{i}
    \fmfright{o}
    \fmf{photon}{i,v1}
    \fmf{photon}{v2,o}
    \fmf{fermion,left,tension=0.25}{v1,v2,v1}
\end{fmfgraph}}}}\left(\frac{\left(X^{{\raisebox{-1ex}{\begin{fmfgraph}(100,100)
    \fmfleft{i}
    \fmfrightn{o}{2}
    \fmf{photon}{i,v}
    \fmf{fermion}{o1,v,o2}
\end{fmfgraph}}}}\right)^{2}}{ \left(X^{{\raisebox{-1ex}{\begin{fmfgraph}(100,60)
    \fmfleft{i}
    \fmfright{o}
    \fmf{fermion}{i,o}
\end{fmfgraph}}}}\right)^{2}}\right) \\
& = \mathbb{I} - x B_+^{{\raisebox{-2ex}{\begin{fmfgraph}(200,120)
    \fmfleft{i}
    \fmfright{o}
    \fmf{photon}{i,v1}
    \fmf{photon}{v2,o}
    \fmf{fermion,left,tension=0.25}{v1,v2,v1}
\end{fmfgraph}}}}(\mathbb{I}) + O(x^2) \\
& = \mathbb{I} - x \raisebox{-1.5ex}{\begin{fmfgraph}(200,120)
    \fmfleft{i}
    \fmfright{o}
    \fmf{photon}{i,v1}
    \fmf{photon}{v2,o}
    \fmf{fermion,left,tension=0.25}{v1,v2,v1}
\end{fmfgraph}}+ O(x^2)
\end{align*}

\begin{align*}
  X^{{\raisebox{-1ex}{\begin{fmfgraph}(100,60)
    \fmfleft{i}
    \fmfright{o}
    \fmf{fermion}{i,o}
\end{fmfgraph}}}} & = \mathbb{I} - x B_+^{{\raisebox{-2ex}{\begin{fmfgraph}(200,120)
    \fmfleft{i}
    \fmfright{o}
    \fmf{fermion}{i,v1,v2,o}
    \fmf{photon,left,tension=0.25}{v1,v2}
\end{fmfgraph}}}}\left(\frac{\left(X^{{\raisebox{-1ex}{\begin{fmfgraph}(100,100)
    \fmfleft{i}
    \fmfrightn{o}{2}
    \fmf{photon}{i,v}
    \fmf{fermion}{o1,v,o2}
\end{fmfgraph}}}}\right)^{2}}{X^{{\raisebox{-1ex}{\begin{fmfgraph}(100,60)
    \fmfleft{i}
    \fmfright{o}
    \fmf{photon}{i,o}
\end{fmfgraph}}}} X^{{\raisebox{-1ex}{\begin{fmfgraph}(100,60)
    \fmfleft{i}
    \fmfright{o}
    \fmf{fermion}{i,o}
\end{fmfgraph}}}}}\right) \\
& = \mathbb{I} - x B_+^{{\raisebox{-2ex}{\begin{fmfgraph}(200,120)
    \fmfleft{i}
    \fmfright{o}
    \fmf{fermion}{i,v1,v2,o}
    \fmf{photon,left,tension=0.25}{v1,v2}
\end{fmfgraph}}}}(\mathbb{I}) + O(x^2) \\
& = \mathbb{I} - x \raisebox{-1.5ex}{\begin{fmfgraph}(200,120)
    \fmfleft{i}
    \fmfright{o}
    \fmf{fermion}{i,v1,v2,o}
    \fmf{photon,left,tension=0.25}{v1,v2}
\end{fmfgraph}} + O(x^2)
\end{align*}

Next work out the coefficient of $x^2$.
\begin{align*}
X^{{\raisebox{-1ex}{\begin{fmfgraph}(100,100)
    \fmfleft{i}
    \fmfrightn{o}{2}
    \fmf{photon}{i,v}
    \fmf{fermion}{o1,v,o2}
\end{fmfgraph}}}} 
& = \mathbb{I} +  
x B_+^{{\raisebox{-1ex}{\begin{fmfgraph}(150,150)
    \fmfleft{i}
    \fmfrightn{o}{2}
    \fmf{photon}{i,v0}
    \fmf{fermion}{o1,v1,v0,v2,o2}
    \fmffreeze
    \fmf{photon}{v1,v2}
\end{fmfgraph}}}}
\left(\frac{\left(X^{{\raisebox{-1ex}{\begin{fmfgraph}(100,100)
    \fmfleft{i}
    \fmfrightn{o}{2}
    \fmf{photon}{i,v}
    \fmf{fermion}{o1,v,o2}
\end{fmfgraph}}}}\right)^{3}}{X^{{\raisebox{-1ex}{\begin{fmfgraph}(100,60)
    \fmfleft{i}
    \fmfright{o}
    \fmf{photon}{i,o}
\end{fmfgraph}}}} \left(X^{{\raisebox{-1ex}{\begin{fmfgraph}(100,60)
    \fmfleft{i}
    \fmfright{o}
    \fmf{fermion}{i,o}
\end{fmfgraph}}}}\right)^2}\right)  \\
& \qquad + x^2 B_+^{{\raisebox{-1ex}{\begin{fmfgraph}(300,200)
    \fmfleft{i}
    \fmfrightn{o}{2}
    \fmf{photon}{i,v0}
    \fmf{fermion}{o1,v1,v3,v0,v4,v2,o2}
    \fmffreeze
    \fmf{photon}{v1,v4}
    \fmf{photon}{v3,v2}
\end{fmfgraph}}}}
\left(\frac{\left(X^{{\raisebox{-1ex}{\begin{fmfgraph}(100,100)
    \fmfleft{i}
    \fmfrightn{o}{2}
    \fmf{photon}{i,v}
    \fmf{fermion}{o1,v,o2}
\end{fmfgraph}}}}\right)^{5}}{\left(X^{{\raisebox{-1ex}{\begin{fmfgraph}(100,60)
    \fmfleft{i}
    \fmfright{o}
    \fmf{photon}{i,o}
\end{fmfgraph}}}}\right)^2 \left(X^{{\raisebox{-1ex}{\begin{fmfgraph}(100,60)
    \fmfleft{i}
    \fmfright{o}
    \fmf{fermion}{i,o}
\end{fmfgraph}}}}\right)^4}\right)
+ O(x^3)\\
& = \mathbb{I} + x B_+^{{\raisebox{-1ex}{\begin{fmfgraph}(150,150)
    \fmfleft{i}
    \fmfrightn{o}{2}
    \fmf{photon}{i,v0}
    \fmf{fermion}{o1,v1,v0,v2,o2}
    \fmffreeze
    \fmf{photon}{v1,v2}
\end{fmfgraph}}}}\left(\frac{\left(\mathbb{I} + x\raisebox{-2.5ex}{\begin{fmfgraph}(200,200)
    \fmfleft{i}
    \fmfrightn{o}{2}
    \fmf{photon}{i,v0}
    \fmf{fermion}{o1,v1,v0,v2,o2}
    \fmffreeze
    \fmf{photon}{v1,v2}
\end{fmfgraph}}\right)^3}{\left(\mathbb{I} - x \raisebox{-1.5ex}{\begin{fmfgraph}(200,120)
    \fmfleft{i}
    \fmfright{o}
    \fmf{photon}{i,v1}
    \fmf{photon}{v2,o}
    \fmf{fermion,left,tension=0.25}{v1,v2,v1}
\end{fmfgraph}}\right)\left(\mathbb{I} - x \raisebox{-1.5ex}{\begin{fmfgraph}(200,120)
    \fmfleft{i}
    \fmfright{o}
    \fmf{fermion}{i,v1,v2,o}
    \fmf{photon,left,tension=0.25}{v1,v2}
\end{fmfgraph}}\right)^2}\right) + x^2 B_+^{{\raisebox{-1ex}{\begin{fmfgraph}(300,200)
    \fmfleft{i}
    \fmfrightn{o}{2}
    \fmf{photon}{i,v0}
    \fmf{fermion}{o1,v1,v3,v0,v4,v2,o2}
    \fmffreeze
    \fmf{photon}{v1,v4}
    \fmf{photon}{v3,v2}
\end{fmfgraph}}}}(\mathbb{I}) + O(x^3) \\
& = \mathbb{I} + x\raisebox{-2.5ex}{\begin{fmfgraph}(200,200)
    \fmfleft{i}
    \fmfrightn{o}{2}
    \fmf{photon}{i,v0}
    \fmf{fermion}{o1,v1,v0,v2,o2}
    \fmffreeze
    \fmf{photon}{v1,v2}
\end{fmfgraph}} + 
x^2B_+^{{\raisebox{-1ex}{\begin{fmfgraph}(150,150)
    \fmfleft{i}
    \fmfrightn{o}{2}
    \fmf{photon}{i,v0}
    \fmf{fermion}{o1,v1,v0,v2,o2}
    \fmffreeze
    \fmf{photon}{v1,v2}
\end{fmfgraph}}}}\left(3\raisebox{-2.5ex}{\begin{fmfgraph}(200,200)
    \fmfleft{i}
    \fmfrightn{o}{2}
    \fmf{photon}{i,v0}
    \fmf{fermion}{o1,v1,v0,v2,o2}
    \fmffreeze
    \fmf{photon}{v1,v2}
\end{fmfgraph}}+ \raisebox{-1.5ex}{\begin{fmfgraph}(200,120)
    \fmfleft{i}
    \fmfright{o}
    \fmf{photon}{i,v1}
    \fmf{photon}{v2,o}
    \fmf{fermion,left,tension=0.25}{v1,v2,v1}
\end{fmfgraph}}+ 2\raisebox{-1.5ex}{\begin{fmfgraph}(200,120)
    \fmfleft{i}
    \fmfright{o}
    \fmf{fermion}{i,v1,v2,o}
    \fmf{photon,left,tension=0.25}{v1,v2}
\end{fmfgraph}}\right)
+ x^2\raisebox{-3ex}{\begin{fmfgraph}(300,200)
    \fmfleft{i}
    \fmfrightn{o}{2}
    \fmf{photon}{i,v0}
    \fmf{fermion}{o1,v1,v3,v0,v4,v2,o2}
    \fmffreeze
    \fmf{photon}{v1,v4}
    \fmf{photon}{v3,v2}
\end{fmfgraph}} + O(x^3) \\
& = \mathbb{I} + x\raisebox{-2.5ex}{\begin{fmfgraph}(200,200)
    \fmfleft{i}
    \fmfrightn{o}{2}
    \fmf{photon}{i,v0}
    \fmf{fermion}{o1,v1,v0,v2,o2}
    \fmffreeze
    \fmf{photon}{v1,v2}
\end{fmfgraph}} + 
x^2\Bigg(
\raisebox{-3ex}{\begin{fmfgraph}(300,200)
    \fmfleft{i}
    \fmfrightn{o}{2}
    \fmf{photon}{i,v0}
    \fmf{fermion}{o1,v1,v3,v0,v4,v2,o2}
    \fmffreeze
    \fmf{photon}{v1,v2}
    \fmf{photon}{v3,v4}
\end{fmfgraph}}
+\raisebox{-3ex}{\begin{fmfgraph}(300,200)
    \fmfleft{i}
    \fmfrightn{o}{2}
    \fmf{photon}{i,v0}
    \fmf{fermion}{o1,v4,v1,v3,v0,v2,o2}
    \fmffreeze
    \fmf{photon}{v1,v2}
    \fmf{photon,right,tension=0.25}{v3,v4}
\end{fmfgraph}}
+\raisebox{-3ex}{\begin{fmfgraph}(300,200)
    \fmfleft{i}
    \fmfrightn{o}{2}
    \fmf{photon}{i,v0}
    \fmf{fermion}{o1,v1,v0,v3,v2,v4,o2}
    \fmffreeze
    \fmf{photon}{v1,v2}
    \fmf{photon,left,tension=0.25}{v3,v4}
\end{fmfgraph}} \\
& \qquad 
+\raisebox{-3ex}{\begin{fmfgraph}(300,200)
    \fmfleft{i}
    \fmfrightn{o}{2}
    \fmf{photon}{i,v0}
    \fmf{fermion}{o1,v1,v4,v3,v0,v2,o2}
    \fmffreeze
    \fmf{photon}{v1,v2}
    \fmf{photon,right,tension=0.25}{v3,v4}
\end{fmfgraph}}
+\raisebox{-3ex}{\begin{fmfgraph}(300,200)
    \fmfleft{i}
    \fmfrightn{o}{2}
    \fmf{photon}{i,v0}
    \fmf{fermion}{o1,v1,v0,v3,v4,v2,o2}
    \fmffreeze
    \fmf{photon}{v1,v2}
    \fmf{photon,left,tension=0.25}{v3,v4}
\end{fmfgraph}}
+\raisebox{-3ex}{\begin{fmfgraph}(300,200)
    \fmfleft{i}
    \fmfrightn{o}{2}
    \fmf{photon}{i,v0}
    \fmf{fermion}{o1,v1,v0,v2,o2}
    \fmffreeze
    \fmf{photon}{v1,v3}
    \fmf{photon}{v4,v2}
    \fmf{fermion,left,tension=0.25}{v3,v4,v3}
\end{fmfgraph}}
+\raisebox{-3ex}{\begin{fmfgraph}(300,200)
    \fmfleft{i}
    \fmfrightn{o}{2}
    \fmf{photon}{i,v0}
    \fmf{fermion}{o1,v1,v3,v0,v4,v2,o2}
    \fmffreeze
    \fmf{photon}{v1,v4}
    \fmf{photon}{v3,v2}
\end{fmfgraph}}\Bigg) + O(x^3)
\end{align*}

\begin{align*}
X^{{\raisebox{-1ex}{\begin{fmfgraph}(100,60)
    \fmfleft{i}
    \fmfright{o}
    \fmf{photon}{i,o}
\end{fmfgraph}}}} 
& = \mathbb{I} - x B_+^{{\raisebox{-2ex}{\begin{fmfgraph}(200,120)
    \fmfleft{i}
    \fmfright{o}
    \fmf{photon}{i,v1}
    \fmf{photon}{v2,o}
    \fmf{fermion,left,tension=0.25}{v1,v2,v1}
\end{fmfgraph}}}}\left(\frac{\left(X^{{\raisebox{-1ex}{\begin{fmfgraph}(100,100)
    \fmfleft{i}
    \fmfrightn{o}{2}
    \fmf{photon}{i,v}
    \fmf{fermion}{o1,v,o2}
\end{fmfgraph}}}}\right)^{2}}{ \left(X^{{\raisebox{-1ex}{\begin{fmfgraph}(100,60)
    \fmfleft{i}
    \fmfright{o}
    \fmf{fermion}{i,o}
\end{fmfgraph}}}}\right)^{2}}\right) \\
& = \mathbb{I} - x \raisebox{-1.5ex}{\begin{fmfgraph}(200,120)
    \fmfleft{i}
    \fmfright{o}
    \fmf{photon}{i,v1}
    \fmf{photon}{v2,o}
    \fmf{fermion,left,tension=0.25}{v1,v2,v1}
\end{fmfgraph}} - x^2B_+^{{\raisebox{-2ex}{\begin{fmfgraph}(200,120)
    \fmfleft{i}
    \fmfright{o}
    \fmf{photon}{i,v1}
    \fmf{photon}{v2,o}
    \fmf{fermion,left,tension=0.25}{v1,v2,v1}
\end{fmfgraph}}}}\left(2\raisebox{-2.5ex}{\begin{fmfgraph}(200,200)
    \fmfleft{i}
    \fmfrightn{o}{2}
    \fmf{photon}{i,v0}
    \fmf{fermion}{o1,v1,v0,v2,o2}
    \fmffreeze
    \fmf{photon}{v1,v2}
\end{fmfgraph}} + 2\raisebox{-1.5ex}{\begin{fmfgraph}(200,120)
    \fmfleft{i}
    \fmfright{o}
    \fmf{fermion}{i,v1,v2,o}
    \fmf{photon,left,tension=0.25}{v1,v2}
\end{fmfgraph}}
\right) + O(x^3) \\
& = \mathbb{I} - x \raisebox{-1.5ex}{\begin{fmfgraph}(200,120)
    \fmfleft{i}
    \fmfright{o}
    \fmf{photon}{i,v1}
    \fmf{photon}{v2,o}
    \fmf{fermion,left,tension=0.25}{v1,v2,v1}
\end{fmfgraph}} 
- x^2\bigg(\raisebox{-4ex}{\begin{fmfgraph}(300,300)
    \fmfleft{i}
    \fmfright{o}
    \fmftop{t}
    \fmfbottom{b}
    \fmf{photon}{i,v1}
    \fmf{photon}{v2,o}
    \fmf{phantom}{t,v3}
    \fmf{phantom}{b,v4}
    \fmf{fermion}{v1,v4,v2,v3,v1}
    \fmffreeze
    \fmf{photon}{v3,v4}
\end{fmfgraph}}
+ \raisebox{-3ex}{\begin{fmfgraph}(300,200)
    \fmfleft{i}
    \fmfright{o}
    \fmf{photon}{i,v1}
    \fmf{photon}{v2,o}
    \fmf{fermion,left,tension=0.25}{v1,v2}
    \fmf{fermion}{v2,v3,v4,v1}
    \fmf{photon,left,tension=0.25}{v3,v4}
\end{fmfgraph}}
+ \raisebox{-3ex}{\begin{fmfgraph}(300,200)
    \fmfleft{i}
    \fmfright{o}
    \fmf{photon}{i,v1}
    \fmf{photon}{v2,o}
    \fmf{fermion,left,tension=0.25}{v2,v1}
    \fmf{fermion}{v1,v3,v4,v2}
    \fmf{photon,left,tension=0.25}{v3,v4}
\end{fmfgraph}}
\bigg) + O(x^3)
\end{align*}

\begin{align*}
  X^{{\raisebox{-1ex}{\begin{fmfgraph}(100,60)
    \fmfleft{i}
    \fmfright{o}
    \fmf{fermion}{i,o}
\end{fmfgraph}}}} & = \mathbb{I} - x B_+^{{\raisebox{-2ex}{\begin{fmfgraph}(200,120)
    \fmfleft{i}
    \fmfright{o}
    \fmf{fermion}{i,v1,v2,o}
    \fmf{photon,left,tension=0.25}{v1,v2}
\end{fmfgraph}}}}\left(\frac{\left(X^{{\raisebox{-1ex}{\begin{fmfgraph}(100,100)
    \fmfleft{i}
    \fmfrightn{o}{2}
    \fmf{photon}{i,v}
    \fmf{fermion}{o1,v,o2}
\end{fmfgraph}}}}\right)^{2}}{X^{{\raisebox{-1ex}{\begin{fmfgraph}(100,60)
    \fmfleft{i}
    \fmfright{o}
    \fmf{photon}{i,o}
\end{fmfgraph}}}} X^{{\raisebox{-1ex}{\begin{fmfgraph}(100,60)
    \fmfleft{i}
    \fmfright{o}
    \fmf{fermion}{i,o}
\end{fmfgraph}}}}}\right) \\
& = \mathbb{I} - x \raisebox{-1.5ex}{\begin{fmfgraph}(200,120)
    \fmfleft{i}
    \fmfright{o}
    \fmf{fermion}{i,v1,v2,o}
    \fmf{photon,left,tension=0.25}{v1,v2}
\end{fmfgraph}} - x^2B_+^{{\raisebox{-2ex}{\begin{fmfgraph}(200,120)
    \fmfleft{i}
    \fmfright{o}
    \fmf{fermion}{i,v1,v2,o}
    \fmf{photon,left,tension=0.25}{v1,v2}
\end{fmfgraph}}}}\left(2\raisebox{-2.5ex}{\begin{fmfgraph}(200,200)
    \fmfleft{i}
    \fmfrightn{o}{2}
    \fmf{photon}{i,v0}
    \fmf{fermion}{o1,v1,v0,v2,o2}
    \fmffreeze
    \fmf{photon}{v1,v2}
\end{fmfgraph}}+ \raisebox{-1.5ex}{\begin{fmfgraph}(200,120)
    \fmfleft{i}
    \fmfright{o}
    \fmf{photon}{i,v1}
    \fmf{photon}{v2,o}
    \fmf{fermion,left,tension=0.25}{v1,v2,v1}
\end{fmfgraph}}+ \raisebox{-1.5ex}{\begin{fmfgraph}(200,120)
    \fmfleft{i}
    \fmfright{o}
    \fmf{fermion}{i,v1,v2,o}
    \fmf{photon,left,tension=0.25}{v1,v2}
\end{fmfgraph}}\right)+ O(x^3) \\
& = \mathbb{I} - x \raisebox{-1.5ex}{\begin{fmfgraph}(200,120)
    \fmfleft{i}
    \fmfright{o}
    \fmf{fermion}{i,v1,v2,o}
    \fmf{photon,left,tension=0.25}{v1,v2}
\end{fmfgraph}} - x^2\left(
\raisebox{-3ex}{\begin{fmfgraph}(300,200)
    \fmfleft{i}
    \fmfright{o}
    \fmf{fermion}{i,v3,v1,v4,v2,o}
    \fmf{photon,left,tension=0.25}{v1,v2}
    \fmf{photon,right,tension=0.25}{v3,v4}
\end{fmfgraph}}
+\raisebox{-3ex}{\begin{fmfgraph}(300,200)
    \fmfleft{i}
    \fmfright{o}
    \fmf{fermion}{i,v1}
    \fmf{fermion,right,tension=0.25}{v1,v2}
    \fmf{fermion}{v2,o}
    \fmf{photon}{v1,v3}
    \fmf{photon}{v4,v2}
    \fmf{fermion,right,tension=0.25}{v3,v4,v3}
\end{fmfgraph}}
+\raisebox{-3ex}{\begin{fmfgraph}(300,200)
    \fmfleft{i}
    \fmfright{o}
    \fmf{fermion}{i,v1,v3,v4,v2,o}
    \fmf{photon,left,tension=0.25}{v1,v2}
    \fmf{photon,right,tension=0.25}{v3,v4}
\end{fmfgraph}}
\right)+ O(x^3)
\end{align*}

The fact that 
\[
\raisebox{-4ex}{\begin{fmfgraph}(300,300)
    \fmfleft{i}
    \fmfright{o}
    \fmftop{t}
    \fmfbottom{b}
    \fmf{photon}{i,v1}
    \fmf{photon}{v2,o}
    \fmf{phantom}{t,v3}
    \fmf{phantom}{b,v4}
    \fmf{fermion}{v1,v4,v2,v3,v1}
    \fmffreeze
    \fmf{photon}{v3,v4}
\end{fmfgraph}} \qquad \text{and}\qquad \raisebox{-3ex}{\begin{fmfgraph}(300,200)
    \fmfleft{i}
    \fmfright{o}
    \fmf{fermion}{i,v3,v1,v4,v2,o}
    \fmf{photon,left,tension=0.25}{v1,v2}
    \fmf{photon,right,tension=0.25}{v3,v4}
\end{fmfgraph}}
\]
appear with coefficient 1 and not 2 is due to the two insertion trees
contributing a 2 to the denominator in Definition \ref{B+ def}.

\end{example}

By \emph{analytic Dyson-Schwinger equations} we will mean the result
of applying the renormalized Feynman rules to combinatorial
Dyson-Schwinger equations.  These are the Dyson-Schwinger equations
which a physicist would recognize.  The counting variable $x$ becomes
the physicists' coupling constant (which we will also denote $x$, but
which might be more typically denoted $\alpha$ or $g^2$ depending on
the theory).  The Feynman rules also introduce one or more scale
variables $L_j$ which come from the external momenta $q_i$ and the
fixed momentum values $\mu_i$ used to renormalize by subtracting.  
In the case of one scale variable we have $L = \log q^2/\mu^2$.   See
Example \ref{bkerfc setup}.  Note
that in the case of more than one scale the $L_j$ are not just $\log
q_i^2/\mu^2$, but also include other expressions in the $q_i$ and the
$\mu_i$, such as ratios of the $q_i$ (such ratios are not
properly speaking scales, but there is no need for a more appropriate
name for them since we will quickly move to the case of one scale
where this problem does not come up). 

The functions of $L_j$ and $x$ appearing in analytic Dyson-Schwinger
equations are called Green functions, particularly in the case where
the Green functions are the result of applying the renormalized
Feynman rules to the series of all graphs with a given external leg
structure.

We can begin to disentangle the analytic and combinatorial information
in the following way.  Suppose we have a
combinatorial Dyson-Schwinger equation, potentially a system.  Suppose the series in Feynman graphs
appearing in the Dyson-Schwinger equation are denoted $X^r$ with $r
\in \mathcal{R}$ some index set.  Denote $G^r$ the corresponding
Green functions.  

For each factor $(X^r)^s$ in the argument to some $B_+^\gamma$ take the formal integrand and multiply it by
$(G^r)^s$.
For the scale arguments to these $G^r$ use the momenta of the edges
where the graphs of $X^r$ are inserted.  Then subtract this integral
at the fixed external momenta $\mu_i$ as when renormalizing a single Feynman integral.  Then the analytic Dyson-Schwinger
equation has the same form as the combinatorial one but
with $G^r$ replacing $X^r$ and with the expression described above
replacing $B_+^\gamma$.  Example \ref{bkerfc setup} illustrates this
procedure.  

In the case with more than one scale the Green functions
may depend on ratios of the different momenta, and we can progress no
further in simplifying the setup.  
Fortunately, in the case with only one scale, which suffices to describe the
general case in view of Chapter \ref{first reduction}, we can further disentangle the analytic and combinatorial information
as follows, see \cite{etude} for more details.  

Suppose we have a
combinatorial Dyson-Schwinger equation and a single scale.  For each primitive
graph $\gamma$ appearing as a $B^\gamma_+$ we have a formal integral
expression \[\int_{\mathbb{R}^{D|v|}}\mathrm{Int} \prod_{k\in v}d^D k\] coming from the
unrenormalized Feynman rules.  Number the edges, say from $1$
to $n$.  Raise the factor associated to the $i$th edge to $1+\rho_i$
where $\rho_i$ is a new variable.  We now have
an analytically regularized integral which can be evaluated for
suitable values of $\rho_i$.
%Now make the single subtraction at fixed external momenta which
%is all that is necessary to renormalize the Feynman integral for a
%primitive graph.  
Finally set all external momenta to 1.
Call the resulting function of $\rho_1, \ldots, \rho_n$
the Mellin transform $F_\gamma(\rho_1, \ldots, \rho_n)$ associated to
$\gamma$.  We are interested in $F_\gamma$ near the origin.

Then, another way to see the analytic Dyson-Schwinger equation as
coming from the combinatorial Dyson-Schwinger equation by replacing
$X^r$ with $G^r$ and 
$B^\gamma_+$ with $F_\gamma$.  The factor with exponent $\rho_i$
indicates the argument for the recursive
appearance of the $X^j$ which is inserted at the insertion place
corresponding to edge $i$.  This will be made precise for the cases of
interest in the following section, and will be motivated by Example
\ref{motivate rho}.

\begin{example}\label{bkerfc setup}
Broadhurst and Kreimer in \cite{bkerfc} discuss the
Dyson-Schwinger equation for graphs from massless Yukawa theory where
powers of the one loop fermion self energy 
\raisebox{-1.5ex}{\begin{fmfgraph}(300,120)
    \fmfleft{i}
    \fmfright{o}
    \fmf{fermion}{i,v1,v2,o}
    \fmf{dashes,left,tension=0.25}{v1,v2}
    \end{fmfgraph}}
are inserted into itself.  The result is that they consider any graph made of
nestings and chainings of this one primitive, for example
\[
\raisebox{-8ex}{\begin{fmfgraph}(1200,480)
    \fmfleft{i}
    \fmfright{o}
    \fmf{fermion}{i,v1,v2,v3,v4,v5,v6,v7,v8,o}
    \fmf{dashes,left,tension=0.25}{v1,v8}
    \fmf{dashes,left,tension=0.25}{v2,v5}
    \fmf{dashes,left,tension=0.25}{v3,v4}
    \fmf{dashes,left,tension=0.25}{v6,v7}
  \end{fmfgraph}}.
\]
A graph like
\[
\raisebox{-4ex}{\begin{fmfgraph}(800,480)
    \fmfleft{i}
    \fmfright{o}
    \fmf{fermion}{i,v1,v2,v3,v4,v5,v6,o}
    \fmf{dashes,left,tension=0.25}{v1,v6}
    \fmf{dashes,left,tension=0.25}{v2,v4}
    \fmf{dashes,left,tension=0.25}{v5,v3}
  \end{fmfgraph}}
\]
is not allowed.
These graphs are in one-to-one correspondence with planar rooted
trees.  The combinatorial
Dyson-Schwinger equation is
\[
  X(x) = \mathbb{I} - x B_+\left(\frac{1}{X(x)}\right).
\]
The Mellin transform associated to the single one loop primitive
\[
  \raisebox{-1.5ex}{\begin{fmfgraph*}(300,120)
    \fmfleft{i}
    \fmfright{o}
    \fmflabel{$q$}{i}
    \fmflabel{$q$}{o}
    \fmf{fermion}{i,v1}
    \fmf{fermion}{v2,o}
    \fmf{fermion,label=$k$}{v1,v2}
    \fmf{dashes,label=$k+q$,left,tension=0.25}{v1,v2}
    \end{fmfgraph*}}
\] 
is,
according to the Feynman rules of Yukawa theory,
\[
  F(\rho_1, \rho_2) = \frac{1}{q^2}\int d^4 k \frac{k \cdot q}{(k^2)^{1+\rho_1}
  ((k+q)^2)^{1+\rho_2}}\bigg|_{q^2 = 1}.
\]
However we are only inserting in the insertion place corresponding to
$\rho_1$ so the Mellin transform we're actually interested in is
\[
  F(\rho) = \frac{1}{q^2}\int d^4 k \frac{k \cdot q}{(k^2)^{1+\rho}
  (k+q)^2}\bigg|_{q^2 = 1}.
\]

Next combine these two facts as described above
to get that the Green function satisfies the analytic Dyson-Schwinger equation
\[
G(x, L) = 1 - \left(\frac{x}{q^2}\int d^4 k \frac{k \cdot q}{k^2 G(x,
  \log (k^2/\mu^2))(k+q)^2}  - \cdots \bigg|_{q^2 = \mu^2}\right)
\]
where $L = \log(q^2/\mu^2)$ and 
$\cdots$ stands for the same integrand evaluated as specified.
  This is the same as what we would have
obtained from applying the Feynman rules directly to the combinatorial
Dyson-Schwinger equation.  

%Lets conclude the example by calculating $F(\rho)$ for this example.
%First rewrite the integrand.
%\begin{align*}
%  & \frac{1}{q^2}\int d^4 k \frac{k \cdot q}{(k^2)^{1+\rho}
%  (k+q)^2} - \cdots \bigg|_{q^2 = \mu^2} \\
%  & = \frac{1}{2 q^2}\int d^4 k
%   \left(
%      \frac{1}{(k^2)^{1+\rho}} -  \frac{1}{(k^2)^\rho(k+q)^2} - q^2 \frac{1}{(k^2)^{1+\rho} (k+q)^2}  \right) - \cdots \bigg|_{q^2 = \mu^2} \\
%\end{align*}
%using
%\[
%  \frac{k\cdot q}{(k+q)^2} = \frac{1}{2} - \frac{k^2 + q^2}{2(k+q)^2}.
%\]
%Next calculate each term in turn.
%
%\begin{align*}
%  \int d^4 k
%      \frac{1}{(k^2)^{1+\rho}}  & = 2\pi^2 \int_0^\infty d |k| |k|^{3-2-2\rho}
%      \\
%      & = \pi^2\frac{(k^2){1-\rho}}{1-\rho} \bigg|_0^\infty \\
%      & = 
%\end{align*}

\end{example}

\section{Setup}\label{our setup}

We will restrict our attention to Dyson-Schwinger equations of the
following form.

\subsection{Single equations}\label{pre single}
Fix $s\in \mathbb{Z}$.  The case $s=0$ is not of particular interest
since it corresponds to the strictly simpler linear situation
discussed in \cite{linetude}.  However, to include $s=0$ as well, we
will make the convention that $\sgn(0) = 1$.

Let $Q = X^{-s}$.  We call $Q$ the combinatorial invariant charge.  Applying the Feynman rules
to $Q$ gives the usual physicists' invariant charge.  
%Sometimes conventions
%differ so that the Feynman rules applied to $Q$ here is the square of what gets called the invariant
%charge.

Consider the Dyson-Schwinger equation
\begin{equation}\label{comb dse}
  X(x) = \mathbb{I} - \sgn(s) \sum_{k\geq 1}\sum_{i=0}^{t_k}x^k
  B_+^{k,i}(X Q^{k}).
\end{equation}
This includes Example \ref{bkerfc setup} where $s=2$ and there is only
one $B_+$ having $k=1$.

 Let
$
  F_{k,i}(\rho_1, \ldots, \rho_{n})
$
be the Mellin transform associated to the primitive
$B_+^{k,i}(\mathbb{I})$.
In view of Chapter \ref{first reduction} we're primarily interested in
the case where $n=1$ at which point we'll assume that the Mellin transforms of the primitives each have a
simple pole at $\rho=0$, which is the case in physical examples.   
We expand the Green functions in a series in $x$ and in $L$ (which
will in general be merely an asymptotic expansion in $x$) using the
following notation
\begin{equation}\label{expand}
  G(x,L) = 1 -\sgn(s) \sum_{k \geq 1}\gamma_k(x) L^k \qquad \gamma_k(x) =
  \sum_{j \geq k} \gamma_{k,j} x^j
\end{equation}

%The idea is that the analytic Dyson-Schwinger equation reads, for $s>0$
%\begin{align*}
%  & G(x,L) \\
%  & = 1 - \sum_{k\geq 1}\sum_{i=0}^{t_k}x^k
%   F_{k,i}(\log_{\ell_1}(G(x, \log(\ell_1/\mu^2))),\ldots, \log_{\ell_{ks-1}}(G(x, \log(\ell_{ks-1}/\mu^2))))
%\end{align*}
%where $\ell_j$ is the base of $\rho_j$, that is
%$\exp(-(\partial_{\rho_j}F_{k,i})/F_{k,i})$, and $\mu^2$ is the scale
%we use in the subtractions used to renormalize.  We can make similar
%definitions for $s<0$ and more than one scale as for vertex
%insertions.  

The idea is to follow the prescriptions of the previous section to
obtain the analytic Dyson-Schwinger equation, then simplify the
resulting expression by following the following steps.  See Example
\ref{motivate rho} for a worked example.  First, expand $G$ as a
series in $L$.  Second, convert the resulting logarithms of the
integration variables into derivatives via the identity
$\partial_{\rho}^k y^{-\rho}|_{\rho=0} = (-1)^k \log^k(y)$.  The
choice of name for the new variable $\rho$ is not coincidental.  Third,
switch the order of integration and derivation.  The result then is a
complicated expression in derivatives of the Mellin transforms of the
primitives. 

However, 
to avoid the need for additional notation and for appropriate
assumptions on the $F_{k,i}$, instead of following this path we will
instead define our analytic Dyson-Schwinger equations to be the final
result of this procedure.

\begin{definition}\label{dot def}
For a single scale $\mu^2$,
the analytic Dyson-Schwinger equation associated to \eqref{comb dse} is 
\begin{align*}
  G(x, L) 
   = 1 - \sgn(s)\sum_{k \geq
  1}\sum_{i=0}^{t_k}x^k&G(x,\partial_{-\rho_1})^{-\sgn(s)}\cdots
  G(x,\partial_{-\rho_{n_k}})^{-\sgn(s)} \\
  &  (e^{-L(\rho_1+\cdots+\rho_{n_k})}-1)F^{k,i}(\rho_1, \ldots, \rho_{n_k})
  \bigg|_{\rho_1 = \cdots = \rho_{n_k}=0}
\end{align*}
where $n_k = \sgn(s)(sk-1)$.
%In general we have
%\begin{multline*}
%  G(x, L_1, \ldots, L_j) = 1 -\sgn(s) \sum_{k \geq
%  1}\sum_{i=0}^{t_k}x^k \widetilde{G}(x, v_1)^{-\sgn(s)} \cdots
%  \widetilde{G}(x, v_{\sgn(s)(sk-1)})^{-\sgn(s)} \\
%  (e^{-L(\rho_1+\cdots+\rho_{n_{k,i}})}-1)F^{k,i}(\rho_1, \ldots, \rho_{n_{k,i}})
%  \bigg|_{\rho_1 = \cdots = \rho_{n_{k,i}}=0}
%\end{multline*}
%where $j+1$ is the number of external momenta, $n_{k,i}$ is the number
%of edges which are insertion places or are adjacent to vertex
%insertion places in
%$B_+^{k,i}(\mathbb{I})$, and $\widetilde{G}(x,v_i)$ is a sum of
%expressions $G(x,a_1, \cdots, a_j)$ where $a_i$ is either some $L_m$
%or some $\partial_{-\rho_m}$ and the sum runs over such expressions where
%there is an insertion of $B_+^{k,i}(\mathbb{I})$ into itself at edge
%or vertex
%$v_i$ with the external edge corresponding to $L_i$ mapped to the
%external edge corresponding to $L_m$ or to the internal edge
%corresponding to $\rho_m$ according to the value of $a_i$.
\end{definition}

We only need one subtraction because in view of the discussion at the
end of Subsection \ref{renormalization} all the integrals of interest
are log divergent.

In view of the following chapters we need not concern ourselves with
the complexity of the general definition as we will further reduce to the case where
there is only one symmetric insertion place and a single scale giving
\[
G(x,L) = 1 - \sgn(s)\sum_{k \geq 1}\sum_{i=0}^{t_k} x^k G(x,\partial_{-\rho})^{1-sk}(e^{-L\rho}-1)F^{k,i}(\rho)\bigg|_{\rho=0}
\]
or rewritten
\begin{equation}\label{dot eq}
  \gamma \cdot L = \sum_{k \geq 1} x^k (1-\sgn(s)\gamma \cdot
  \partial_{-\rho})^{1-sk}(e^{-L\rho}-1)F^{k}(\rho)\bigg|_{\rho=0}
\end{equation}
where $\gamma \cdot U = \sum \gamma_k U^k$, $F^k(\rho) = \sum_{i=0}^{t_k}F^{k,i}(\rho)$.

%We will additionally assume the following.
%\begin{itemize}
% \item $F_{k,i}(\ldots, -\log_{\ell_j}(\sum_{t \geq 0}a_t H_t),
%   \ldots) = \sum_{t \geq 0}a_t F_{k,i}(\ldots, -\log_{\ell_j}H_t,
%   \ldots)$
% \item $F_{k,i}(\ldots, -\log_{\ell_j}(x H),
%   \ldots) = x F_{k,i}(\ldots, -\log_{\ell_j}H,
%   \ldots)$
% \item $F_{k,i}(\ldots, -\log_{\ell_j}(\log \ell_i H),
%   \ldots) = \partial_{\rho_j} F_{k,i}(\ldots, \rho_j-\log_{\ell_j}H,
%   \ldots)|_{\rho_j=0}$
%\end{itemize}
%where the $a_t \in \mathbb{R}$ and the $H$ may depend on $x$, $\mu^2$,
%$q^2$, and the integration variables (the bases of the $\ell_j$).  The
%obscurity of this setup and notation can be explained by a motivating
%example in which we see that the first of the above properties is
%simply linearity, the second is the independence of $x$ from the
%integration variables, and the third is the fact that
%$\partial_{\rho}^k y^{-\rho}|_{\rho=0} = (-1)^k \log^k(y)$. 

The connection between the different forms of the analytic
Dyson-Schwinger equation and the notational messiness of the original
presentation can be explained by a motivating example.

\begin{example}\label{motivate rho}
Let us return to Example \ref{bkerfc setup}.  The analytic
Dyson-Schwinger equation is
\[
  G(x, L) = 1 - \frac{x}{q^2}\int d^4 k \frac{k \cdot q}{k^2 G(x,
  \log (k^2/\mu^2))(k+q)^2}  - \cdots \bigg|_{q^2 = \mu^2}
\]
where $L = \log(q^2/\mu^2)$.

Substitute in the Ansatz
\[
  G(x,L) = 1 - \sum_{k \geq 1}\gamma_k(x) L^k
\]
to get
\begin{align*}
  \sum_{k \geq 1}\gamma_k(x)L^k & = \frac{x}{q^2}\int d^4 k
  \sum_{\ell_1 + \cdots + \ell_s = \ell} \frac{(k \cdot q)
    \gamma_{\ell_1}(x) \cdots \gamma_{\ell_s}(x) \log^\ell
    (k^2/\mu^2)}{k^2 (k+q)^2}  - \cdots \bigg|_{q^2 = \mu^2} \\
  & =  \frac{x}{q^2}
  \sum_{\ell_1 + \cdots + \ell_s = \ell}
    \gamma_{\ell_1}(x) \cdots \gamma_{\ell_s}(x) \int d^4 k \frac{(k
      \cdot q) \log^\ell
    (k^2/\mu^2)}{k^2 (k+q)^2}  - \cdots \bigg|_{q^2 = \mu^2} \\
  & = \frac{x}{q^2}
  \sum_{\ell_1 + \cdots + \ell_s = \ell}
    \gamma_{\ell_1}(x) \cdots \gamma_{\ell_s}(x)   \int d^4 k \frac{(k
      \cdot q) (-1)^\ell \partial_{\rho}^\ell
      (k^2/\mu^2)^{-\rho}|_{\rho=0}}{k^2 (k+q)^2}  - \cdots
    \bigg|_{q^2 = \mu^2} \\
  & = \frac{x}{q^2}
  \sum_{\ell_1 + \cdots + \ell_s = \ell}
    \gamma_{\ell_1}(x) \cdots \gamma_{\ell_s}(x)  (-1)^\ell
    \\ 
    & \qquad \cdot \partial_{\rho}^\ell  (\mu^2)^\rho \int d^4 k \frac{k
      \cdot q}{(k^2)^{1+\rho} (k+q)^2}  - \cdots
    \bigg|_{q^2 = \mu^2} \Bigg|_{\rho=0}\\ 
    & = x\left(1-\sum_{k \geq 1}\gamma_k(x)\partial_{-\rho}^k\right)^{-1}\frac{(\mu^2)^\rho}{q^2}\int d^4 k \frac{k
      \cdot q}{(k^2)^{1+\rho} (k+q)^2}  - \cdots
    \bigg|_{q^2 = \mu^2} \Bigg|_{\rho=0} \\
    & = x\left(1-\sum_{k \geq 1}\gamma_k(x)\partial_{-\rho}^k\right)^{-1}\frac{(\mu^2)^\rho}{(q^2)^\rho}\int d^4 k_0 \frac{k_0
      \cdot q_0}{(k_0^2)^{1+\rho} (k_0+q_0)^2}  - \cdots
    \bigg|_{q^2 = \mu^2} \Bigg|_{\rho=0} \\
    & \qquad \text{where $q=rq_0$ with $r\in \mathbb{R}$, $r^2=q^2$,
      $q_0^2=1$ and $k=rk_0$} \\
    & = x\left(1-\sum_{k \geq
        1}\gamma_k(x)\partial_{-\rho}^k\right)^{-1}(e^{-L\rho} -
    1)F(\rho) \bigg|_{\rho=0}
\end{align*}
using $\partial_{\rho}^k y^{-\rho}|_{\rho=0} = (-1)^k \log^k(y)$.
Thus using the notation $\gamma \cdot U = \sum \gamma_k U^k$ we can
write
\[
\gamma \cdot L = x(1-\gamma \cdot
  \partial_{-\rho})^{-1}(e^{-L\rho}-1)F(\rho)|_{\rho=0}
\]
\end{example}

\begin{example}
To see an example of a two variable Mellin transform (a slightly
different example can be found in \cite{etude}) consider again the
graph
\[
\gamma = \mbox{\begin{fmfgraph*}(400,40)
    \fmfleftn{i}{2}
    \fmfrightn{o}{2}
    \fmf{plain}{i1,v1}
    \fmf{plain}{i2,v1}
    \fmf{plain,left,tension=0.25,label=$k+p$}{v1,v2}
    \fmf{plain,left,tension=0.25,label=$k$}{v2,v1}
     \fmf{plain}{v2,o1}
    \fmf{plain}{v2,o2}
  \end{fmfgraph*}} 
\]
with the momenta associated to the two right hand external edges
summing to $p$.  As an integral the Mellin transform of $\gamma$ is
\[
\int d^D k \frac{1}{(k^2)^{1+\rho_1}((k+q)^2)^{1+\rho_2}} \bigg|_{q^2=1}.
\]
By the calculations of Subsection \ref{renormalization}
\[
  \int d^4 k \frac{1}{(k^2)^{1+\rho_1}((k+q)^2)^{1+\rho_2}}
  \bigg|_{q^2=1}
  = \frac{\pi^2\Gamma(\rho_1+\rho_2)}{\Gamma(1+\rho_1)\Gamma(1+\rho_2)}(q^2)^{-\rho_1-\rho_2} \frac{\Gamma(-\rho_1)\Gamma(-\rho_2)}{\Gamma(2-\rho_1-\rho_2)}
\]
So the Mellin transform is
\[
F_{\gamma}(\rho_1,\rho_2)=\frac{\pi^2\Gamma(\rho_1+\rho_2)}{\Gamma(1+\rho_1)\Gamma(1+\rho_2)}(q^2)^{-\rho_1-\rho_2} \frac{\Gamma(-\rho_1)\Gamma(-\rho_2)}{\Gamma(2-\rho_1-\rho_2)}.
\]
Upon subtracting at $q^2=\mu^2$ then we get
\[
  ((q^2)^{-\rho_1-\rho_2} -
(\mu^2)^{-\rho_1-\rho_2})F_{\gamma}(\rho_1, \rho_2)
= (e^{-L(\rho_1+\rho_2)} - 1) (\mu^2)^{-\rho_1-\rho_2}F_{\gamma}(\rho_1, \rho_2)
\]
So the only dependence on $q$ is the dependence on $L$ which is
showing up in the correct form for Definition \ref{dot def}.  The
extra powers of $\mu^2$ would get taken care of by the recursive
iteration as in Example \ref{motivate rho}.
\end{example}

\subsection{Systems}\label{pre system}
Now suppose we have a system of Dyson-Schwinger equations
\begin{equation}\label{system combDSE}
  X^r(x) = \mathbb{I} - \sgn(s_r)\sum_{k \geq 1}\sum_{i=0}^{t^r_k}x^k
  B_+^{k,i;r}(X^rQ^{k})
\end{equation}
for $r \in \mathcal{R}$ with $\mathcal{R}$ a finite set and
where
\begin{equation}\label{DSE Q}
  Q = \prod_{r\in \mathcal{R}}X^r(x)^{-s_r}.
\end{equation}

The
fact that the system can be written in terms of the invariant charge
$Q$ in this form is 
typical of realistic quantum field theories.  For example, in QED (see
Example
\ref{qed comb system}) 
\[
  Q = \frac{\left(X^{{\raisebox{-1ex}{\begin{fmfgraph}(100,100)
    \fmfleft{i}
    \fmfrightn{o}{2}
    \fmf{photon}{i,v}
    \fmf{fermion}{o1,v,o2}
\end{fmfgraph}}}}\right)^{2}}{\left(X^{{\raisebox{-1ex}{\begin{fmfgraph}(100,60)
    \fmfleft{i}
    \fmfright{o}
    \fmf{photon}{i,o}
\end{fmfgraph}}}}\right) \left(X^{{\raisebox{-1ex}{\begin{fmfgraph}(100,60)
    \fmfleft{i}
    \fmfright{o}
    \fmf{fermion}{i,o}
\end{fmfgraph}}}}\right)^{2}}.
\]

 Suppose a theory $T$ has a
single vertex $v \in \mathcal{R}$ with external legs $e_i \in
\mathcal{R}$ appearing with multiplicity $m_i$, $i=1,\ldots,n$ where
the external legs (made of half-edges types under our definitions) are
viewed as full edge types, hence as being in $\mathcal{R}$, by simply
taking the full edge type which contains the given half-edge types
(hence ignoring whether the edge is the front or back half of an
oriented edge type).  Let $\mathrm{val}(v)$ be the valence of the
vertex type $v$.  Then we define
\begin{equation}\label{physical Q}
  Q = \left(\frac{(X^v)^2}{\prod_{i=1}^n (X^{e_i})^{m_i}}\right)^{1/(\mathrm{val}(v)-2)}
\end{equation}
For theories with more than one vertex we form such a quotient for each
vertex. We are again (see section \ref{B+}) saved
by the Slavnov-Taylor identities which tell us that 
these quotients agree,  giving a unique invariant charge
\cite[section 2]{anatomy}.  Then $X^rQ^{k}$ is exactly what can be
inserted into a graph with 
external leg structure $r$ and $k$ loops.
\begin{prop}
  Suppose $Q$ is as defined as in the
  previous paragraph.  Let $G$ be a 1PI Feynman graph with external leg
  structure $r$ and $k>0$ loops.
  Then $X^rQ^{k}$ is exactly what can be inserted into $G$ in the
  sense that we can write $X^rQ^{k} = \prod_{j \in
    \mathcal{R}}(X^j)^{t_j}$ so that $G$ has $t_j$ vertices of type
  $j$ for $j$ a vertex type and $G$ has $-t_j$ edges
  of type $j$ for $j$ an edge type.
\end{prop}

\begin{proof}
In view of \eqref{physical Q} for $r$ a vertex type it suffices to prove that we can write
\[Q^{k+(\mathrm{val}(r)-2)/2} = \prod_{j \in
    \mathcal{R}}(X^j)^{\widetilde{t_j}}\] so that $G$ has $\widetilde{t_j}$ vertices of type
  $j$ for $j$ a vertex type and $G$ has $-2\widetilde{t_j}$ half edges
  in edge type $j$ (including external half edges) for $j$
  an edge type.
For $r$ an edge type it likewise suffices to prove that we can do the
same where we define $\mathrm{val}(r)=2$.

This holds for some $k$ by viewing a graph as made from a set
consisting of
vertices each attached to their adjacent half edges.  

To see that $k$
is correct note that since $G$ has 1 connected component,
$e-v+1=\ell$, where $e$ is the number of internal edges of $G$, $v$ the number
of vertices of $G$, and $\ell$ the loop number of $G$.  Letting $h$ be the number of
half edges (including external half edges) of $G$ we have
\begin{equation}\label{top graph th}
  \frac{h}{2} - v + 1 - \frac{\mathrm{val}(r)}{2} = \ell.
\end{equation}
Each $Q$ contributes $\mathrm{val}(r)/(\mathrm{val}(r)-2)$ edge
insertions and $2/(\mathrm{val}(r)-2)$ vertex insertions; so each $Q$
contributes 
\[
  \frac{\mathrm{val}(r)}{\mathrm{val}(r)-2} -
  \frac{2}{\mathrm{val}(r)-2} = 1
\]
to \eqref{top graph th}.  So if $k$ is so that
$Q^{k+(\mathrm{val}(r)-2)/2}$ counts the half edges and vertices of
$G$ as described above then
\[
  \ell = k+\frac{\mathrm{val}(r)-2}{2} + 1 - \frac{\mathrm{val}(r)}{2} = k.
\]
So $k$ is the loop number as required.

\end{proof}

The specific form of $Q$ from \eqref{physical Q} will
only be used in the renormalization group derivation of the first
recursion, section \ref{rg section}.
%For the scattering-type formula
%derivation of the first recursion we will use the weaker fact that 

%\begin{prop}\label{up and down}
%  Vertex types appear with nonnegative exponents and edge types appear
%  with nonpositive exponents in $Q$.
%\end{prop}

Write
$
  F_{k,i;r}(\rho_1, \ldots, \rho_{n_{k,i;r}})
$
for the Mellin transform associated to the primitive
$B_+^{k,i;r}(\mathbb{I})$.  Again assume a simple pole at the origin.
We can then write the analytic Dyson-Schwinger equations as in the
single equation case. 

\begin{definition}
The analytic Dyson-Schwinger equations associated to \eqref{system
  combDSE} are 
\begin{multline*}
  G^r(x, L_1, \ldots, L_j) \\= 1 - \sgn(s_r)\sum_{k \geq
  1}\sum_{i=0}^{t^r_k}x^k  G^r(x, \partial_{-\rho^r_1})^{-\sgn(s_r)} \cdots
  G^r(x, \partial_{-\rho^r_{\sgn(s_r)(s_rk-1)}})^{-\sgn(s_r)} \\
  \prod_{t \in \mathcal{R} \smallsetminus \{r\}}
  G^t(x, \partial_{-\rho^t_1})^{-\sgn(s_t)} \cdots
  G^t(x, \partial_{-\rho^t_{\sgn(s_t)(s_tk)}})^{-\sgn(s_t)} \\
  (e^{-L(\rho_1+\cdots+\rho_{n_{k,i;r}})}-1)F^{k,i;r}(\rho_1, \ldots, \rho_{n_{k,i;r}})
  \bigg|_{\rho_1 = \cdots = \rho_{n_{k,i;r}}=0}
\end{multline*}
where the $\rho_i^j$ run over the $\rho_{k}$ so that the $i$th factor
of $G^j$ is inserted at $\rho^j_i$.
%\begin{multline*}
%  G^r(x, L_1, \ldots, L_j) \\= 1 - \sgn(s_r)\sum_{k \geq
%  1}\sum_{i=0}^{t^r_k}x^k  \widetilde{G}^r(x, v^r_1)^{-\sgn(s_r)} \cdots
%  \widetilde{G}^r(x, v^r_{\sgn(s_r)(s_rk-1)})^{-\sgn(s_r)} \\
%  \prod_{t \in \mathcal{R} \smallsetminus \{r\}}
%  \widetilde{G}^t(x, v^t_1)^{-\sgn(s_t)} \cdots
%  \widetilde{G}^t(x, v^t_{\sgn(s_t)(s_tk)})^{-\sgn(s_t)} \\
%  (e^{-L(\rho_1+\cdots+\rho_{n_{k,i}})}-1)F^{k,i;r}(\rho_1, \ldots, \rho_{n_{k,i;r}})
%  \bigg|_{\rho_1 = \cdots = \rho_{n_{k,i}}=0}
%\end{multline*}
%where $j+1$ is the number of external momenta for $r$, $n_{k,i}$ is the number
%of edges which are insertion places or are adjacent to vertex
%insertion places in
%$B_+^{k,i;r}(\mathbb{I})$, and $\widetilde{G}^t(x,v_i)$ is a sum of
%expressions $G^t(x,a_1, \cdots, a_{j_t})$ where $a_i$ is either some $L_m$
%or some $\partial_{-\rho_m}$ and the sum runs over such expressions where
%there is an insertion of $B_+^{k,i;r}(\mathbb{I})$ into itself at edge
%or vertex
%$v_i$ with the external edge corresponding to $L_i$ mapped to the
%external edge corresponding to $L_m$ or to the internal edge
%corresponding to $\rho_m$ according to the value of $a_i$.
\end{definition}

The following notation will be used for expanding the analytic
Dyson-Schwinger equations as (in general asymptotic) series about the
origin,
\begin{equation}\label{system expand}
  G^r(x,L) = 1 - \sgn(s_r)\sum_{k \geq 1}\gamma^r_k(x) L^k \qquad \gamma^r_k(x) =
  \sum_{j \geq k} \gamma^r_{k,j} x^j
\end{equation}

In view of the following chapters we will reduce to the case
\begin{align*}
& G^r(x,L) = \\
& 1 - \sgn(s_r)\sum_{k \geq 1}\sum_{i=0}^{t_k} x^k
G^r(x,\partial_{-\rho})^{1-s_rk}\prod_{j \in \mathcal{R}\smallsetminus
  \{r\}}G^j(x,\partial_{-\rho})^{-s_jk}(e^{-L\rho}-1)F^{k,i}(\rho)\bigg|_{\rho=0}
\end{align*}
or rewritten
\begin{multline}\label{system dot eq}
  \gamma^r \cdot L = 
  \sum_{k \geq 1}\sum_{i=0}^{t_k} x^k (1-\sgn(s_r)\gamma^r \cdot
  \partial_{-\rho})^{1-sk}\prod_{j \in \mathcal{R}\smallsetminus
  \{r\}}(1-\sgn(s_j)\gamma^j \cdot
  \partial_{-\rho})^{-s_jk} \\(e^{-L\rho}-1)F^{k,i}(\rho)\bigg|_{\rho=0}
\end{multline}
where $\gamma^j \cdot U = \sum \gamma^j_k U^k$.

\end{fmffile}

\chapter{The first recursion}\label{first recursion}

There are two approaches to deriving the first recursion, neither of
which is completely self contained.  The first goes directly through the
renormalization group equation, and the second through the
Connes-Kreimer scattering-type formula \cite{ckII}.

\section{From the renormalization group equation}\label{rg section}

This is primarily an exercise in converting from usual physics
conventions to ours.

Using the notation of section \ref{pre system} the renormalization
group equation, see for instance \cite[Section 3.4]{cdial} or \cite{g}, reads
\begin{align*}
  \left(\frac{\partial}{\partial L} + \beta(x)\frac{\partial}{\partial
  x} - \sum_{e \text{ adjacent to } v}\gamma^e(x)\right)x^{(\mathrm{val}(v)-2)/2}G^v(x, L) & =
0 \qquad \text{for $v$ a vertex type} \\
\left(\frac{\partial}{\partial L} + \beta(x)\frac{\partial}{\partial
  x} - 2\gamma^e(x)\right)G^e(x, L) & =
0 \qquad \text{for $e$ an edge type}
\end{align*}
where $\beta(x)$ is the $\beta$ function of the theory, $\gamma^e(x)$ is
the anomalous dimension for $G^e(x,L)$ (both of which will be defined
in our notation below), and $\mathrm{val}(v)$ is the
valence of $v$.  The edge case and vertex case can be unified by
writing $\mathrm{val}(e) = 2$ and taking the edges adjacent to $e$ to
be two copies of $e$ itself (one for each half edge making $e$).
Our scale variable $L$ already has a log taken so $\partial_L$ often
appears as $\mu \partial_\mu$ in the literature where $\mu$ is the
scale before taking logarithms.  The use of $x^{(\mathrm{val}(v)-2)/2}G^v(x,L)$ in the vertex case in place of what is more typically simply
$G^v(x,L)$ comes about because by taking the coupling constant to
count the loop number rather than having the Feynman rules associate a
coupling constant factor to each vertex we have divided out by the
coupling constant factor for one vertex, that is by $x^{(\mathrm{val}(v)-2)/2}$.  As a result our series begin with a constant term
even for vertices.  

To see that it makes sense that a vertex $v$ 
contributes a factor of $x^{(\mathrm{val}(v)-2)/2}$ recall that that for
a graph $G$ with one connected component and external leg structure
$r$ we have \eqref{top graph th} which reads
\[
\frac{h}{2}-t+1 - \frac{\mathrm{val}(r)}{2}=\ell
\] where $h$ is the
number of half edges of $G$, $t$ the number of vertices, $\ell$ the
loop number, and where we take $\mathrm{val}(r) = 2$ for if $r$ is an
edge type.  Suppose vertices, but not edges, contribute some power of
$x$.  Then a vertex $v$ contributes $\mathrm{val}(v)/2 - 1$ to the
left hand side of \eqref{top graph th}, so it is consistent that $v$
also contribute the same power of $x$.  The whole graph $G$ then has
$x^{\ell+\mathrm{val}(r)/2-1}$ as expected.  If the power of $x$
associated to a vertex depends only on its valence, then this is the
only way to make the counting work.

Returning to $\beta$ and $\gamma$, define
\begin{equation}\label{beta}
  \beta(x) = \partial_L x\phi_R(Q) |_{L=0}
 \end{equation}
and
\begin{equation}\label{gamma}
  \gamma^e(x) = -\frac{1}{2}\partial_L G^e(x,L)|_{L=0} = \frac{1}{2}\gamma_1^e
\end{equation}
for $e$ an edge, that is $s_e > 0$ as discussed in subsection \ref{pre system}.
 The factor of $x$ in $\beta$ comes again from our normalization of
 the coupling constant powers to serve to count the loop number
 (recall from the discussion surrounding \eqref{physical Q} that $Q$
 contributes a $1$ to \eqref{top graph th} and so, in view of the
 previous paragraph, $Q$ is short one power
 of $x$), while the factor of $1/2$ in
\eqref{gamma} is usual.  
The
sign in \eqref{gamma} comes from the fact that our conventions have
the Green functions for the edges with a negative sign, while the
second equality uses the explicit expansion \eqref{system expand}.
Note that this $\beta$-function is not the Euler $\beta$
function, $\Gamma(x)\Gamma(y)/\Gamma(x+y)$; rather it encodes the flow
of the coupling constant depending on the energy scale.  Another way
to look at matters is that the $\beta$-function
measures the nonlinearity in a theory, specifically it is essentially the
coefficient of $L$ in
the invariant charge, as in the definition above.

In the case of an edge type $e$ we obtain quickly from
\eqref{beta}, \eqref{gamma}, \eqref{DSE Q}, and \eqref{system expand}, that
\begin{align*}
  0 & = \left(\frac{\partial}{\partial L} + \beta(x)\frac{\partial}{\partial
  x} - \gamma^e_1(x)\right)G^e(x, L) \\
& = \left(\frac{\partial}{\partial L} + \sum_{j \in \mathcal{R}}|s_j| \gamma^j_1(x)x\frac{\partial}{\partial
  x} - \gamma^e_1(x)\right)G^e(x, L).
\end{align*}

In the case of a vertex type $v$ compute as follows.
\begin{align*}
  0 & = \left(\frac{\partial}{\partial L} + \beta(x)\frac{\partial}{\partial
      x} - \sum_{e \text{ adjacent to } r}\gamma^e(x)\right)x^{(\mathrm{val}(v)-2)/2}G^v(x, L) \\
  & = x^{(\mathrm{val}(v)-2)/2} \frac{\partial}{\partial L}G^v(x,L) + x^{(\mathrm{val}(v)-2)/2}
  \beta(x)\frac{\partial}{\partial x}G^v(x,L) \\
  & \qquad + \frac{\mathrm{val}(v)-2}{2}x^{(\mathrm{val}-4)/2}\beta(x)G^v(x,L) - x^{(\mathrm{val}(v)-2)/2}
  \sum_{e \text{ adjacent to } r}\gamma^e(x)G^v(x, L) \\
  & = x^{(\mathrm{val}(v)-2)/2} \frac{\partial}{\partial L}G^r(x,L) + x^{(\mathrm{val}(v)-2)/2}
  \beta(x)\frac{\partial}{\partial x}G^r(x,L) \\
  & \qquad +
  x^{(\mathrm{val}-2)/2}\frac{\mathrm{val}(v)-2}{2}\frac{1}{\mathrm{val}(v)-2}
  \left(2 \gamma_1^v(x) + \sum_{e \text{ adjacent to } r}\gamma^e(x)\right)G^v(x,L) \\
  & \qquad - x^{(\mathrm{val}(v)-2)/2}
  \sum_{e \text{ adjacent to } r}\frac{1}{2}\gamma_1^e(x)G^v(x, L) \\
  & \qquad \text{from \eqref{beta}, \eqref{physical Q}, and
    \eqref{system expand}} \\
  & = x^{(\mathrm{val}(v)-2)/2} \left(\frac{\partial}{\partial L}G^v(x,L) +
  \beta(x)\frac{\partial}{\partial x}G^v(x,L) + \gamma^v_1
  G^v(x,L)\right) 
\end{align*}
Dividing by $x^{(\mathrm{val}(v)-2)/2}$ and using \eqref{DSE Q} and
\eqref{system expand} we have
\[
  \left(\frac{\partial}{\partial L} + \sum_{j \in \mathcal{R}}|s_j| \gamma^j_1(x)x\frac{\partial}{\partial
  x} + \gamma^v_1(x)\right)G^v(x, L) = 0.
\]

In both cases extracting the coefficient of $L^{k-1}$ and rearranging gives
\begin{thm}\label{systems first recursion}
  \[
  \gamma^r_k(x) =\frac{1}{k}\left( \sgn(s_r)\gamma^r_1(x) - \sum_{j
      \in \mathcal{R}}|s_j|\gamma^j_1(x)x
      \partial_x\right)\gamma^r_{k-1}(x).
   \]
\end{thm}

Specializing to the single equation case gives
\begin{thm}\label{single first recursion}
  \[
  \gamma_k = \frac{1}{k} \gamma_1(x)(\sgn(s)-|s|x \partial_x)\gamma_{k-1}(x).
  \]
\end{thm}

Note that the signs in the above do not match with \cite{radii} because here the sign
conventions have that the $X^r$ have their graphs appear with a
negative sign precisely if $r$ is an edge type, whereas in
\cite{radii} there was a negative sign in all cases.

\section{From $S\star Y$}

\begin{definition}
$Y$ is the grading operator on $\h$.
  $Y(\gamma) = |\gamma|\gamma$ for $\gamma \in \h$.
\end{definition}

\begin{definition}
  Let 
  \[
  \sigma_1 =\partial_L \phi_R(S \star Y)
 |_{L=0}
 \]
 and
 \[\sigma_n = \frac{1}{n!}m^{n-1}(\underbrace{\sigma_1 \otimes \cdots \otimes
  \sigma_1}_{n \text{ times}})\Delta^{n-1}
\]
\end{definition}

\begin{lemma}\label{S star Y linear}
  $S\star Y$ is zero off $\Hlin$
\end{lemma}

\begin{proof}
  First $S\star Y(\mathbb{I}) = \mathbb{I}\cdot 0 = 0$.
  Suppose $\Gamma_1, \Gamma_2 \in \h \setminus \mathbb{Q}\mathbb{I}$.
  Since $S$ is a homomorphism and $Y$ is a derivation,
  \begin{align*}
    S\star Y (\Gamma_1 \Gamma_2) & = \sum
    S(\gamma_1'\gamma_2')Y(\gamma_1''\gamma_2'') \\
    & = \left(\sum S(\gamma_1')\gamma_1''\right)\left(\sum
    S(\gamma_2')Y(\gamma_2''\right) + \left(\sum
    S(\gamma_1')Y(\gamma_1'')\right)\left(\sum
    S(\gamma_2')\gamma_2''\right) \\
    & = 0
  \end{align*}
  since by definition $S\star\id(\Gamma_1) = S\star\id(\Gamma_2) =
  0$.  Here we used the Sweedler notation,
  $\sum \gamma_j' \otimes \gamma_j'' = \Delta(\Gamma_j)$.
\end{proof}

\begin{lemma}\label{breaking apart game}
  \begin{align*}
  \Delta([x^k]X^r) & = \sum_{j=0}^{k} [x^j]X^rQ^{k-j} \otimes [x^{k-j}]X^r \\
  \Delta([x^k]X^rQ^\ell) & = \sum_{j=0}^{k} [x^j]X^rQ^{k+\ell-j} \otimes [x^{k-j}]X^rQ^\ell
  \end{align*}
  where $[\cdot]$ denotes the coefficient operator as in
Definition \ref{coefficient of}.
\end{lemma}

\begin{proof}
  The proof follows by induction.  Note that both equations read
  $\mathbb{I} \otimes \mathbb{I} = \mathbb{I} \otimes \mathbb{I}$ when
  $k = 0$.
  For a given value of $k>0$ the second equality follows from the
  first for all $0 \leq \ell \leq k$ using the
  multiplicativity of $\Delta$ and the fact that partitions of $k$
  into $m$ parts each part then partitioned into two parts are
  isomorphic with partitions of $k$ into two parts with each part then
  partitioned into $m$ parts.

  Consider then the first equation with $k>0$.  By Assumption \ref{b+
  1 cocycle}, for
  all $1 \leq \ell$, $\sum_{i=0}^{t^r_\ell}B_+^{\ell,i;r}$ is a
  Hochschild 1-cocycle.  Thus using \eqref{system combDSE}
  \begin{align*}
    \Delta([x^k]X^r) & = \Delta\left(- \sgn(s_r)\sum_{1 \leq \ell \leq
  k}\sum_{i=0}^{t^r_\ell} 
  B_+^{\ell,i;r}([x^{k-\ell}]X^rQ^{\ell})\right) \\
    & = - \sgn(s_r)\sum_{1 \leq \ell \leq
  k}\sum_{i=0}^{t^r_\ell}(\id \otimes
  B_+^{\ell,i;r})\Delta([x^{k-\ell}]X^rQ^{\ell}) \\
  & \qquad - \sgn(s_r)\sum_{1
  \leq \ell \leq 
  k}\sum_{i=0}^{t^r_\ell}(B_+^{\ell,i;r}([x^{k-\ell}]X^rQ^{\ell})\otimes
  \mathbb{I}) \\
  & = - \sgn(s_r)\sum_{1 \leq \ell \leq
  k}\sum_{i=0}^{t^r_\ell}(\id \otimes
  B_+^{\ell,i;r})\left(\sum_{j=0}^{k-\ell} [x^j]X^rQ^{k-j} \otimes
  [x^{k-\ell-j}]X^rQ^\ell\right) \\
  &\qquad + [x^k]X^r \otimes \mathbb{I} \\
  & =  \sum_{j=0}^{k-1}\left( [x^j]X^rQ^{k-j} \otimes
  -\sgn(s_r)\sum_{1 \leq \ell \leq
  k-j}\sum_{i=0}^{t^r_\ell}B_+^{\ell,i;r}([x^{k-\ell-j}]X^rQ^\ell)\right)
  \\
  &\qquad + [x^k]X^r \otimes \mathbb{I} \\
  &  =  \sum_{j=0}^{k-1} [x^j]X^rQ^{k-j}\otimes [x^{k-j}]X^r + [x^k]X^r \otimes \mathbb{I}
  \end{align*}
The result follows.
\end{proof}

Note that $\Delta(X^r) = \sum_{k=0}^\infty X^rQ^k\otimes (\text{terms
  of degree $k$ in $X^r$})$.

%  This proof is another example of how the Slavnov-Taylor identities
%  save us by serving to take the teeth out of overlapping divergences.

\subsection{Single equations}

\begin{prop}\label{sigma to gamma n}
  $\sigma_n(X) = \sgn(s)\gamma_n(x)$
\end{prop}

\begin{proof}
For $n=1$ this appears as equation (25) of \cite{ckII} and equation
(12) of \cite{bk30}.  For $n>1$ expand the scattering type formula
\cite[(14)]{ckII}.  The sign is due to our sign conventions, see \eqref{expand}.
\end{proof}

Rephrasing Lemma \ref{breaking apart game} we have
\begin{cor}\label{inserting game}
  Suppose $\Gamma x^k$ appears in $X$ with coefficient $c$ and $Z \otimes \Gamma x^k$
  consists of all terms in
  $\Delta X$ with $\Gamma$ on the right hand side.  
  Then $Z= cXQ^k$.
\end{cor}

\begin{prop}\label{expression for delta lin}
  \[
  (\Plin \otimes \id)\Delta X = X \otimes X -s X \otimes x \partial_x
  X
  \]
\end{prop}

\begin{proof}
  The Corollary implies that every graph appearing on the right hand side of
  $\Delta X$ also appears in $X$ and vice versa.
  Suppose $\Gamma x^k$ appears in $X$ and $Z \otimes \Gamma x^k$
  consists of all terms in
  $\Delta X$ with $\Gamma$ on the right hand side.

  By Corollary \ref{inserting game} $XQ^k = Z$.  So in $(\Plin \otimes
  \id)\Delta X$ we have the corresponding terms $\Plin(XQ^k) \otimes
  \Gamma$.

  Compute
  \begin{align*}
    \Plin(XQ^k) & = \Plin X + \Plin Q^k \\
    & = \Plin X + k\Plin Q \\
    & = \Plin X - k s \Plin X \\
    & = X -ks X
  \end{align*}

  Thus
  \[
  (\Plin \otimes \id)\Delta X = X \otimes X -s X \otimes x \partial_x
  X
  \]
\end{proof}

\begin{thm}\label{single first recursion 2}
  \[
  \gamma_k = \frac{1}{k} \gamma_1(x)(\sgn(s)-|s|x \partial_x)\gamma_{k-1}(x)
  \]
\end{thm}

\begin{proof}
  \begin{align*}
    \gamma_k & = \sgn(s)\sigma_k(X) \qquad \text{by Proposition \ref{sigma to
    gamma n}} \\
    & = \frac{\sgn(s)}{k!}m^{k-1}(\underbrace{\sigma_1 \otimes \cdots \otimes
  \sigma_1}_{k \text{ times}})\Delta^{k-1} (X) \\
    & = \frac{\sgn(s)}{k}m\left(\sigma_1 \otimes \frac{1}{(k-1)!}m^{k-2}(\underbrace{\sigma_1 \otimes \cdots \otimes
  \sigma_1}_{k-1 \text{ times}})\Delta^{k-2}\right)\Delta(X) \\
    & = \frac{\sgn(s)}{k} m(\sigma_1 \Plin \otimes \sigma_{k-1})\Delta(X) \qquad \text{by
    Lemma \ref{S star Y linear}}\\
    & = \frac{1}{k} \sgn(s)\sigma_1(X)\sigma_{k-1}(X) - s \sigma_1(X)x \partial_x
    \sigma_{k-1}(X) \qquad \text{by Proposition \ref{expression for
    delta lin}} \\
    & = \frac{1}{k} \gamma_1(x)(\sgn(s)\gamma_{k-1}(x) - |s|x\partial_x\gamma_{k-1}(x))
  \end{align*}
\end{proof}

\subsection{Systems of equations}

\begin{prop}\label{systems sigma to gamma n}
  $\sigma_n(X^r) = \sgn(s_r)\gamma_n^r(x)$
\end{prop}

\begin{proof}
The arguments of \cite{ckII} and \cite{bk30} do not depend on how or
whether the Green functions depend on other Green functions, so the
same arguments as in the single equation case applied.  The sign comes
from our conventions, see \eqref{system expand}.  Note that
$\beta$ in \cite{ckII} is the operator 
associated to the $\beta$-function only in the single equation case,
otherwise it is simply the anomalous dimension.
\end{proof}

As in the single equation case we can rewrite Lemma \ref{breaking
  apart game} to get
\begin{cor}\label{system inserting game}
  Suppose $\Gamma x^k$ appears in $X^r$ with coefficient $c$ and $Z \otimes \Gamma x^k$
  consists of all terms in
  $\Delta X^r$ with $\Gamma$ on the right hand side.  Then $Z = cX^rQ^k$.  
\end{cor}

%\begin{proof}
%  The proof follows as in the single equation case with the
%  observation that if $\gamma_i$ has external leg structure $r_i$ then
%  we can write 
%  \[
%    \Lambda = \lambda_1\cdots\lambda_m
%  \]
%  with $\lambda_i$ appearing in
%  $X^{r_i}Q^{|\gamma_i|}/X^{r_{i+1}}$ for $1 \leq i \leq m-1$ and $\lambda_m$ appearing
%  in $X^{r_m}Q^{|\gamma_m|}$.  This makes sense since the
%  decomposition of $\Gamma$ gives that the image of $B_+^{\gamma_i+1}$
%  appears in $X^{r_i} Q^{|\gamma_i|}$ hence that $X^{r_{i+1}}$ is a factor of
%  either the numerator or the denominator of $X^{r_i} Q^{|\gamma_i|}$.
%  But, using Proposition \ref{up and down}, if $r_{i+1}$ is a vertex
%  type then $X^{r_{i+1}}$ is a factor of the numerator of $X^{r_i}
%  Q^{|\gamma_i|}$ while if $r_{i+1}$ is an edge type then
%  $X^{r_{i+1}}$ is a factor of the denominator.  Consequently in each
%  $X^{r_i}Q^{|\gamma_i|}/X^{r_{i+1}}$ vertex types appear only in the
%  numerator and edge types only in the denominator, implying that the
%  decomposition of $\Lambda$ makes sense.  
%\end{proof}

\begin{prop}\label{systems expression for delta lin}
  \[
  (\Plin \otimes \id)\Delta X^r = X^r \otimes X^r - \sum_{j\in \mathcal{R}}s_j X^j \otimes x \partial_x
  X^r
  \]
\end{prop}

\begin{proof}
  As in the single equation case every graph appearing on the right hand side of
  $\Delta X^r$ also appears in $X^r$ and vice versa.
  Suppose $\Gamma x^k$ appears in $X^r$ and $Z \otimes \Gamma x^k$
  consists of all terms in
  $\Delta X$ with $\Gamma$ on the right hand side.

  By Corollary \ref{system inserting game} $X^rQ^k = Z$, and 
  \begin{align*}
    \Plin(X^rQ^k) & = \Plin X^r + \Plin Q^k \\
    & = \Plin X^r + k\Plin Q \\
    & = \Plin X^r - k \sum_{j \in \mathcal{R}} s_j \Plin X^j \\
    & = X^r -k \sum_{j \in \mathcal{R}} s_j X^j
  \end{align*}
  The result follows.
\end{proof}

\begin{thm}\label{systems first recursion 2}
  \[
  \gamma^r_k = \frac{1}{k} \left(\sgn(s_r)\gamma^r_1(x)^2 -\sum_{j \in
  \mathcal{R}}|s_j| \gamma^j_1(x) x \partial_x \gamma^r_{k-1}(x)\right)
  \]
\end{thm}

\begin{proof}
  \begin{align*}
    \gamma^r_k & = \sgn(s_r)\sigma_k(X^r) \qquad \text{by Proposition \ref{systems sigma to
    gamma n}} \\
    & = \frac{\sgn(s_r)}{k}m(\sigma_1 \Plin \otimes \sigma_{k-1})\Delta(X^r) \qquad
    \text{as in the single equation case} \\
    & = \frac{}{k}\left(\sgn(s_r)\sigma_1(X^r)\sigma_{k-1}(X^r) - \sum_{j \in \mathcal{R}}s_j \sigma_1(X^j)x \partial_x
    \sgn(s_r)\sigma_{k-1}(X^r)\right) \\ & \qquad \text{by Proposition \ref{systems expression for
    delta lin}} \\
    & = \frac{1}{k}\left(\sgn(s_r)\gamma^r_1(x)\gamma^r_{k-1}(x) - \sum_{j \in \mathcal{R}}|s_j|\gamma^j_1(x)x\partial_x\gamma^r_{k-1}(x))\right).
  \end{align*}
\end{proof}

As in the previous section the signs do not match with \cite{radii}
because here the sign conventions have that the $X^r$ have their
graphs appear with a 
negative sign precisely if $r$ is an edge type, whereas in
\cite{radii} there was a negative sign in all cases.

\section{Properties}

The following observation is perhaps obvious to the physicists, but
worth noticing
\begin{lemma}\label{what appears}
  As a series in $x$, the lowest term in $\gamma^r_{k}$ is of order at
  least $k$.  If $\gamma^r_{1,1} \neq \ell
  \sum_{j\in\mathcal{R}}s_j\gamma^j_{1,1}$, for $\ell = 0, \ldots,
  k-1$ then the lowest term in $\gamma^r_k$
  is exactly order $k$.
\end{lemma}

Note that in the single equation case, the condition to get lowest
term exactly of order $k$ is simply $\gamma_{1,1}\neq 0$.

\begin{proof}
  Expanding the combinatorial Dyson-Schwinger equation, \eqref{comb
    dse} or \eqref{system combDSE}, in $x$ we see immediately that the $x^0$ term is exactly
  $\mathbb{I}$.  The Feynman rules are independent of $x$ so the $x^0$
  term in the analytic Dyson-Schwinger equation is $1-1 = 0$ due to
  the fact that we renormalize by subtractions.

  Then inductively the $\gamma_k^r$ recursion, Theorem \ref{systems
    first recursion} or \ref{systems
    first recursion 2}, gives that as a
  series in $x$, $\gamma_k^r$ has no nonzero term before
  \[
    \frac{x^k}{k}\left(\sgn(s_r)\gamma_{1,1}^r - (k-1)\sum_{j \in \mathcal{R}}|s_j|\gamma_{1,1}^j\right)\gamma^r_{k-1,k-1}
  \]
  The result follows.
\end{proof}

We also can say that $\gamma^r_{k,j}$ is homogeneous in the
coefficients of the Mellin transforms in the sense indicated below.  This will not be used in the following.
For simplicity we will 
only give it in the single equation case with one insertion place.

Expand $
  F_{k,i}(\rho) = \sum_{j \geq -1} m_{j,k,i} \rho^j
$.

%Specializing \cite[(26)]{etude} to this  ***single equation*** case we get the recursion
%\begin{equation}\label{higher recursion}
%  \gamma_k(x) = \frac{1}{k} \gamma_1(x)(1+rx \partial_x)\gamma_{k-1}(x)
%\end{equation}
%independently of the $p_i(k)$.

%Using the dot notation, $\gamma \cdot U = \sum \gamma_k U^k$, of \cite{etude} we have
%\begin{equation}\label{dot DSE}
%  \gamma \cdot L = \sum_k \sum_ix^k (1 + \gamma \cdot
%  \partial_{-\rho})^{-rk+1}(1-e^{-L\rho})F_{k,i}(\rho) |_{\rho=0}
%\end{equation}
Recall \eqref{dot eq}
\[
  \gamma \cdot L = \sum_{k \geq 1} x^k (1-\sgn(s)\gamma \cdot
  \partial_{-\rho})^{1-sk}(e^{-L\rho}-1)F^{k}(\rho)\bigg|_{\rho=0}
\]

Taking one $L$ derivative and setting $L$ to $0$ we get
\begin{equation}\label{1 recursion}
  \gamma_1 = -\sum_k \sum_i x^k(1 - \sgn(s)\gamma \cdot
  \partial_{-\rho})^{1-sk} \rho F_{k,i}(\rho) |_{\rho=0}
\end{equation}

\begin{prop}
  Writing $\gamma_{k, j} = \sum c_{k, j, j_1, \cdots j_u,
    \overline{\ell}, \overline{i}} m_{j_1,
    \ell_1, i_1} \cdots m_{j_u, \ell_u, i_u}$
  for $j \geq k$ we have that $j_1 + \cdots + j_u = j - k$
\end{prop}

\begin{proof}
  The proof proceeds by induction.  Call $j_1 + \cdots + j_u$ the
  $m$-degree of $\gamma_{k,j}$.  First note that $\gamma_{1,1} =
  \sum_i m_{0,1,i}$  from (\ref{1 recursion}).

  Assume the result holds for $k, j < n$.

  Then from Theorem \ref{single first recursion} or \ref{single first
    recursion 2} $\gamma_{1, n}$ is a
  sum over $s$ of terms of the form
  \begin{equation}\label{gamma1n piece}
    C \gamma_{\ell_1, t_1} \cdots \gamma_{\ell_u, s_u} m_s
  \end{equation}
  where $\ell_1+ \cdots + \ell_u = s$ and $t_1 + \cdots t_u = n-1$.
  By the induction hypothesis $\gamma_{\ell_i, t_i}$ has $m$-degree
  $t_i - \ell_i$, so (\ref{gamma1n piece}) has $m$-degree $\sum t_i
  - \sum \ell_i + s = n-1 -s +s = n-1$ as desired.

  Next from  Theorem \ref{single first recursion} or \ref{single first
    recursion 2} $\gamma_{k, j}$, for $k,j \leq n$, is a
  sum over $1 \leq i \leq n$ of terms of the form
  \begin{equation}\label{gammanj piece}
    C \gamma_{1, i} \gamma_{k-1, j-i}
  \end{equation}
  By the induction hypothesis $\gamma_{1, i}$ has $m$-degree $i-1$ and
  $\gamma_{k-1, j-1}$ has $m$-degree $j-i-k+1$ so (\ref{gammanj
    piece}) has $m$-degree $j-k$ as desired.
\end{proof}

\chapter{Reduction to one insertion place}\label{first reduction}

\section{Colored insertion trees}
{}From now on we will need to carry around some additional information
with our Feynman graphs.
Namely we want to keep track of two different kinds of insertion,
normal insertion, and a modified insertion which inserts symmetrically
into all insertion places.  Symmetric insertion does not analytically
create overlapping divergences, but simply marking each subgraph
by how it was inserted may be ambiguous as in the example below.  We will use insertion trees to retain the information of how
a graph was formed by insertions.

\begin{fmffile}{redandblackegs}
\setlength{\unitlength}{0.05mm}
\fmfset{arrow_len}{2mm}
\fmfset{wiggly_len}{2mm}
\fmfset{thin}{0.5pt}
\fmfset{curly_len}{1.5mm}
\fmfset{dot_len}{1mm}
\fmfset{dash_len}{1.5mm}

In examples without overlaps, and even
in simple overlapping cases, it suffices to label the divergent
subgraphs with one of two colors, black for normal insertion and red
for symmetric insertion.
To see that coloring does not suffice in the
general case consider the graph 
  \[
    \mbox{\begin{fmfgraph}(300,300)
        \fmfleftn{i}{4}
        \fmfrightn{o}{4}
        \fmftop{t}
        \fmfbottom{b}
        \fmf{plain}{i2,v1}
        \fmf{plain}{i3,v1}
        \fmf{phantom,tension=2}{t,v2}
        \fmf{plain}{o2,v3}
        \fmf{plain}{o3,v3}
        \fmf{phantom,tension=2}{b,v4}
        \fmf{plain}{v1,v2,v3,v4,v1}
        \fmf{plain,left=0.5,tension=0.25}{v2,v4,v2}
      \end{fmfgraph}}
  \]
There are three proper subdivergent graphs; give them the
  following names for easy reference
  \begin{align*}
    A & = \raisebox{-1.5ex}{\begin{fmfgraph}(120,120)
        \fmftopn{t}{2}
        \fmfbottomn{b}{2}
        \fmf{plain}{t1,v1,t2}
        \fmf{plain}{b1,v2,b2}
        \fmf{plain,left=0.5,tension=0.25}{v1,v2,v1}
        \end{fmfgraph}}\\
      B & = \raisebox{-2ex}{\begin{fmfgraph}(200,200)
          \fmfleftn{i}{4}
          \fmfrightn{o}{2}
          \fmf{plain}{i2,v1,i3}
          \fmf{plain}{o1,v2,v1,v3,o2}
          \fmf{plain,left=0.5,tension=0.25}{v2,v3,v2}
          \end{fmfgraph}} \\
        C & = \raisebox{-2ex}{\begin{fmfgraph}(200,200)
          \fmfrightn{i}{4}
          \fmfleftn{o}{2}
          \fmf{plain}{i2,v1,i3}
          \fmf{plain}{o1,v2,v1,v3,o2}
          \fmf{plain,left=0.5,tension=0.25}{v2,v3,v2}
          \end{fmfgraph}}
  \end{align*}
  Then if $A$ is red while $B$ and $C$ are black then this could
  represent $A$ inserted symmetrically into \mbox{\begin{fmfgraph}(200,40)
      \fmfrightn{i}{2}
      \fmfleftn{o}{2}
      \fmf{plain}{i1,v1,i2}
      \fmf{plain}{o1,v2,o2}
      \fmf{plain,left,tension=0.25}{v1,v3,v2,v3,v1}
      \end{fmfgraph}} or it could represent $B$ inserted into  \mbox{\begin{fmfgraph}(200,40)
      \fmfrightn{i}{2}
      \fmfleftn{o}{2}
      \fmf{plain}{i1,v1,i2}
      \fmf{plain}{o1,v2,o2}
      \fmf{plain,left,tension=0.25}{v1,v2,v1}
      \end{fmfgraph}} while $B$ itself is made of $A$ symmetrically
    inserted into \mbox{\begin{fmfgraph}(200,40)
      \fmfrightn{i}{2}
      \fmfleftn{o}{2}
      \fmf{plain}{i1,v1,i2}
      \fmf{plain}{o1,v2,o2}
      \fmf{plain,left,tension=0.25}{v1,v2,v1}
      \end{fmfgraph}} and likewise for $C$.
\end{fmffile}

\begin{definition}
  A \emph{decorated rooted tree} is a finite
  rooted tree (not embedded in the
  plane) with a map from its vertices to a fixed, possibly infinite, set of decorations.
\end{definition}

The polynomial algebra over $\mathbb{Q}$ generated by (isomorphism
classes) of decorated rooted trees forms a Hopf algebra as follows.
\begin{definition}
  The \emph{(Connes-Kreimer) Hopf algebra of decorated rooted trees}, $\h_{CK}$, consists
  of the $\mathbb{Q}$ span of forests of decorated rooted trees with disjoint
  union as multiplication, including the empty forest $\mathbb{I}$.
  The coproduct on $\h_{CK}$ is the
  algebra homomorphism defined
  on a tree by
  \[
  \Delta(T) = \sum_c P_c(T) \otimes R_c(T)
  \]
  where the sum runs over ways to cut edges of $T$ so that each path
  from the root to a leaf is cut at most once, $R_c(T)$ is the
  connected component of the result connected to the original root, and
  $P_c(T)$ is the forest of the remaining components.  
  The antipode is defined
  recursively from $S \star \id = e\eta$ (as in the Feynman graph
  situation), 
\end{definition}

See \cite{ck0} for more details on $\h_{CK}$.
Insertion trees are decorated rooted trees where each element in the decoration set
consists of an ordered triple of a primitive of $\h$
(potentially a sum), an insertion
place in the primitive of the parent of the current vertex, and a bijection from the external edges of the Feynman graph to
the half edges of the insertion place.  The second and third elements of
the triple serve to unambiguously define an insertion as in Subsection
\ref{operations subsec}.  Often the insertion information will be left out if it is unambiguous.

\begin{definition}
  For a 1PI Feynman graph $G$ in a given theory let $F(G)$ be the forest
  of insertion trees which give $G$.
\end{definition}

From $F(G)$, or even just one tree of $F(G)$, we can immediately
recover $G$ simply by doing the specified insertions.  The result of
the insertion defined by a particular parent and child pair of
vertices is unambiguous since all the insertion information is
included in the decoration.  The choice of order to do the insertions
defined by an insertion tree does not affect the result due to the
coassociativity of the Feynman graph Hopf algebra.

Extend $F$ to
$F:\h \rightarrow \h_{CK}$ as an algebra homomorphism.  In fact it is
an injective Hopf algebra morphism by the following proposition.

\begin{prop}
  $F(\Delta(G)) = \Delta(F(G))$.
\end{prop}

\begin{proof}
  Let $\gamma$ be a (not necessarily connected) divergent subgraph of $G$.  Since $G$ can be
  made by inserting $\gamma$ into $G/\gamma$, then among
  $F(G)$ we can find each tree
  of $F(\gamma)$ grafted into each tree of $F(G/\gamma)$.  Cutting
  edges where $F(\gamma)$ is grafted into $F(G/\gamma)$ we see that $F(\gamma)
  \otimes F(G/\gamma)$ appears in $\Delta(F(G))$.  The coefficients
  are the same since each insertion place for $\gamma$ in $G/\gamma$
  which gives $G$ we have a grafting with this insertion information
  and vice versa.  Finally every cut of $F(G)$ consists of a forest of
  insertion trees, which by doing the insertions gives a divergent
  subgraph of $G$.  The result follows.
\end{proof}

Now we wish to extend this situation by coloring the edges of the
insertion trees.

\begin{definition}
  Let $T$ be a decorated rooted tree with edge set $E$.
  Define an \emph{insertion coloring map} to be a map $f:E 
  \rightarrow \{\textrm{black}, \textrm{red}\}$.  If $T$ is an
  insertion tree when call $T$ with $f$ a
  \emph{colored insertion tree}.
\end{definition}

\begin{definition}
  For a colored insertion tree define the coproduct
  to be as before
  with the natural colorings.
\end{definition}

To translate back to Feynman graphs think of the edge as coloring the
graph defined by the insertion tree below it.  The result is a Feynman
graph with colored proper subgraphs.  The coproduct in the tree case
forgets the color of the cut edges.  Correspondingly in the Feynman
graph case the color of the graphs, but not their subgraphs, on the left hand sides of
the tensor product are forgotten.

\begin{prop}
 Colored insertion trees form a Hopf algebra with the above
  coproduct which agrees with $\h_{CK}$  upon forgetting the colors.
\end{prop}

\begin{proof}
  Straightforward.
\end{proof}

Call the Hopf algebra of colored insertion trees $\h_c$.  In view of
the above $R:\h \hookrightarrow \h_c$ by taking $R:\h \hookrightarrow
\h_{CK}$ and coloring all edges black.

Analytically, black insertion follows the usual
Feynman rules, red insertion follows the symmetric insertion rules as
defined in subsection \ref{symmetric insertion}.

\begin{definition}
  For $\gamma$ a primitive element of $\h$ or $\h_c$,
  write $R_+^{\gamma} : \h_{c} \rightarrow \h_{c}$ for the operation
  of adding a root decorated with $\gamma$ with the edges connecting
  it colored red.
  Also write
  $B_{+}^\gamma :\h_{c} \rightarrow \h_{c}$ for ordinary insertion of Feynman graphs translated to
  insertion trees with new edges colored black.  Note that this is not the usual $B_+$ on rooted
  trees in view of overlapping divergences.
\end{definition}

When working directly with Feynman graphs $R_+^\gamma$ corresponds to
insertion with the inserted graphs colored red and no overlapping
divergences.

Another way of understanding the importance of Definition \ref{B+ def} and Theorem
\ref{b+ 1 cocycle} is that $\sum_{i=0}^{t^r_k}B_+^{k,i;r}$ is the same
whether interpreted as specified above by $B_+$ on Feynman graphs
translated to $\h_{c}$, or directly on $\h_{c}$ simply by adding a
new root labelled by $\gamma$ and the corresponding insertion places
without consideration for overlapping divergences.

\begin{lemma}
  $R_+^\gamma$ is a Hochschild 1-cocycle for $\h_c$.
\end{lemma}

\begin{proof}
  The standard $B_+$ of adding a root is a Hochschild 1-cocycle in
  $\h_{CK}$, see \cite[Theorem 2]{ck0}.  Edges attached to the
  root on the right hand side of the tensors are red on both sides of
  the 1-cocycle identity.  The remaining edge colors must also satisfy
  the 1-cocycle property which we can see by attaching this
  information to the decoration of the node which is further from the root.
\end{proof}

\section{Dyson-Schwinger equations with one insertion place}

To reduce to one insertion place we need to show that we can write
Dyson-Schwinger equations in which only involves $R_+$s but which,
results in the same series $X^r$ which contains only black insertions.
We can achieve this recursively, while viewing $\h \hookrightarrow \h_c$.

Suppose our combinatorial Dyson-Schwinger equation is as in
\eqref{system combDSE}
\[
X^r(x) = \mathbb{I} - \sgn(s_r)\sum_{k \geq 1}\sum_{i=0}^{t^r_k}x^k
  B_+^{k,i;r}(X^rQ^{k}).
\]
Then, using $[\cdot]$ to denote the coefficient operator as in
Definition \ref{coefficient of}, define
\begin{align*}
  q^r_1 & = -\sgn(s_r)[x]X^r = \sum_{i=0}^{t^r_i}B_+^{1,i;r}(\mathbb{I}) \\
  q^r_n & = -\sgn(s_r)[x^n]X^r +\sgn(s_r) \sum_{k=1}^{n-1} R_+^{q^r_k}([x^{n-k}]X^rQ^k)\\
  & = \sum_{k = 1}^{n}\sum_{i=0}^{t^r_k}
  B_+^{k,i;r}([x^{n-k}]X^rQ^k)  +\sgn(s_r) \sum_{k=1}^{n-1} R_+^{q^r_k}([x^{n-k}]X^rQ^k)
\end{align*}

In order to know that the $q^r_n$ are well defined we need to
know that they are primitive.

\begin{prop}
  $q^r_n$ is primitive 
  %and $R_+^{q_n^r}$ is a Hochschild 1-cocycle 
  for $r\in\mathcal{R}$ and $n\geq 1$.
\end{prop}

\begin{proof}
  First note that $B_+(\mathbb{I})$ is primitive for any $B_+$ and the
  sum of primitives is primitive, so $q_1^r$ is primitive for
  each $r \in \mathcal{R}$.  
  %Also $R_+^{q_1^r} = \sum_{i=0}^{t^r_i}
  %B_+^{1,i;r}$ which is a 1-cocycle.

  Then inductively for $n > 1$
  \begin{align*}
    \Delta(q_n^r)  
    & = \sum_{k = 1}^{n}\sum_{i=0}^{t^r_k}
  (\id \otimes B_+^{k,i;r})(\Delta[x^{n-k}]X^rQ^k) - \sum_{k=1}^{n-1}
  (\id \otimes R_+^{q^r_k})(\Delta[x^{n-k}]X^rQ^k) \\
    & \qquad +  \sum_{k = 1}^{n}\sum_{i=0}^{t^r_k}
   B_+^{k,i;r}([x^{n-k}]X^rQ^k)\otimes \mathbb{I} - \sum_{k=1}^{n-1}
   R_+^{q^r_k}([x^{n-k}]X^rQ^k)\otimes \mathbb{I} \\
   & = \sum_{k = 1}^{n}\sum_{i=0}^{t^r_k}\sum_{\ell=0}^{n-k}
   \left([x^\ell]X^rQ^k \otimes B_+^{k,i;r}([x^{n-\ell-k}]X^rQ^k)\right) \\
   & \qquad - \sum_{k=1}^{n-1}
   \sum_{\ell=0}^{n-k}\left([x^\ell]X^rQ^k  \otimes
   R_+^{q^r_k}([x^{n-\ell-k}]X^rQ^k)\right)
   + q^r_n \otimes \mathbb{I}  \\
   & =  \mathbb{I} \otimes q^r_n + q^r_n \otimes \mathbb{I} \\
      & + \sum_{\ell=1}^{n-1} \sum_{k=1}^{n-\ell} \left([x^\ell]X^rQ^{n-\ell}
   \otimes \left(\sum_{i=0}^{t_k^r}
   B_+^{k,i;r}([x^{n-\ell-k}]X^rQ^k) -
   R_+^{q^r_k}([x^{n-\ell-k}]X^rQ^k)\right)\right) \\
   &=  \mathbb{I} \otimes q^r_n  + q^r_n \otimes \mathbb{I} -\sum_{\ell=1}^{n-1}
   \left([x^\ell]X^rQ^{n-\ell} \otimes (q_\ell^r - q_\ell^r)\right) \\
   & =\mathbb{I} \otimes q_n^r + q_n^r \otimes \mathbb{I}.
 \end{align*}
\end{proof}

\begin{thm}
  \[
  X^r = 1 - \sgn(s_r)\sum_{k\geq 1} x^k R_+^{q^r_k}(X^rQ^k).
  \]
\end{thm}

\begin{proof}
  The constant terms of both sides of the equation match and for
  $n\geq 1$
  \begin{align*}
    -\sgn(s_r) [x^n]\sum_{k\geq 1} x^k R_+^{q^r_k}(X^rQ^k) 
    & = - \sgn(s_r)\sum_{k= 1}^n x^k R_+^{q^r_k}([x^{n-k}]X^rQ^k) \\
    & = -\sgn(s_r)q^r_k - \sgn(s_r)\sum_{k= 1}^{n-1} x^k R_+^{q^r_k}([x^{n-k}]X^rQ^k) \\
    & = [x^n]X^r.
  \end{align*}
\end{proof}

The interpretation of the Theorem is that we can reduce to considering
only red insertion, that is to a single symmetric insertion place.

In simple cases we can avoid the not only the insertion trees, but also
the subgraph coloring, and literally
reduce to a single insertion place in the original Hopf algebra.
However this cannot work with different types of insertions or with
vertex insertions where each vertex can not take an arbitrary number
of inserted graphs.  Consequently such simple examples can only arise
with a single type of edge insertion as in the following example.

\begin{fmffile}{firstreductioneggraphs}
\setlength{\unitlength}{0.1mm}
\fmfset{arrow_len}{2mm}
\fmfset{wiggly_len}{2mm}
\fmfset{thin}{0.5pt}
\fmfset{curly_len}{1.5mm}
\fmfset{dot_len}{1mm}
\fmfset{dash_len}{1.5mm}

\begin{example}\label{first reduction eg}
Suppose we have the Dyson-Schwinger equation
\[
  X = 1 - xB_+^{\frac{1}{2}\raisebox{-0.4ex}{{\begin{fmfgraph}(50,30)
\fmfleftn{i}{1}
\fmfrightn{o}{1}
\fmf{plain}{i1,in1}
\fmf{plain}{in2,o1}
\fmf{plain,left,tension=0.25}{in1,in2,in1}
\end{fmfgraph}}}}\left(\frac{1}{X^2}\right).
\]
where we insert into both internal edges.  In this case we need not
resort to red insertion in order to reduce to one insertion place.

Let 
\[
q_1 = \frac{1}{2}\raisebox{-1ex}{\begin{fmfgraph}(100,60)
\fmfkeep{1loop}
\fmfleftn{i}{1}
\fmfrightn{o}{1}
\fmf{plain}{i1,in1}
\fmf{plain}{in2,o1}
\fmf{plain,left,tension=0.25}{in1,in2,in1}
\end{fmfgraph}}
\]
where we only insert into the bottom edge and let
\[
  X_1 = 1 - xB_+^{q_1}\left(\frac{1}{X_1^2}\right)
\]
Then to order $x^3$ we have that
\[
  X = 1 - x\frac{1}{2}\raisebox{-1ex}{\fmfreuse{1loop}}
  -x^2\frac{1}{2}\raisebox{-1ex}{\begin{fmfgraph}(100,60)
\fmfkeep{2loops}
\fmfleftn{i}{1}
\fmfrightn{o}{1}
\fmf{plain}{i1,in1,in3}
\fmf{plain}{in4,in2,o1}
\fmf{plain,left,tension=0.25}{in1,in2}
\fmf{plain,left,tension=0.25}{in3,in4,in3}
\end{fmfgraph}}
  -x^3\left(
\frac{1}{8}\raisebox{-1ex}{\begin{fmfgraph}(100,60)
\fmfkeep{3loops1plus1}
\fmfleftn{i}{1}
\fmfrightn{o}{1}
\fmf{plain}{i1,in1,in3}
\fmf{plain}{in4,in2,o1}
\fmf{plain}{in1,in5}
\fmf{plain}{in6,in2}
\fmf{plain,left,tension=0.25}{in5,in6,in5}
\fmf{plain,left,tension=0.25}{in3,in4,in3}
\fmffreeze
\fmfshift{(0,.5h)}{in5}
\fmfshift{(0,.5h)}{in6}
\end{fmfgraph}}
+\frac{1}{2}\raisebox{-1ex}{\begin{fmfgraph}(100,60)
\fmfkeep{3loopsrainbow}
\fmfleftn{i}{1}
\fmfrightn{o}{1}
\fmf{plain}{i1,in1,in3,in5}
\fmf{plain}{in6,in4,in2,o1}
\fmf{plain,left,tension=0.25}{in1,in2}
\fmf{plain,left,tension=0}{in3,in4}
\fmf{plain,left,tension=0.25}{in5,in6,in5}
\end{fmfgraph}}
+\frac{1}{4}\raisebox{-1ex}{\begin{fmfgraph}(100,60)
\fmfkeep{3loopseyes}
\fmfleftn{i}{1}
\fmfrightn{o}{1}
\fmf{plain}{i1,in1,in3}
\fmf{plain}{in4,in2,o1}
\fmf{plain}{in5,in6}
\fmf{plain,left,tension=0.25}{in1,in2}
\fmf{plain,left,tension=0.25}{in3,in5,in3}
\fmf{plain,left,tension=0.25}{in6,in4,in6}
\end{fmfgraph}}
\right)
\]
and
\[
  X_1 = 1 - x\frac{1}{2}\raisebox{-1ex}{\fmfreuse{1loop}}
  -x^2\frac{1}{2}\raisebox{-1ex}{\fmfreuse{2loops}}
  -x^3\left(\frac{3}{8}\raisebox{-1ex}{\fmfreuse{3loopseyes}}+\frac{1}{2}\raisebox{-1ex}{\fmfreuse{3loopsrainbow}}\right)
\]
so
\[
  q_2 = 0 \qquad \text{and} \qquad q_3 =
  \frac{1}{8}\raisebox{-1ex}{\fmfreuse{3loops1plus1}} -
  \frac{1}{16}\raisebox{-1ex}{\fmfreuse{3loopseyes}} - \frac{1}{16}\raisebox{-1ex}{\fmfreuse{3loopseyes}}
\]
where in the first graph of $q_3$ we insert only in the bottom edge of
the bottom inserted bubble, in the second graph we insert only in the
bottom edge of the leftmost inserted bubble, and in the third graph we
insert only in the bottom edge of the rightmost inserted bubble.

Note that $q_3$ is primitive.
Let
\[
  X_2 = 1 - xB_+^{q_1}\left(\frac{1}{X_1^2}\right) - x^3B_+^{q_3}\left(\frac{1}{X^8}\right)
\]

The order $x^4$ we have
\begin{align*}
  X = & 1 - x\frac{1}{2}\raisebox{-1ex}{\fmfreuse{1loop}}
  -x^2\frac{1}{2}\raisebox{-1ex}{\fmfreuse{2loops}}
  -x^3\left(\frac{1}{8}\raisebox{-1ex}{\fmfreuse{3loops1plus1}}+\frac{1}{2}\raisebox{-1ex}{\fmfreuse{3loopsrainbow}}
  + \frac{1}{4}\raisebox{-1ex}{\fmfreuse{3loopseyes}}\right) \\
  & -x^4\left(\frac{1}{8}\raisebox{-1ex}{\begin{fmfgraph}(100,60)
\fmfkeep{4loops1plus2}
\fmfleftn{i}{1}
\fmfrightn{o}{1}
\fmf{plain}{i1,in1,in3}
\fmf{plain}{in8,in2,o1}
\fmf{plain}{in1,in5}
\fmf{plain}{in6,in2}
\fmf{plain}{in4,in7}
\fmf{plain,left,tension=0.5}{in5,in6,in5}
\fmf{plain,left,tension=0.25}{in3,in4,in3}
\fmf{plain,left,tension=0.25}{in7,in8,in7}
\fmffreeze
\fmfshift{(0,.5h)}{in5}
\fmfshift{(0,.5h)}{in6}
\end{fmfgraph}}
+ \frac{1}{4}\raisebox{-1ex}{\begin{fmfgraph}(100,60)
\fmfkeep{4loops1plusrainbow}
\fmfleftn{i}{1}
\fmfrightn{o}{1}
\fmf{plain}{i1,in1,in3,in7}
\fmf{plain}{in8,in4,in2,o1}
\fmf{plain}{in1,in5}
\fmf{plain}{in6,in2}
\fmf{plain,left,tension=0.5}{in5,in6,in5}
\fmf{plain,left,tension=0}{in3,in4}
\fmf{plain,left,tension=0.25}{in7,in8,in7}
\fmffreeze
\fmfshift{(0,.5h)}{in5}
\fmfshift{(0,.5h)}{in6}
\end{fmfgraph}}
+ \frac{1}{2}\raisebox{-1ex}{\begin{fmfgraph}(100,60)
\fmfkeep{4loopsrainbow}
\fmfleftn{i}{1}
\fmfrightn{o}{1}
\fmf{plain}{i1,in1,in3,in5,in7}
\fmf{plain}{in8,in6,in4,in2,o1}
\fmf{plain,left,tension=0.25}{in1,in2}
\fmf{plain,left,tension=0}{in3,in4}
\fmf{plain,left,tension=0}{in5,in6}
\fmf{plain,left,tension=0.25}{in7,in8,in7}
\end{fmfgraph}}
+ \frac{1}{8}\raisebox{-1ex}{\begin{fmfgraph}(100,60)
\fmfkeep{4loops3inarow}
\fmfleftn{i}{1}
\fmfrightn{o}{1}
\fmf{plain}{i1,in1,in3}
\fmf{plain}{in4,in2,o1}
\fmf{plain}{in5,in6}
\fmf{plain}{in7,in8}
\fmf{plain,left,tension=0.25}{in1,in2}
\fmf{plain,left,tension=0.25}{in3,in5,in3}
\fmf{plain,left,tension=0.25}{in6,in7,in6}
\fmf{plain,left,tension=0.25}{in8,in4,in8}
\end{fmfgraph}}
+ \frac{1}{4}\raisebox{-1ex}{\begin{fmfgraph}(100,60)\fmfkeep{4loops2eyes}
\fmfleftn{i}{1}
\fmfrightn{o}{1}
\fmf{plain}{i1,in1,in3,in5}
\fmf{plain}{in6,in4,in2,o1}
\fmf{plain}{in7,in8}
\fmf{plain,left,tension=0.25}{in1,in2}
\fmf{plain,left,tension=0.25}{in5,in7,in5}
\fmf{plain,left,tension=0}{in3,in4}
\fmf{plain,left,tension=0.25}{in8,in6,in8}
\end{fmfgraph}}\right. \\
& \qquad \left.
+ \frac{1}{8}\raisebox{-1ex}{\begin{fmfgraph}(100,60)
\fmfkeep{4loopsinner1plus1}
\fmfleftn{i}{1}
\fmfrightn{o}{1}
\fmf{plain}{i1,in7,in1,in3}
\fmf{plain}{in4,in2,in8,o1}
\fmf{plain}{in1,in5}
\fmf{plain}{in6,in2}
\fmf{plain,left,tension=0}{in7,in8}
\fmf{plain,left,tension=0.3}{in5,in6,in5}
\fmf{plain,left,tension=0.3}{in3,in4,in3}
\fmffreeze
\fmfshift{(0,.31h)}{in5}
\fmfshift{(0,.31h)}{in6}
\end{fmfgraph}}
+ \frac{1}{4}\raisebox{-1ex}{\begin{fmfgraph}(100,60)
\fmfkeep{4loopsrighteye}
\fmfleftn{i}{1}
\fmfrightn{o}{1}
\fmf{plain}{i1,in1,in3}
\fmf{plain}{in6,in4,in2,o1}
\fmf{plain}{in7,in8,in5}
\fmf{plain,left,tension=0.25}{in1,in2}
\fmf{plain,left,tension=0.25}{in5,in6,in5}
\fmf{plain,left,tension=0.25}{in3,in7,in3}
\fmf{plain,left,tension=0}{in8,in4}
\end{fmfgraph}}
+ \frac{1}{4}\raisebox{-1ex}{\begin{fmfgraph}(100,60)
\fmfkeep{4loopslefteye}
\fmfleftn{i}{1}
\fmfrightn{o}{1}
\fmf{plain}{i1,in1,in3,in5}
\fmf{plain}{in4,in2,o1}
\fmf{plain}{in6,in7,in8}
\fmf{plain,left,tension=0.25}{in1,in2}
\fmf{plain,left,tension=0.25}{in5,in6,in5}
\fmf{plain,left,tension=0}{in3,in7}
\fmf{plain,left,tension=0.25}{in8,in4,in8}
\end{fmfgraph}}
\right)
\end{align*}
and
\begin{align*}
X_2 = & 1 - x\frac{1}{2}\raisebox{-1ex}{\fmfreuse{1loop}}
  -x^2\frac{1}{2}\raisebox{-1ex}{\fmfreuse{2loops}}
  -x^3\left(\frac{1}{2}\raisebox{-1ex}{\fmfreuse{3loopsrainbow}}
  + \frac{3}{8}\raisebox{-1ex}{\fmfreuse{3loopseyes}}\right) \\
  & -x^4\left(\frac{3}{8}\raisebox{-1ex}{\fmfreuse{4loops2eyes}} +
  \frac{1}{4}\raisebox{-1ex}{\fmfreuse{4loops3inarow}} +
  \frac{1}{2}\raisebox{-1ex}{\fmfreuse{4loopsrainbow}} +
  \frac{3}{8}\raisebox{-1ex}{\fmfreuse{4loopslefteye}} +
  \frac{3}{8}\raisebox{-1ex}{\fmfreuse{4loopsrighteye}} \right.\\
  & \qquad
  \left. +\frac{1}{8}\raisebox{-1ex}{\fmfreuse{4loopsinner1plus1}} - \frac{1}{8}\raisebox{-1ex}{\fmfreuse{4loops2eyes}}\right)
\end{align*}
where the first 2 lines come from inserting $X_2$ into $q_1$ and the
third line comes from inserting $X_2$ into $q_3$.

Consequently let
\[
  q_4 = \frac{1}{8}\raisebox{-1ex}{\fmfreuse{4loops1plus2}}
  - \frac{1}{8}\raisebox{-1ex}{\fmfreuse{4loops3inarow}} -
  \frac{1}{4}\raisebox{-1ex}{\fmfreuse{4loops1plusrainbow}} +
  \frac{1}{8}\raisebox{-1ex}{\fmfreuse{4loopsrighteye}} + \frac{1}{8}\raisebox{-1ex}{\fmfreuse{4loopslefteye}}
\]
which we can check is primitive.  Continue likewise.
\end{example}

\end{fmffile}

\chapter{Reduction to geometric series}\label{second reduction}

\section{Single equations}

Let $D=\sgn(s)\gamma\cdot \partial_{-\rho}$ and $F_k(\rho) =
\sum_{i=0}^{t_k}F_{k,i}(\rho)$ so the Dyson-Schwinger
equation \eqref{dot eq} reads
\[
  \gamma \cdot L = \sum_{k \geq 1}x^k (1-D)^{1-sk}(e^{-L\rho}-1)F_{k}(\rho)\bigg|_{\rho=0}
\]
Only terms $L^jx^k$ with $k\geq j\geq 1$ occur by Lemma \ref{what
  appears} so this series lies in $(\mathbb{R}[L])[[x]]$.
Then we have the following

\begin{thm}\label{geo thm}
  There exists unique $r_{k}, r_{k, i}\in \mathbb{R}$, $k\geq 1$,
  $1\leq i < k$ such
  that
  \begin{align*}
    &\sum_k x^k (1 - D)^{1-sk}(e^{-L\rho}-1)F_{k}(\rho) \bigg|_{\rho=0}  \\
   & =  \sum_k x^k (1
    -D)^{1-sk}(e^{-L\rho}-1)\left(\frac{r_{k}}{\rho(1-\rho)} +
    \sum_{1 \leq i < k}\frac{r_{k,i}L^i}{\rho}\right) \bigg|_{\rho=0} 
  \end{align*}
\end{thm}

\begin{proof}
  For $\ell \geq 0$ the series in $x$
  \[
  x^k(1 -D)^{1-sk} \rho^\ell|_{\rho=0} 
  \]
  has no term of degree less than $k+\ell$ since $\gamma_i(x)$ has no
  term of degree less than $i$ by Lemma \ref{what appears}.  It
  follows that
\[
  x^k(1-D)^{1-sk}(e^{-L\rho}-1)\frac{1}{\rho(1-\rho)} \bigg|_{\rho=0}
  = -Lx^k + O(x^{k+1})
\]  
and
\[
  x^k(1-D)^{1-sk}(e^{-L\rho}-1)\frac{L^i}{\rho} \bigg|_{\rho=0}
  = -L^{i+1}x^k + O(x^{k+1})
\]  

Now expand 
  \[
    F_{k,i} = \sum_{j=-1}^{\infty} f_{k,i, j}\rho^j.
  \]
and define $r_n$, $r_{n,i}$ recursively in $n$ so
 \begin{align*}
    &\sum_k x^k (1 - D)^{1-sk}(e^{-L\rho}-1)F_{k}(\rho) \bigg|_{\rho=0}  \\
   & =  \sum_k x^k (1
    -D)^{1-sk}(e^{-L\rho}-1)\left(\frac{r_{k}}{\rho(1-\rho)} +
    \sum_{1\leq i < k}\frac{r_{k,i}L^i}{\rho}\right) \bigg|_{\rho=0}
    + O(x^{n+1}).
  \end{align*}
 This is possible since as noted above the coefficient of $x^n$ in
 $\sum_k x^k (1 - D)^{1-sk}(e^{-L\rho}-1)F_{k}(\rho) |_{\rho=0}$ is a
 polynomial in $L$ with degree at most $n-1$.
\end{proof}

The meaning of this theorem is that we can modify the Mellin
transforms of the primitives to be geometric series at order $L$.  The higher
powers of $\rho$ in the Mellin transform of a primitive at $k$ loops
become part of the coefficients of primitives at higher loops.  Note
that there are now terms at each loop order even if this was not
originally the case.  

\begin{example}\label{bkerfc phi3}
Consider the case $s=2$ with a single $B_+$ at order 1 as in Example
\ref{bkerfc setup}.   Write
\[
  F = \sum_{j=-1}^{\infty} f_{j}\rho^j.
\]
then computation gives
\begin{align*}
  r_1 = & f_{-1} \\
  r_2 = & f_{-1}^2 - f_{-1}f_0 \\
  r_{2,1} = & 0 \\
  r_3 = & 2f_{-1}^3 + f_{-1}^2(-4f_0 + f_1) + f_{-1}f_0^2 \\
  r_{3,1}= &-f_{-1}^3 + f_{-1}^2f_0 \\
  r_{3,2} =& 0\\
  r_4 = & 2f_{-1}^4 + f_{-1}^3(-12f_0+6f_1-f_2) + f_{-1}^2(9f_0^2 -
  3f_0f_1) - f_{-1}f_0^3 \\
  r_{4,1} = & -f_{-1}^4 + f_{-1}^3(6f_0-2f_1) - 3f_{-1}^2f_0^2 \\
  r_{4,2} = & \frac{7}{6}f_{-1}^4-\frac{7}{6}f_{-1}^3f_0 \\
  r_{4,3} = & 0\\
  r_5 = & -10f_{-1}^5 + f_{-1}^4(-6f_0 + 18f_1 - 8f_2 + f_3) +
  f_{-1}^3(40f_0^2 -32f_0f_1 + 4f_0f_2 + 2f_1^2) \\
  & + f_{-1}^2(-16f_0^3 +
  6f_0^2f_1) + f_{-1}f_0^4 \\
  \vdots
  %r_6 = & -18f_{-1}^6 + f_{-1}^5(124f_0 - 12f_1 - 24f_2 + 10f_3 -f_4) \\
  %& + f_{-1}^4(-2f_0^2 -136f_0f_1 +50f_0f_2 - 5f_0f_3 +25f_1^2 -
  %5f_1f_2) \\
  %& + f_{-1}^3(-100f_0^3 + 100f_0^2f_1 - 10f_0^2f_2 - 10f_0f_1^2) +
  %f_{-1}^2(25f_0^4 - 10f_0^3f_1) - f_{-1}f_0^5
\end{align*}

These identities are at present still a mystery.  Even the coefficients
of $f_{-1}^k$ in $r_k$ do not appear in Sloane's encyclopedia of
integer sequences \cite{OEIS} in any straightforward manner.  In the
case
\[
  F(\rho) = \frac{-1}{\rho(1-\rho)(2-\rho)(3-\rho)},
\]
as in the $\phi^3$ example from \cite{bkerfc}, the above specializes
to the also mysterious sequence
\begin{align*}
  r_1 = & -\frac{1}{6} & & & & \\
  r_2 = & -\frac{5}{6^3} & r_{2,1} = & 0 & & \\
  r_3 = & -\frac{14}{6^5} & r_{3,1} = &
  \frac{-5}{6^4} & r_{3,2} = & 0 \\
  r_4 = & \frac{563}{6^7} & r_{4,1} = &
  \frac{-173}{6^6} & r_{4,2} = & \frac{-35}{6^6} \\
  r_5 = & \frac{13030}{6^9} & \vdots &&\vdots & \\
  r_6 = & -\frac{194178}{6^{11}} & & & &
\end{align*}

\end{example}

Note that even if the coefficients of the original Mellin transforms are all of one
sign the $r_{k}$ may unfortunately not be so.

\section{Systems}

As in the single equation case we can reduce to geometric series
Mellin transforms at order $L$.
\begin{thm}\label{systems geo thm}
  There exists unique $r^j_{k}, r_{k,i}^j \in \mathbb{R}$, $k\geq
  1$, $1 \leq i < k$, $j\in \mathcal{R}$ such
  that
  \begin{align*}
    & \sum_{k \geq 1}\sum_{i=0}^{t_k} x^k (1-\sgn(s_r)\gamma^r \cdot
    \partial_{-\rho})^{1-s_rk}\prod_{j \in \mathcal{R}\smallsetminus
      \{r\}}(1-\sgn(s_j)\gamma^j
    \partial_{-\rho})^{-s_jk}
    (e^{-L\rho}-1)F^{k,i}(\rho)\bigg|_{\rho=0} \\
    & = \sum_{k \geq 1}\sum_{i=0}^{t_k} x^k (1-\sgn(s_r)\gamma^r \cdot
  \partial_{-\rho})^{1-s_rk}\prod_{j \in \mathcal{R}\smallsetminus
  \{r\}}(1-\sgn(s_j)\gamma^j 
  \partial_{-\rho})^{-s_jk} \\ & \qquad
(e^{-L\rho}-1)\left(\frac{r^r_{k,i}}{\rho(1-\rho)} + \sum_{1 \leq i <
  k}\frac{r^r_{k,i}L^i}{\rho}\right)\bigg|_{\rho=0}
  \end{align*}
\end{thm}

\begin{proof}
The proof follows as in the single equation case with the observation
that for $\ell\geq 0$
\[
 x^k \prod_{j \in
    \mathcal{R}}(1 + \gamma^j \cdot
  \partial_{-\rho})^{-s_jk+1}\rho^\ell |_{\rho=0}
\]
still has lowest term $x^{k+\ell}$.
\end{proof}

\chapter{The second recursion}\label{second recursion}

\section{Single equations}
Reducing to geometric series Mellin transforms at order $L$ allows us
to write a 
tidy recursion for $\gamma_1$.  Again let $D=\sgn(s)\gamma\cdot \partial_{-\rho}$ and $F_k(\rho) =
\sum_{i=0}^{t_k}F_{k,i}(\rho)$.  By Theorem \ref{geo thm} we have
\begin{equation}\label{new dot eq}
 \gamma \cdot L = \sum_k x^k (1
    -D)^{1-sk}(e^{-L\rho}-1)\left(\frac{r_{k}}{\rho(1-\rho)} +
    \sum_{1 \leq i < k}\frac{r_{k,i}L^i}{\rho}\right) \bigg|_{\rho=0} 
\end{equation}
Taking the coefficients of $L$ and $L^2$ gives
\begin{align*}
  \gamma_1 & = \sum_k x^k (1 - D)^{1-sk}
  \left(\frac{-r_k}{1-\rho}\right)\bigg|_{\rho=0} \\
  \gamma_2 & = \sum_k x^k (1 - D)^{1-sk} \left(\rho
  \frac{r_{k}}{2(1-\rho)} - r_{k,1}\right)\bigg|_{\rho =0}
\end{align*}
So
\[
 \gamma_1 + 2\gamma_2 = \sum_k x^k (1 - D)^{1-sk} (-r_k -
 2r_{k,1})\bigg|_{\rho=0} = \sum_k p(k) x^k = P(x)
\]
where $p(k) =
-r_{k}-2r_{k,1}$.  Then from Theorem \ref{single first recursion} or \ref{single first
    recursion 2}
\[
 \gamma_1 = P(x) - 2 \gamma_2 = P(x) - \gamma_1(\sgn(s)-|s|x\partial_x)\gamma_1
\]
giving

\begin{thm}\label{second recursion thm}
\[
\gamma_1(x) = P(x) - \gamma_1(x)(\sgn(s)-|s|x\partial_x)\gamma_1(x)
\]
or at the level of coefficients
\[
  \gamma_{1,n} = p(n) + \sum_{j=1}^{n-1} (|s|j-\sgn(s))\gamma_{1,j}\gamma_{1,n-j}
\]
\end{thm}

Notice that in defining the $r_{k}$ and $r_{k,i}$ we only used a
geometric series in the first case.  Specifically, we used
$1/(\rho(1-\rho))$ for $r_{k}$ but $1/\rho$ for $r_{k,i}$.  We could
have used $1/\rho$ in all cases; then one $L$ derivative would give
$\gamma_1(x) = \sum r_kx^k$ so all the information of $\gamma_1$ is in
the $r_k$, we learn nothing recursively.  The choice of a geometric series at order $L$ was made to
capture the fact that conformal invariance tells us that the Mellin
transform will be symmetrical when $\rho \mapsto 1-\rho$, and it also entirely
captures examples such as the Yukawa example from \cite{bkerfc} and
Example \ref{bkerfc setup}.  On the other hand choosing to use a
geometric series for the $r_{k,i}$ as well would not have resulted in a tidy
recursion for $\gamma_1$ using these techniques.  We hope that
the choice here gives an appropriate balance between representing the
underlying physics and giving tractable results all without putting
too much of the information into $P(x)$.

Another important question is how to interpret $P(x)$.  In cases like
the Yukawa example of \cite{bkerfc} where the various reductions are
unnecessary, then $P(x)$ is simply the renormalized Feynman rules
applied to the primitives.  In that particular example there is only
one primitive, and $P(x) = cx$ for appropriate $c$.  In the general
case we would like to
interpret $P(x)$ as a modified version of the renormalized
Feynman rules applied to the primitives.  For the first reduction this
is a reasonable interpretation since that reduction simply makes new
primitives, either within the Hopf algebra of Feynman graphs or more
generally.  For the second reduction the idea is that the geometric
series part of each Mellin transform is the primary part due to
conformal invariance.  At order $L$ the rest of the Mellin transform
gets pushed into higher loop orders, while at order $L^2$ the
reduction is a bit more crass.  This information together gives the
$r_k$ and the $r_{k,1}$ and hence gives $P(x)$.  So again, in view of
the previous paragraph, we view this as a modified version of the
Feynman rules applied to the primitives.

\section{Systems}

%Assume $\rho F_{k,i}(\rho) = r_{k,i;r}/(1-\rho)$.  Recall \eqref{system dot eq}
%\begin{align*}
%  \gamma^r \cdot L =
%  \sum_{k \geq 1}\sum_{i=0}^{t_k} & x^k 
%  (1-\sgn(s_r)\gamma^r \cdot
%  \partial_{-\rho})^{1-s_rk} \\
%  & \prod_{j \in \mathcal{R}\smallsetminus
%  \{r\}}(1-\sgn(s_j)\gamma^j \cdot
%  \partial_{-\rho})^{-s_jk}(e^{-L\rho}-1)F^{k,i}(\rho)\bigg|_{\rho=0}
%\end{align*}
Theorem \ref{systems geo thm} gives us
\begin{align*}
  & \gamma^r \cdot L \\
  & = \sum_{k \geq 1}\sum_{i=0}^{t_k} x^k (1-\sgn(s_r)\gamma^r \cdot
  \partial_{-\rho})^{1-s_rk}\prod_{j \in \mathcal{R}\smallsetminus
  \{r\}}(1-\sgn(s_j)\gamma^j 
  \partial_{-\rho})^{-s_jk} \\ & \qquad
(e^{-L\rho}-1)\left(\frac{r^r_{k,i}}{\rho(1-\rho)} + \prod_{1 \leq i <
  k}\frac{r^r_{k,i}L^i}{\rho}\right)\bigg|_{\rho=0}
\end{align*}
As in the single equation case we can find tidy recursions for the $\gamma^r_1$ by
comparing the coefficients of $L$ and $L^2$ in the above. We get
\[
  \gamma^r_1 = - \sum_k x^k 
(1-\sgn(s_r)\gamma^r \cdot
  \partial_{-\rho})^{1-s_rk}\prod_{j \in \mathcal{R}\smallsetminus
  \{r\}}(1-\sgn(s_j)\gamma^j \cdot
  \partial_{-\rho})^{-s_jk}
  \frac{-r_k^r}{1-\rho}|_{\rho=0}
\]
and
\begin{align*}
  2 \gamma^r_2 & = \sum_k  x^k 
 (1-\sgn(s_r)\gamma^r \cdot
  \partial_{-\rho})^{1-s_rk}\prod_{j \in \mathcal{R}\smallsetminus
  \{r\}}(1-\sgn(s_j)\gamma^j \cdot
  \partial_{-\rho})^{-s_jk} \\
  & \qquad \left(\frac{\rho r_{k}^r}{1-\rho} -2 r_{k,1}^r\right)
  |_{\rho =0} \\
  & = - \gamma^r_1 - \sum_{k\geq 1}(r^r_{k} + 2r_{k,1}^r) x^k 
\end{align*}
Thus letting $p^r(k) = -r^r_{k}-2r_{k,1}^r$ and using the first recursion
(Theorem \ref{systems first recursion} or \ref{systems first recursion
  2})
\[
 \gamma^r_1 = \sum_{k\geq 1} p^k(k) x^k - 2 \gamma^r_2 =
 \sum_{k\geq 1} p^r(k) x^k - \sgn(s_r)\gamma^r_1(x)^2 + \sum_{j
      \in \mathcal{R}}|s_j|\gamma^j_1(x)x
    \partial_x\gamma^r_1(x)
\]
giving

\begin{thm}\label{system second recursion thm}
\[
 \gamma^r_1 =  \sum_{k\geq 1} p^r(k) x^k -\sgn(s_r) \gamma^r_1(x)^2 + \sum_{j
      \in \mathcal{R}}|s_j|\gamma^j_1(x)x
    \partial_x\gamma^r_1(x)
\]
or at the level of coefficients
\[
  \gamma^r_{1,n} = p^r(n) +
  \sum_{i=1}^{n-1}(|s_r|i-\sgn(s_r))\gamma^r_{1,i}\gamma^r_{1,n-i} +
  \sum_{\substack{j \in \mathcal{R} \\ j
      \neq r}}\sum_{i=1}^{n-1} (|s_j|i)\gamma^j_{1,n-i}\gamma^r_{1,i}
\]
\end{thm}

\section{Variants}

The value of the reduction to geometric series is that if $F(\rho) =
r/(\rho(1-\rho))$ then $\rho^2 F(\rho) = \rho F(\rho) - 1$.  However
this reduction is rather crass, particularly for higher orders of $L$, so it is worth considering other
special forms of $F$ as in the following example.

\begin{example}
  Consider again the $\phi^3$ example from \cite{bkerfc} as setup in
  Example \ref{bkerfc phi3}.  We have $s=2$ and
  \[
  F(\rho) = \frac{-1}{\rho(1-\rho)(2-\rho)(3-\rho)},
  \]
  so 
  \begin{align*}
    \rho F(\rho) =&  \frac{-1}{(1-\rho)(2-\rho)(3-\rho)}  \\
    = & -\frac{1}{6}\left( 1 + \frac{\rho - \frac{11}{6}\rho^2 +
        \frac{1}{6}\rho^3}{(1-\rho)(2-\rho)(3-\rho)}\right) \\
    = & -\frac{1}{6} + \rho^2F(\rho) - \frac{11}{6}\rho^3F(\rho) +
    \frac{1}{6}\rho^4F(\rho)  
  \end{align*}

  This gives that
  \begin{align*}
    \gamma_1 = &- x (1 - \gamma \cdot
    \partial_{-\rho})^{-1} \rho F(\rho) |_{\rho =0} \\
    = &\frac{1}{6}x(1 - \gamma \cdot
    \partial_{-\rho})^{-1}1|_{\rho = 0} 
    - x(1 - \gamma \cdot
    \partial_{-\rho})^{-1}\rho^2F(\rho)|_{\rho = 0}  
    \\ & + \frac{11}{6}x(1 - \gamma \cdot
    \partial_{-\rho})^{-1}\rho^3F(\rho)|_{\rho = 0}  
     - \frac{1}{6}x(1 - \gamma \cdot
    \partial_{-\rho})^{-1}\rho^4F(\rho)|_{\rho = 0}  \\
    = & \frac{x}{6} - 2\gamma_2 - 11\gamma_3 - 4\gamma_4.
  \end{align*}
  In view of Theorem \ref{single first recursion} or \ref{single
    first recursion 2}, which in this case reads
  \[
  \gamma_k = \frac{1}{k} \gamma_1(x)(1-2x\partial_x)\gamma_{k-1}(x), 
  \]
  we thus get a fourth order differential equation
  for $\gamma_1$ which contains no infinite series and for which
  we completely understand the signs of the coefficients.
\end{example}

%***if the polylog stuff ever worked it would go here too***

\chapter{The radius of convergence}\label{radii}

\section{Single equations}

We see from the second recursion, Theorem \ref{second recursion thm}, that if $\sum p(k)x^k$ is
Gevrey-$n$ but not
Gevrey-$m$ for any $m<n$, then
$\gamma_1$ is at best Gevrey-$n$.

Of most interest for quantum field theory applications is the case
where only finitely many $p(k)$ are nonzero but all are nonnegative
and the case where $p(k) = c^k k!$
giving the Lipatov bound.  In both cases $\sum p(k)x^k$ is Gevrey-1.
Also for positivity reasons we are interested in $s \geq 1$ or $s<0$.  
Thus for
the remainder of this section the following assumptions are in effect.
\begin{assume}
 Assume $|s| \geq 1$ or $s<0$.
Assume $p(k) \geq 0$ for $k \geq 1$ and
\[
   \sum_{k \geq 1} x^k \frac{p(k)}{k!} = f(x)
\]
has radius of convergence $0 < \rho \leq \infty$ and is not
identically zero.% and $f(x) > 0$ for
%$|x|\leq \rho$.
\end{assume}
Under these assumptions $\gamma_1$ is also Gevrey-1 and the radius is the minimum of $\rho$ and
$1/(sa_1)$ (where we view $1/(sa_1)$ as $+ \infty$ in the case $a_1
= 0$) the proof of which is the content of this section.

\begin{definition}
Let $a_n = \gamma_{1,n}/n!$, $\mathbf{A}(x) = \sum_{n \geq 1} a_n
x^n$, and let $\rho_a$ be the radius of convergence of $\mathbf{A}(x)$. 
\end{definition}

Then $a_1 = \gamma_{1,1} = p(1)$ and
\begin{align}
  a_n = & \frac{p(n)}{n!} + \sum_{j=1}^{n-1} (|s|j-\sgn(s))\binom{n}{j}^{-1}
  a_j a_{n-j} \notag\\
  = & \frac{p(n)}{n!} + \frac{1}{2}\sum_{j=1}^{n-1} (|s|j-\sgn(s)
  +|s|(n-j)-\sgn(s))\binom{n}{j}^{-1} a_j a_{n-j} \notag \\
  = & \frac{p(n)}{n!} + \left(|s|\frac{n}{2} -\sgn(s)\right)\sum_{j=1}^{n-1} \binom{n}{j}^{-1}
  a_j a_{n-j} \label{a equation}
\end{align}
Inductively, we see that $a_1$, $a_2$, \ldots are all nonnegative

Note that if $s=1$, $p(1)>0$, and $p(n)=0$ for $n>1$ then $a_1=p(1)$,
$a_n=0$ for $n>1$ solves the recursion.  In this case $\rho_a = \rho =
\infty$, but $0<1/(|s|a_1) < \infty$.  This boundary case is the
only case with this behavior as we see in the following Proposition.

\begin{prop}\label{single b bound}Suppose that either $s\neq 1$, or
  $p(n) > 0$ for some $n>1$.  Then
$\rho_a \leq \min \{\rho, 1/(|s|a_1)\}$ where $1/(|s|a_1) = \infty$ when $a_1=0$
%The radius of convergence of $\sum_{n \geq 1} a_n x^n$ is at most 
%$\min \{\rho, 1/(sa_1)\}$, where
%$1/(sa_1)$ is
%interpreted to mean $+\infty$ in the case $a_1 = 0$.
\end{prop}

\begin{proof}
Take the first and last terms of the sum \eqref{a equation} to get
\begin{equation}\label{pre b bound}
  a_n \geq \frac{p(n)}{n!} +|s|\frac{n-2}{n}a_1a_{n-1}
\end{equation}
for $n \geq 2$.
In particular
\[
  a_n \geq \frac{p(n)}{n!}
\]
so $\rho_a \leq \rho$.  Further if $a_1$ and at least one $a_j$, $j>1$ are nonzero
then by \eqref{pre b bound} $a_n>0$ for all $n>j$, since the
$p(n)$ are assumed nonnegative.  In this case, then, we also have
\[
  \frac{a_{n-1}}{a_n} \leq \frac{n}{(n-2)a_1|s|}
\]
and so $\rho_a \leq 1/(|s|a_1)$.  The inequality $\rho_a \leq 1/(|s|a_1)$ also holds by convention if
$a_1=0$.  Finally suppose $a_1 \neq 0$ but $a_n=0$ for all $n>1$.
Then $p(n) = 0$ for all $n>1$, and, from \eqref{a equation} for
$n=2$, $s=1$.  This is the case we have excluded.  The result follows.
\end{proof}

For the lower bound on the radius we need a few preliminary
results.  First, some simple combinatorial facts.

\begin{lemma}\label{bin comb ineq}
  \[
  \binom{n}{k} \geq \left(\frac{n}{k}\right)^k
  \]
  for $n, k \in \mathbb{Z}$, $n\geq k \geq 0$.
\end{lemma}

\begin{proof}
  \[
  \binom{n}{k} = \frac{n}{k}\frac{n-1}{k-1} \cdots \frac{n-k+1}{1}
  \geq \frac{n}{k} \frac{n}{k} \cdots \frac{n}{k} = \left(\frac{n}{k}\right)^k
  \]
\end{proof}

\begin{lemma}\label{theta}
  Given $0 < \theta < 1$
  \[
    \frac{1}{n} \binom{n}{j} \geq \frac{\theta^{-j+1}}{j}
  \]
  for $1 \leq j \leq \theta n$ and $n \geq 2$.
\end{lemma}

\begin{proof}
  Fix $n$.  Write $j = \lambda n$, $0 < \lambda \leq \theta$.  Then
  using Lemma \ref{bin comb ineq}
  \[
    \frac{1}{n} \binom{n}{j} =
    \frac{1}{n} \binom{n}{\lambda n} \geq  \frac{n^{\lambda
        n-1}}{(\lambda n)^{\lambda n}}
    = \frac{\lambda^{-\lambda n + 1}}{\lambda n} \geq
    \frac{\theta^{-j+1}}{j}
  \]
\end{proof}

Second, we need to understand the behavior of $\sum a_n x^n$ at the
radius of convergence.

\begin{lemma}\label{converge}
  \[
    \mathbf{A}(x) \leq f(x) + x|s|(1+\epsilon)\mathbf{A}'(\theta x) \mathbf{A}(x) +
    \frac{|s|}{2x}\frac{d}{dx}\left(x^2\mathbf{A}^2(\theta^\theta
    x)\right) + P_\epsilon(x)
  \] for all $0 < \theta < 1/e$, $\epsilon > 0$, and $0 < x < \rho_a$,
  where $P_\epsilon(x)$ is a polynomial in $x$ with nonnegative coefficients.
\end{lemma}

\begin{proof}
  Take $0 < \theta < 1/e$ and $\epsilon > 0$.
  \begin{align*}
    a_n = & \frac{p(n)}{n!} + \left(|s|\frac{n}{2}
      -\sgn(s)\right)\sum_{j=1}^{n-1} \binom{n}{j}^{-1}  a_j a_{n-j} \\
    \leq & \frac{p(n)}{n!} + |s|(n+2)\sum_{1
      \leq j \leq \theta n}  \binom{n}{j}^{-1} a_j a_{n-j} +|s|
      \frac{n+2}{2} \sum_{\theta n \leq j \leq n-\theta
      n}\binom{n}{j}^{-1} a_j a_{n-j} \\
    \leq & \frac{p(n)}{n!} + |s|\frac{n+2}{n} \sum_{1
      \leq j \leq \theta n} j\theta^{j-1} a_j a_{n-j}
    +|s| \binom{n}{ \lceil \theta n \rceil}^{-1}\frac{n+2}{2}
      \sum_{\theta n \leq j \leq n-\theta
      n}a_j a_{n-j} \\
      & \qquad \text{by Lemma \ref{theta}} \\
    \leq & \frac{p(n)}{n!} +|s| \frac{n+2}{n}\sum_{1
      \leq j \leq \theta n} j \theta^{j-1} a_j a_{n-j}
    +\frac{|s|}{2} (n+2) \theta^{\theta n}
    \sum_{\theta n \leq j \leq n-\theta
      n} a_j a_{n-j} \\
    & \qquad \text{by Lemma \ref{theta} and since $(x/n)^x$ is
      decreasing for $0<x<n/e$} 
  \end{align*}
  Thus for $n$ sufficiently large that $(n+2)/n \leq 1+\epsilon$ the 
  coefficients of $\mathbf{A}(x)$ are bounded above by the
  coefficients of
  \[
    f(x) + x|s|(1+\epsilon)\mathbf{A}'(\theta x) \mathbf{A}(x) +
    \frac{|s|}{2x}\frac{d}{dx}\left(x^2\mathbf{A}^2(\theta^\theta
    x)\right).
  \]
  Adding a polynomial to dominate the earlier
  coefficients of $\mathbf{A}(x)$ we get that the coefficients of
  $\mathbf{A}(x)$ are bounded above by the coefficients of
\[
    f(x) + x|s|(1+\epsilon)\mathbf{A}'(\theta x) \mathbf{A}(x) +
    \frac{|s|}{2x}\frac{d}{dx}\left(x^2\mathbf{A}^2(\theta^\theta
    x)\right) + P_\epsilon(x).
  \]
  Since all coefficients are nonnegative, for any $0 < x <
  \rho_a$ we have
  \[
  \mathbf{A}(x) \leq f(x) + x|s|(1+\epsilon)\mathbf{A}'(\theta x)
  \mathbf{A}(x) + 
  \frac{|s|}{2x}\frac{d}{dx}\left(x^2\mathbf{A}^2(\theta^\theta
  x)\right)
  + P_\epsilon(x).
  \]
  %which is continuous in $\theta$.  So for fixed $0 < x <
  %\rho_a$ we can let $\theta \rightarrow 0$ giving
  %\[
  %  \mathbf{A}(x) \leq f(x) + xsa_1\mathbf{A}(x)
  %\]
  %so
  %\[
  %  \frac{\mathbf{A}(x)(1-xsa_1)}{f(x)} \leq 1
  %\]
  %for $0 < x < \rho_a$.  
  %The result follows.
\end{proof}

\begin{lemma}\label{A rho finite}
  If $\rho_a < \rho$ and $\rho_a < 1/(|s|a_1)$ then $\mathbf{A}(\rho_a)
  < \infty$.
\end{lemma}

\begin{proof}
Consider Lemma \ref{converge}.  Choose
$\theta > 0$ and $\epsilon > 0$ so that 
\begin{equation}\label{theta bound}
 \rho_a < \frac{1}{|s|(1+\epsilon)\mathbf{A}'(\theta \rho_a)}
\end{equation}
which is possible since $\lim_{\theta \rightarrow 0}\mathbf{A}'(\theta
x) = a_1$ and $\rho_a < 1/(|s|a_1)$.
Letting $x \rightarrow \rho_a$ we see that 
\[
 \lim_{x \rightarrow \rho_a} \mathbf{A}(x) \leq C + \rho_a|s|(1+\epsilon)\mathbf{A}'(\theta \rho_a)\lim_{x
   \rightarrow \rho_a}\mathbf{A}(x)
\]
where $C$ is constant, since $\theta^\theta <
1$, and $\rho_a < \rho$.  So
\[
  (1-\rho_a|s|(1+\epsilon) \mathbf{A}'(\theta \rho_a)) \lim_{x \rightarrow \rho_a}
\mathbf{A}(x) \leq C.
\]
But by \eqref{theta bound}, $1-\rho_a|s|(1+\epsilon) \mathbf{A}'(\theta \rho_a) >
0$, so $\mathbf{A}(\rho_a) < \infty$.
\end{proof}

\begin{lemma}\label{A rho infinite}
If $\rho_a < \rho$ and $\rho_a < 1/(|s|a_1)$ then $\mathbf{A}(x)$ is
unbounded on $0 < x < \rho_a$.
\end{lemma}

\begin{proof}
  Take any
$\epsilon > 0$.  Then there
exists an $N>0$ such that for $n>N$
\[
 a_n  \leq \frac{p(n)}{n!} + |s|a_1a_{n-1} + \epsilon \sum_{j=1}^{n-1}a_ja_{n-j}
\]

Define
\[
  c_n = \begin{cases}
    a_n & \text{if } a_n > \frac{p(n)}{n!} + |s|c_1c_{n-1} +
    \epsilon \sum_{j=1}^{n-1}c_jc_{n-j} \\
    \frac{p(n)}{n!} + |s|c_1c_{n-1} + \epsilon
    \sum_{j=1}^{n-1}c_jc_{n-j} & \text{otherwise (in particular
      when $n>N$)}
    \end{cases}
\]
In particular $c_1 = a_1$.  Let $\mathbf{C}(x) = \sum_{x \geq 1} c_n
x^n$ (which implicitly depends on $\epsilon$) have radius
$\rho_\epsilon$.  Since $a_n \leq c_n$, $\rho_a \geq \rho_\epsilon$.
Rewriting with generating series
\[
  \mathbf{C}(x) = f(x) + |s|a_1x\mathbf{C}(x) + \epsilon
  \mathbf{C}^2(x) + P_\epsilon(x)
\]
where $P_\epsilon(x)$ is some polynomial.
This equation can be solved by the
quadratic formula.  The discriminant is
\[
  \Delta_\epsilon = (1-|s|a_1x)^2 - 4 \epsilon (f(x)+P_\epsilon(x)).
\]
$\rho_\epsilon$ is the closest root to $0$  of $\Delta_\epsilon$.
%Let $\rho_0 = 1/(sa_1)$ which is the root of $\Delta_0$ closest to
%$0$.

  By construction, the coefficient of $x^n$ in $P_\epsilon(x)$ is
  bounded by $a_n$.   Suppose $\mathbf{A}(\rho_a) < \infty$. Thus
  $f(\rho_a) + P_\epsilon(\rho_a) \leq f(\rho_a) +
  \mathbf{A}(\rho_a)$. By the nonnegativity of the coefficients of
  $f$ and $P_\epsilon$ then $f(x)+P_\epsilon(x)$ independently of
  $\epsilon$ for
 for $0 < x \leq
  \rho_a$.  Thus
   \[
    \lim_{\epsilon \rightarrow 0}\Delta_\epsilon =
    (1-|s|a_1\rho_a)^2
  \]
  for $0 < x \leq \rho_a$.
  So 
  \[
  \frac{1}{|s|a_1} > \rho_a \geq  \rho_\epsilon \rightarrow \frac{1}{|s|a_1}
  \]
  as $\epsilon
    \rightarrow 0$
  which is a contradiction, giving that
  $\mathbf{A}(x)$ is unbounded on $0<x<\rho_a$.
\end{proof}

\begin{prop}\label{single c bound}
$\rho_a \geq \min \{\rho, 1/(|s|a_1)\}$, where
$1/(|s|a_1) = \infty$ when $a_1=0$.
\end{prop}

\begin{proof}
Suppose on the contrary that $\rho_a < \rho$ and $\rho_a < 1/(|s|a_1)$
then  Lemmas \ref{A rho finite} and \ref{A rho
  infinite} contradict each other so this cannot be the case.
%  This equation can be solved by the
%quadratic formula.  The discriminant is
%\begin{equation}\label{discriminant}
%  \Delta_\epsilon = (1-sa_1x)^2 - 4 \epsilon (f(x)+P_\epsilon(x))
%\end{equation}
%which has no singularities in $|x| < \rho$.
%So $\rho_\epsilon$ is the closest root to $0$  of
%\eqref{discriminant}.
%Let $\rho_0 = 1/(sa_1)$ which is the root of $\Delta_0$ closest to
%$0$.
%
%Then to get the lower bound on the radius of $\mathbf{A}(x)$ it remains
%only to prove the following lemma.
\end{proof}

%\begin{lemma}
%  With notation and assumptions as above if $a_1 \neq 0$, $\lim_{\epsilon \rightarrow 0}
%  |\rho_\epsilon| \geq \rho_0$ (which is a contradiction to $\rho_a <
%  1/(sa_1)$), while if $a_1 = 0$ we also have a contradiction.
%\end{lemma}
%
%\begin{proof}
%  In view of Lemma \ref{converge} this is a short exercise in analysis.
%
%  By construction the coefficient of $x^n$ in $P_\epsilon(x)$ is
%  bounded by $a_n$.  So $P_\epsilon(x)$ has coefficients
%  which are nonnegative and bounded by those of $\mathbf{A}(x)$.  Thus
%  $f(\rho_a) + P_\epsilon(\rho_a) \leq f(\rho_a) +
%  \mathbf{A}(\rho_a) < M$ for $M>0$ independent of $\epsilon$. By the nonnegativity of the coefficients of
%  $f$ and $P_\epsilon$ then $|f(x)+P_\epsilon(x)| < M$
% for $|x| \leq
%  \rho_a$.
%
%  Suppose $a_1 \neq 0$.  Take any $\eta > 0$.   Consider $|x| \leq \rho_a$.
%  Choose $\delta > 0$ such that
%  $(1-sa_1x)^2 < \delta$ implies $|x - \rho_0| < \eta$.
%  Pick $\epsilon < \delta/(4M)$.  Then
%  $
%    (1-sa_1\rho_\epsilon)^2 = 4\epsilon(f(\rho_\epsilon) +
%    P_\epsilon(\rho_\epsilon)) < \delta
%  $
%  so $|\rho_\epsilon - \rho_0| < \eta$.
%
%  Suppose on the other hand that $a_1 = 0$.  Take $0 < \delta < 1$.
%  Then, since $|\rho_\epsilon|
%  \leq \rho_a$, we get that for $\epsilon < \delta/(4M)$, $1 =
%  4\epsilon(f(\rho_\epsilon) + P_\epsilon(\rho_\epsilon)) < \delta$
%  which is a contradiction.
%\end{proof}

Taking the two bounds together we get the final result

\begin{thm}
Assume $\sum_{k \geq 1} x^k p(k)/k!$ has radius $\rho$.
Then $\sum x^n\gamma_{1,n}/n!$ converges with radius
of convergence $\min \{\rho, 1/(s\gamma_{1,1})\}$, where
$1/(|s|\gamma_{1,1}) = \infty$ if $\gamma_{1,1} = 0$.
\end{thm}

\begin{proof}
Immediate from Lemmas \ref{single b bound} and \ref{single c bound}.
\end{proof}

\section{Systems}
Now suppose we have a system of Dyson-Schwinger equations as in
\eqref{system combDSE}
\[
  X^r(x) = \mathbb{I} - \sgn(s_r)\sum_{k \geq 1}\sum_{i=0}^{t^r_k}x^k
  B_+^{k,i;r}(X^rQ^{k})
\]
for $r \in \mathcal{R}$ with $\mathcal{R}$ a finite set and
where
\[
  Q = \prod_{r\in \mathcal{R}}X^r(x)^{-s_r}
\]
for all $r \in \mathcal{R}$.

To attack the growth of the $\gamma^r_1$ we will again assume that the
series of primitives is Gevrey-1 and that the $s_r$ give nonnegative series.

\begin{assume}
Assume $s_r \geq 1$ or $s_r<0$ for each $r \in \mathcal{R}$.  
Assume
that
\[
   \sum_{k \geq 1} x^k \frac{p^r(k)}{k!} = f^r(x)
\]
has radius $0 < \rho_r \leq \infty$, $p^r(k) > 0$ for $k\geq 1$, and
the $f^r(x)$ are not identically $0$.
\end{assume}
We'll proceed
by similar bounds to before.

\begin{definition}
Let $a_n^r = \gamma^r_{1,n}/n!$ and $\mathbf{A}^r(x) = \sum_{n \geq 1}
a_n^rx^n$.
\end{definition} 
Again the $a_i^r$ are all nonnegative.

Then
\begin{equation}\label{ar equation}
  a_n^r = \frac{p^r(n)}{n!} +
  \sum_{i=1}^{n-1}(|s_r|i-\sgn(s_r))a^r_ia^r_{n-i}\binom{n}{i}^{-1} +
  \sum_{\substack{j \in \mathcal{R} \\ j
      \neq r}}\sum_{i=1}^{n-1} (|s_j|i)a^j_{n-i}a^r_{i}\binom{n}{i}^{-1}
\end{equation}

\begin{prop}\label{system b bound}
For all $r \in \mathcal{R}$, the radius of convergence of
$\mathbf{A}^r(x)$ is at most
\[
  \min \left\{ \rho_r, \frac{1}{\sum_{j \in \mathcal{R}} |s_j|a_1^j}\right\}
\]
interpreting the second possibility to be $\infty$ when $\sum_{j \in \mathcal{R}} |s_j|a_1^j=0$.
\end{prop}

\begin{proof}
Taking the last term in each sum of \eqref{ar equation} we have
\[
  a_n^r \geq \frac{p^r(n)}{n!} + \left(\sum_{j \in \mathcal{R}} |s_j|a_1^j\right)\frac{n-2}{n}a_{n-1}^r
\]
Let $b^r_n$ be the series defined by $b^r_1 = a^r_1$ and equality in the
above recursion. Then argue
as in the single equation case, Proposition \ref{single b bound}, to get 
that the radius of $\mathbf{A}^r(x)$
is at most
\[
  \min \left\{ \rho_r, \frac{1}{\sum_{j \in \mathcal{R}} |s_j|a_1^j}\right\}.
\]
\end{proof}

\begin{prop}\label{system c bound}
The radius of convergence of
$\sum_{r\in\mathcal{R}}\mathbf{A}^r(x)$ is at least
\[
  \min_{r \in \mathcal{R}} \left\{ \rho_r, \frac{1}{\sum_{j \in \mathcal{R}} |s_j|a_1^j}\right\}
\]
interpreting the second possibility to be $\infty$ when $\sum_{j \in \mathcal{R}} |s_j|a_1^j=0$.
\end{prop}

\begin{proof}
The overall structure of the argument is as in the single equation case.

The equivalent of Lemma \ref{converge} for this case
follows from
\begin{align*}
  \sum_{r \in \mathcal{R}} a_n^r & \leq \sum_{r\in\mathcal{R}}
  \frac{p^r(n)}{n!}  + \frac{n+2}{n}\sum_{j \in \mathcal{R}}|s_j|a_1^j\sum_{r \in
    \mathcal{R}}a_{n-1}^r + \sum_{r, j\in\mathcal{R}}\sum_{i=1}^{n-1}(|s_j|
  (i+1))a_{n-i}^ja_i^r \binom{n}{i}^{-1} \\
  & \leq \sum_{r\in\mathcal{R}} \frac{p^r(n)}{n!}  + \frac{n+2}{n}\sum_{j \in
    \mathcal{R}}|s_j|a_1^j\sum_{r \in \mathcal{R}}a_{n-1}^r \\
  & \qquad +
  \max_{j}(|s_j|)\sum_{i=2}^{n-2} (i+1) \binom{n}{i}^{-1}
  \left(\sum_{r\in\mathcal{R}}a^r_{n-i}\right) \left(\sum_{r
      \in\mathcal{R}} a^r_{i}\right) \\
  & = \sum_{r\in\mathcal{R}} \frac{p^r(n)}{n!}  + \frac{n+2}{n}\sum_{j \in
    \mathcal{R}}|s_j|a_1^j\sum_{r \in \mathcal{R}}a_{n-1}^r \\
  & \qquad + \max_{j}(|s_j|) (n+2)
  \sum_{2 \leq i \leq \theta n} \binom{n}{i}^{-1}
  \left(\sum_{r\in\mathcal{R}}a^r_{n-i}\right) \left(\sum_{r
      \in\mathcal{R}} a^r_{i}\right) \\
  & \qquad + \max_{j}(|s_j|) \frac{n+2}{2}
  \sum_{\theta n \leq i \leq n- \theta n} \binom{n}{i}^{-1}
  \left(\sum_{r\in\mathcal{R}}a^r_{n-i}\right) \left(\sum_{r
      \in\mathcal{R}} a^r_{i}\right)
\end{align*}
for $\theta$ as in Lemma \ref{converge} with $\sum_{r \in \mathcal{R}}
\mathbf{A}^r(x)$ in place of $\mathbf{A}(x)$, where $\mathbf{A}^r(x) =
\sum a^r(n)x^n$.  Then continue the argument as in Lemma
\ref{converge} with $\sum_{r \in
  \mathcal{R}}f^r(x)$ in place of $f(x)$ and $\max_{j}(s_j)$ in
place of $s$, and using the second term to get the correct linear
part.

For the argument as in Lemma \ref{A rho infinite}
Take any $\epsilon > 0$ then there
exists an $N > 0$ such that for $n > N$ we get
\[
  a_n^r \leq \frac{p^r(n)}{n!} + \left(\sum_{j \in \mathcal{R}}
    |s_j|a_1^j\right)a^r_{n-1} + \epsilon \sum_{i=1}^{n-1} \sum_{j \in
    \mathcal{R}} a^r_i a^j_{n-i}
\]
Taking $\mathbf{C}^r(x)$ to be the series whose coefficients satisfy
the above recursion with equality in the cases when this gives a result $\geq
a^r_n$ and equal to $a^r_n$ otherwise we get
\[
  \mathbf{C}^r(x) = f^r(x) + \left(\sum_{j \in \mathcal{R}}
    |s_j|a_1^j\right)x\mathbf{C}^r(x) + \epsilon \sum_{j \in
    \mathcal{R}}\mathbf{C}^r(x)\mathbf{C}^j(x) + P_\epsilon^r(x)
\]
where $P_\epsilon^r$ is a polynomial.

Summing over $r$ we get a recursive equation for $\sum_{r \in
  \mathcal{R}}\mathbf{C}^r(x)$ of the same form as in the single
equation case.  Note that since each $\mathbf{C}^r$ is a series with
nonnegative coefficients there can be no cancellation of singularities
and hence the radius of convergence of each $\mathbf{C}^r$ is at least
that of the sum.  
Thus by the analysis of the single equation case we
get a lower bound on the radius of $\sum_r \mathbf{A}^r(x)$ of $\min_{s \in \mathcal{R}} \{\rho_s,
1/\sum_{j \in \mathcal{R}}|s_j|a_1^j\}$.  
\end{proof}

\begin{prop}\label{same rad}
Each $\mathbf{A}^s(x)$, $s \in \mathcal{R}$, has the same
radius of convergence.
\end{prop}

\begin{proof}
Suppose the radius of $\mathbf{A}^s(x)$ was strictly greater
than that of $\mathbf{A}^r(x)$.  Then we can find $\beta > \delta > 0$
such that
\[
   a_n^r > \beta^n > \delta^n > a_n^s
\]
for $n$ sufficiently large.  Pick a $k \ge 1$ such that $a_k^s > 0$.  Then
\[
  \delta^n > a_n^s \geq \frac{|s_r| k! a_k^s}{n\cdots (n-k+1)}
  a_{n-k}^r > \frac{|s_r| k!
    a_k^s}{n\cdots (n-k+1)} \beta^{n-k}
\]
so
\[
  \frac{\delta^k}{|s_r| a_k^s} \left(\frac{\delta}{\beta}\right)^{n-k}
  > \frac{k!}{n \cdots (n-k+1)}
\]
which is false for $n$ sufficiently large, giving a contradiction.
\end{proof}

\begin{thm}
For all $r \in \mathcal{R}$, $\sum x^n\gamma^r_{1,n}/n!$ converges with radius \[\min_{r
  \in \mathcal{R}}\{\rho_r,
1/\sum_{j \in \mathcal{R}}|s_j|\gamma_{1,1}^j\},\] where the second
possibility is interpreted as $\infty$ when $\sum_{j \in
  \mathcal{R}}|s_j|\gamma_{1,1}^j = 0$.
\end{thm}

\begin{proof}
Take $s \in
\mathcal{R}$ such that $\rho_s$ is minimal.

Since we are working with nonnegative series the radius of
$\mathbf{A}^s(x)$ is at least that of
$\sum_{r\in\mathcal{R}} \mathbf{A}^r(x)$.  Hence by Lemmas
\ref{system b bound} and \ref{system c bound} $\mathbf{A}^s(x)$
has radius exactly \[\min_{r \in \mathcal{R}}\{\rho_r,
1/\sum_{j \in \mathcal{R}}|s_j|a_1^j\}.\]

Thus by Lemma \ref{same rad} all the $\sum a_n^s x^n$ have the same radius $\min_{r \in \mathcal{R}}\{\rho_r,
1/\sum_{j \in \mathcal{R}}|s_j|a_1^j\}$
\end{proof}

\section{Possibly negative systems}\label{gamma1 pos}
Let us relax the restriction that $p^r(n) \geq 0$.
It is now difficult to 
make general statements concerning the radius of convergence of the
$\mathbf{A}^r(x)$.  For example consider the system
\begin{align*}
  a^1_n & = \frac{p^1(n)}{n!} + \sum_{j=1}^{n-1}(2j-1)a_j^1 a_{n-j}^1
  \binom{n}{j}^{-1} + \sum_{j=1}^{n-1}ja_j^1
  a_{n-j}^2\binom{n}{j}^{-1} \\
  a^2_n & = \frac{p^2(n)}{n!} + \sum_{j=1}^{n-1}(j+1)a_j^2 a_{n-j}^2
  \binom{n}{j}^{-1} + \sum_{j=1}^{n-1}2ja_j^2
  a_{n-j}^1\binom{n}{j}^{-1}
\end{align*}
so $s_1 = 2$ and $s_2 = -1$.  Suppose also that
\begin{align*}
  p^2(2) & = -4(a_1^2)^2 \\
  a_1^1 & = a^2_1 \\
  p^2(n) & = -2(n-1)!a_1^2 a_{n-1}^1 \\
\end{align*}
Then $a_2^2 = 0$ and inductively $a_n^2 = 0$ for $n\geq 2$ so the system degenerates to
\begin{align*}
  a^1_n & = \frac{p^1(n)}{n!} + \sum_{j=1}^{n-1}(2j-1)a_j^1 a_{n-j}^1
  \binom{n}{j}^{-1} - \frac{n-1}{n}a_1^1a_{n-1}^1 \\
  a^2_n & =
  \begin{cases}
    a^1_1 & \text{if n = 1} \\
    0 & \text{otherwise}
  \end{cases}
\end{align*}
We still have a free choice of $p^1(n)$, and hence control of the
radius of the $a^1$ series.  On the other hand the
$a^2$ series trivially has infinite radius of convergence.

Generally, finding a lower bound on the radii of the solution series,
remains approachable by the preceding methods while control of the
radii from above is no longer apparent.

Precisely, 
\begin{thm}
  The radius of convergence of $\sum_{n \geq 1} x^n\gamma^r_{1,n}/n!$
  is at least
  \[
    \min_{r \in \mathcal{R}}\left\{\rho_r,
    \frac{1}{\left|\sum_{j \in \mathcal{R}}|s_j|\gamma_{1,1}^j\right|}\right\}
  \]
  where the second possibility is interpreted as $\infty$ when
  $\sum_{j \in \mathcal{R}}|s_j|\gamma_{1,1}^j = 0$.
\end{thm}

\begin{proof}
for any $\epsilon > 0$
\begin{align*}
  |a_n^r|
  & \leq \frac{\left|p^r(n)\right|}{n!} + \left|\sum_{j \in \mathcal{R}}
    |s_j|a_1^j\right||a^r_{n-1}| +
    \sum_{i=1}^{n-2}|(|s_r|i-\sgn(s_r))||a^r_i||a^r_{n-i}|\binom{n}{i}^{-1} \\
    & \qquad +
    \sum_{\substack{j \in \mathcal{R} \\ j
        \neq r}}\sum_{i=1}^{n-2}
    |s_j|i|a^j_{n-i}||a^r_{i}|\binom{n}{i}^{-1} \\
  & \leq \frac{\left|p^r(n)\right|}{n!} + \left|\sum_{j \in \mathcal{R}}
    |s_j|a_1^j\right||a^r_{n-1}| + \epsilon \sum_{i=1}^{n-1} \sum_{j \in
    \mathcal{R}} |a^r_i| |a^j_{n-i}|
\end{align*}
So, for a lower bound on the radius we may proceed as in the
nonnegative case using the absolute value of the coefficients.
\end{proof}

\chapter{The second recursion as a differential equation}\label{dechapter}

In this final chapter let us consider the second recursion derived in
Chapter \ref{second recursion} as a differential equation rather than
as a recursive equation.  That is, in the system case
\begin{equation}\label{system}
  \gamma_1^r(x) = P_r(x) - \sgn(s_r)\gamma_1^r(x)^2 + \left(\sum_{j \in \mathcal{R}}|s_j|\gamma_1^j(x)\right) x \partial_x \gamma_1^r(x)
\end{equation}
as $r$ runs over $\mathcal{R}$, the residues of the theory.
While in the single equation case
\begin{equation}\label{de}
  m \gamma_1(x) = P(x) - \sgn(s)\gamma_1(x)^2 + |s|\gamma_1(x) x \partial_x \gamma_1(x)
\end{equation}
The parameter $m$ was added to keep the QED example in the most
natural form, however it is not interesting since since we can remove
it by the transformation $\gamma_1(x) \mapsto m \gamma_1(x)$, $P(x)
\mapsto m^2 P(x)$.

No non-trivial results will be proved in this chapter, we will simply
discuss some features of some important examples.  More substantial
results will appear in \cite{vBKUY}. 

As a consequence of the renormalization group origin of the first
recursion discussed in section \ref{rg section} the $\beta$-function
for the system shows up as the 
coefficient of $(\gamma_1^r)'(x)$, namely 
\[
  \beta(x) = x\sum_{j \in \mathcal{R}} |s_j| \gamma_1^j(x)
\]
in the system case and 
\[
  \beta(x) = x|s|\gamma_1(x)
\]
in the single
equation case.  Consequently this differential equation is well suited
to improving our understanding of the $\beta$-function.  

In particular
in the single equation case, we see immediately from \eqref{de} that
any zeroes of $\beta(x)$ must occur either where $P(x)=0$ or where
$\gamma_1'(x)$ is infinite.  The second of these possibilities does
not turn out to be physically reasonable as we will discuss in more
detail below.  
The system case is not quite so simple.  Assume $\beta(x)=0$.  If we
rule out infinite $(\gamma_1^r)'(x)$, then we can only conclude that
for each $r \in \mathcal{R}$
\[
  \gamma_1^r(x) + \sgn(s_r)\gamma_1^r(x)^2 - P_r(x) = 0.
\]

In order to extract further information in both the single equation
and the system case we will proceed to examine plots of the vector
field of $(\gamma_1^r)'(x)$, first in some toy single equation cases,
second in the case of QED reduced to one equation, and finally in the
2 equation example of $\phi^4$.

\section{Toys}

First let us consider a family of examples which are simpler than those which
occur in full quantum field theories,  namely the family where $m=1$
and $P(x) = x$. 

\subsection{The case $s=2$}\label{erfc}
If we set $s=2$ we get the situation explored in \cite{bkerfc} which
describes the piece of massless Yukawa theory consisting of nestings
and chainings of the one loop fermion self energy into itself as
discussed in Example \ref{bkerfc setup}.  The second recursion
viewed as a differential equation is
\[
  \gamma_{1}(x) = x - \gamma_1(x)(1-2x\partial_x)\gamma_{1}(x).
\]

Broadhurst and Kreimer \cite{bkerfc} solved this Dyson-Schwinger equation
by clever rearranging and recognizing the resulting asymptotic
expansion.  The solution, written in a slightly different form, is
given implicitly by 
\[
  \exp\left(\frac{(1+\gamma_1(x))^2}{2x}\right)\sqrt{-x} +
  \mathrm{erf}\left(\frac{1+\gamma_1(x)}{\sqrt{-2x}}\right)\frac{\sqrt{\pi}}{\sqrt{2}}
  = C
\]
with integration constant $C$.

\begin{figure}
\epsfig{file=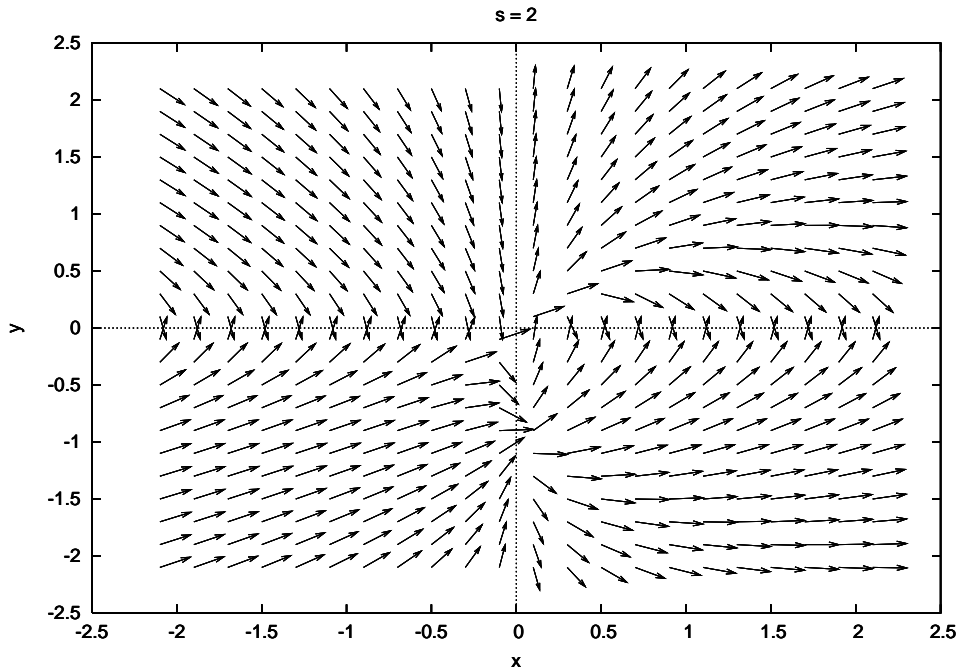, width=13cm}
\caption{The vector field of $\gamma_1'(x)$ with $s=2$, $m=1$, and
  $P(x)=x$.}
\label{s2quadgp}
\end{figure}

We can proceed to look at the vector field of $\gamma_1'(x)$, see Figure
\ref{s2quadgp}.

\begin{figure}
\epsfig{file=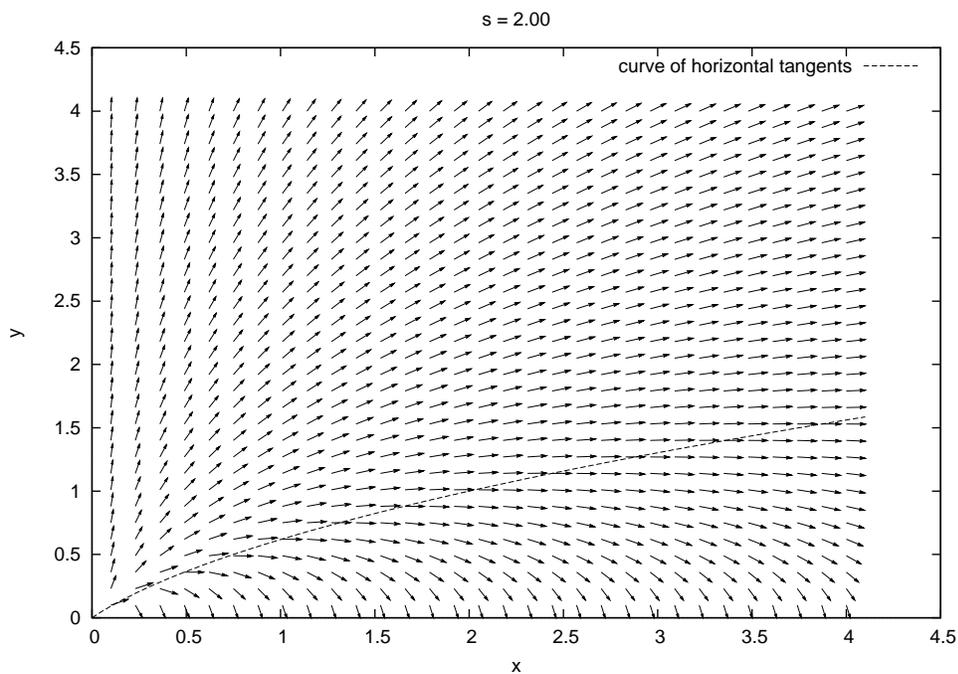,
    width=13cm}
\caption{Solutions which die in finite time
along with the curve where
$\gamma_1'(x)=0$.}
\label{s2horiz}
\end{figure}

\begin{figure}
\epsfig{file=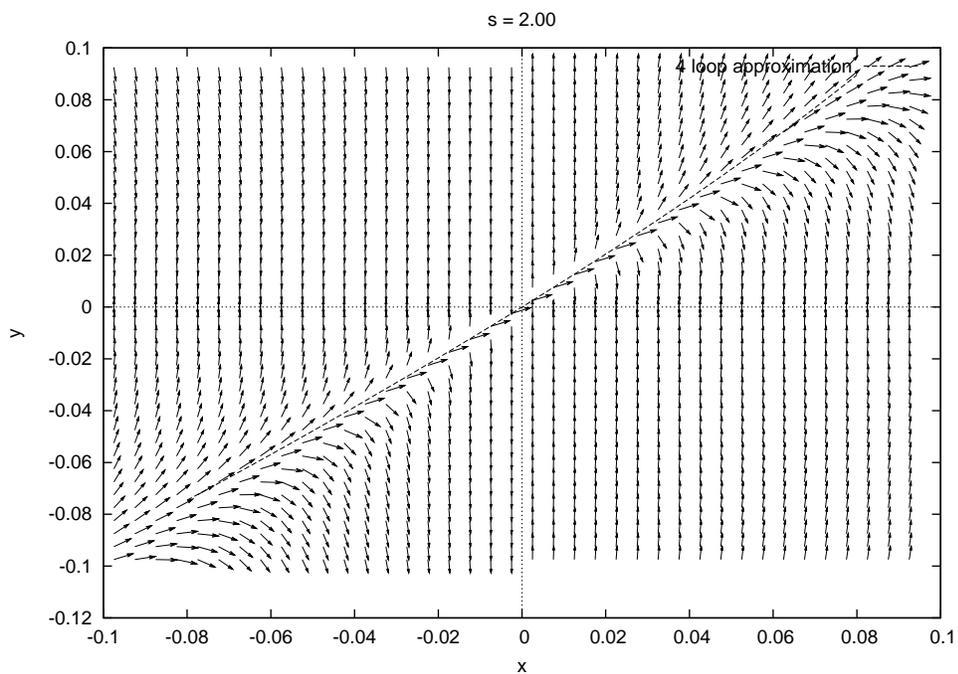, width=13cm}
\caption{The four loop approximation near the origin.}
\label{s2serieszoom}
\end{figure}

We are primarily interested in the behavior in the first quadrant.
Of particular interest are possible zeros of solutions since, in this
simple single equation situation, $x\gamma_1(x) = \beta(x)$ where
$\beta(x)$ is the $\beta$-function of the system.  

From the figures we notice a family of solutions which come down to
hit the $x$ axis with vertical tangent.  These solutions have no real
continuation past this point.  These solutions are consequently
unphysical.
It is not clear
from the figure whether all solutions have this behavior.  One of the
major goals of \cite{vBKUY} is to find conditions guaranteeing
the existence of a separatrix.

Viewing the vector field near the origin can be quite
misleading, since it appears to have both types of behavior simply
because all the solutions have the same asymptotic expansion at the
origin.  Additionally this implies that the
apparent, but potentially false, separatrix is well matched by the first four
terms of the asymptotic expansion as illustrated in Figure \ref{s2serieszoom}.  Of course
given that we have a recursive equation and an implicit solution we can easily calculate the
asymptotic series out to hundreds of terms \cite{bkerfc}, and the
use of a four loop approximation is merely meant to be illustrative.

Another simple observation is that we can derive the
equation for the curve where the solutions are horizontal by solving
for $\gamma_1'(x)$
\[
  \gamma_1'(x) = \frac{\gamma_1(x) + \gamma_1^2(x) - x}{2x\gamma_1(x)}
\]
and then solving the numerator to get the curve
\[
  y = \frac{-1 + \sqrt{1+4x}}{2}
\]
illustrated in Figure \ref{s2horiz}.

\subsection{Other cases}
Let us return to general $s$ while maintaining the assumption $m=1$,
$P(x)=x$.

The case $s=0$ is degenerate, giving the algebraic equation
$\gamma_1(x) = x - \gamma_1(x)^2$ with solutions
\[
\gamma_1(x) = \frac{-1 \pm \sqrt{1+4x}}{2}.
\]
From now on we will assume $s \neq 0$.

We can obtain implicit solutions for a few other isolated values of
$s$ using Maple
\begin{align*}
s=1: & \gamma_1(x)=x + x
W\left(C\exp\left(-\frac{1+x}{x}\right)\right), \\
s=\frac{3}{2}: & 
    A\left(X\right) -
    x^{1/3}2^{1/3}A'\left(X\right)=
    C\left(B\left(X\right) -
    x^{1/3}2^{1/3}B'\left(X\right)\right)
    \text{ where } X=\frac{1+\gamma_1(x)}{2^{2/3}x^{2/3}}, \\
s=2: & \exp\left(\frac{(1+\gamma_1(x))^2}{2x}\right)\sqrt{-x} +
  \mathrm{erf}\left(\frac{1+\gamma_1(x)}{\sqrt{-2x}}\right)\frac{\sqrt{\pi}}{\sqrt{2}}
  = C, \\
s=3: &
  (\gamma_1(x)+1)A\left(X\right) - 2^{2/3}A'\left(X\right)=
  C\left((\gamma_1(x)+1)B\left(X\right) -
    2^{2/3}B'\left(X\right)\right) \\ & \qquad \text{ where } X = \frac{(1+\gamma_1(x))^2 +
  2x}{2^{4/3}x^{2/3}},
\end{align*}
where $A$ is the Airy Ai function, $B$ the Airy Bi function and $W$
the Lambert W function.

\begin{figure}
\subfloat[$s=1$]{\epsfig{file=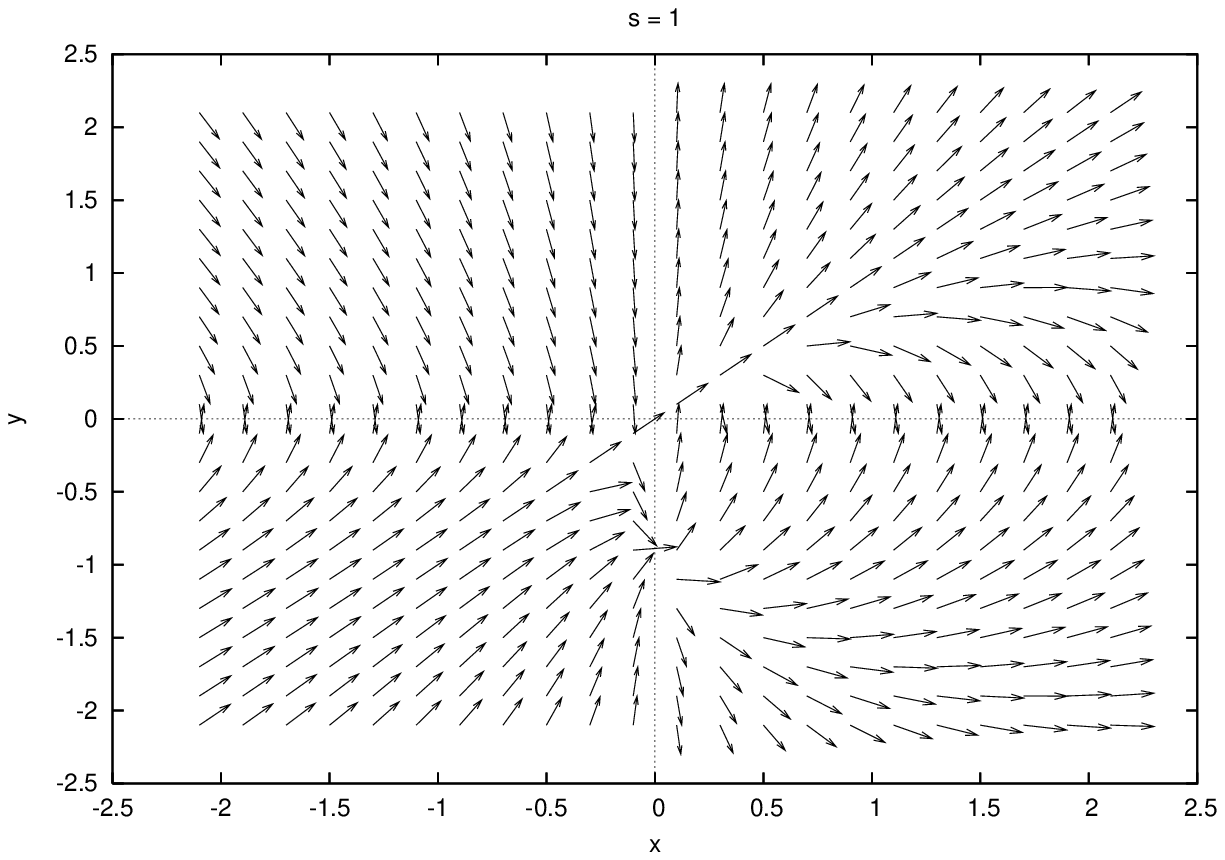, width=8cm}}
\subfloat[$s=3/2$]{\epsfig{file=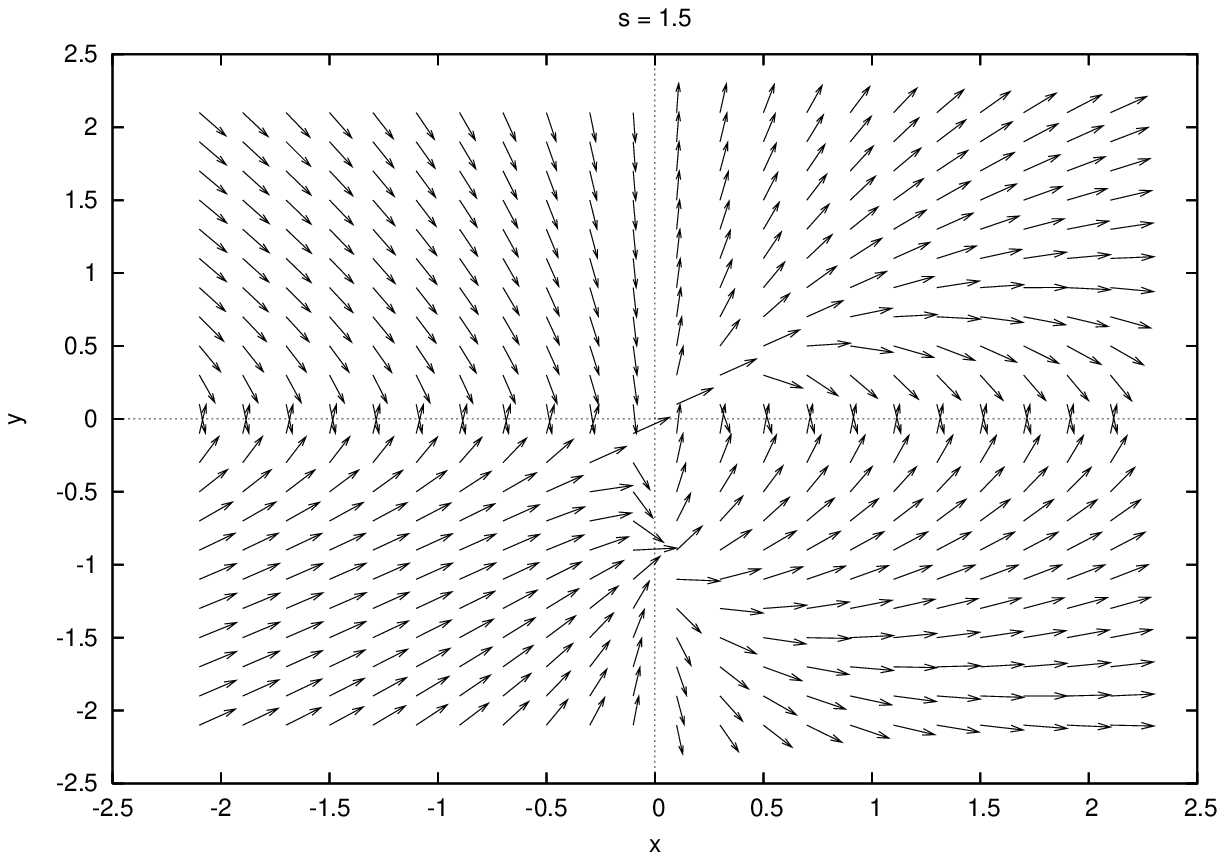, width=8cm}}\\
\subfloat[$s=2$]{\epsfig{file=s2quadgp.eps, width=8cm}}
\subfloat[$s=3$]{\epsfig{file=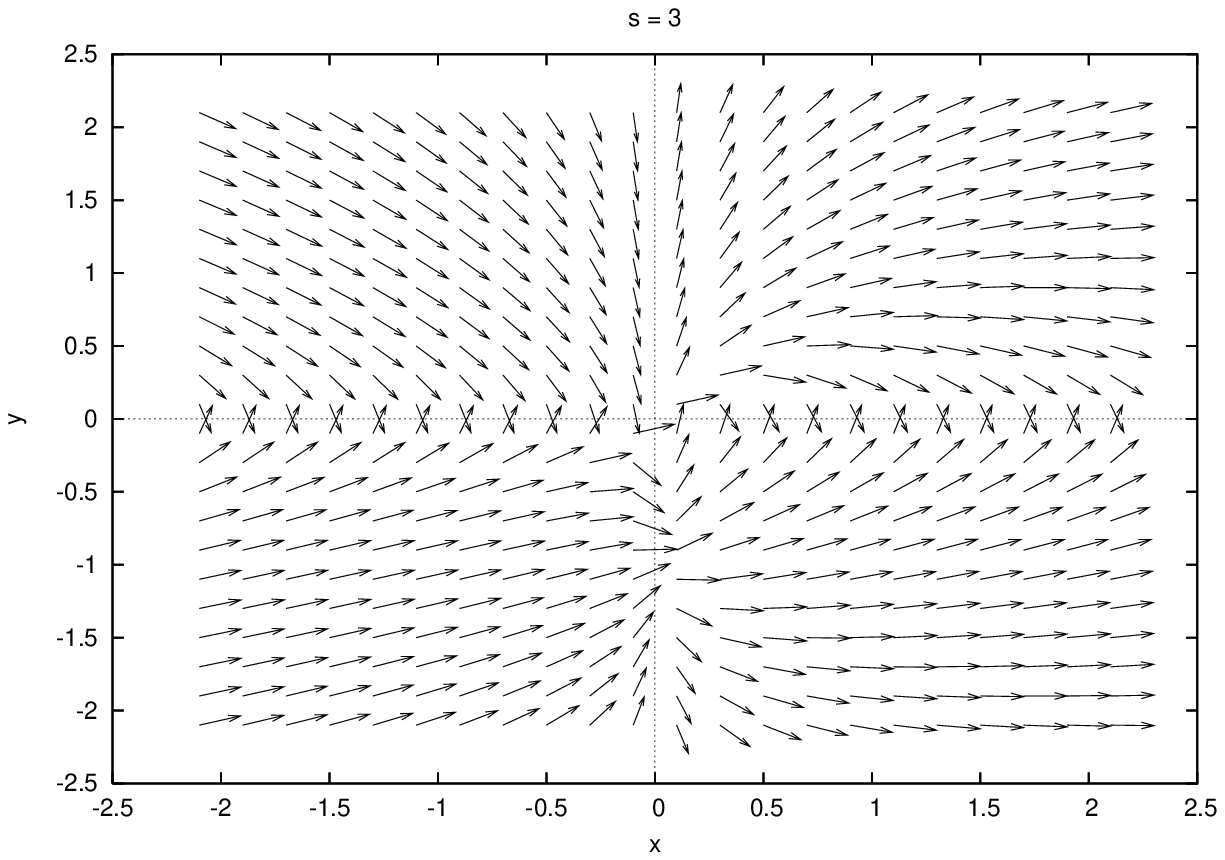, width=8cm}}
\caption{The vector field of $\gamma_1'(x)$ with $m=1$ and $P(x)=x$,
  showing the dependence on $s>0$.}\label{alltogether}
\end{figure}

\begin{figure}
\epsfig{file=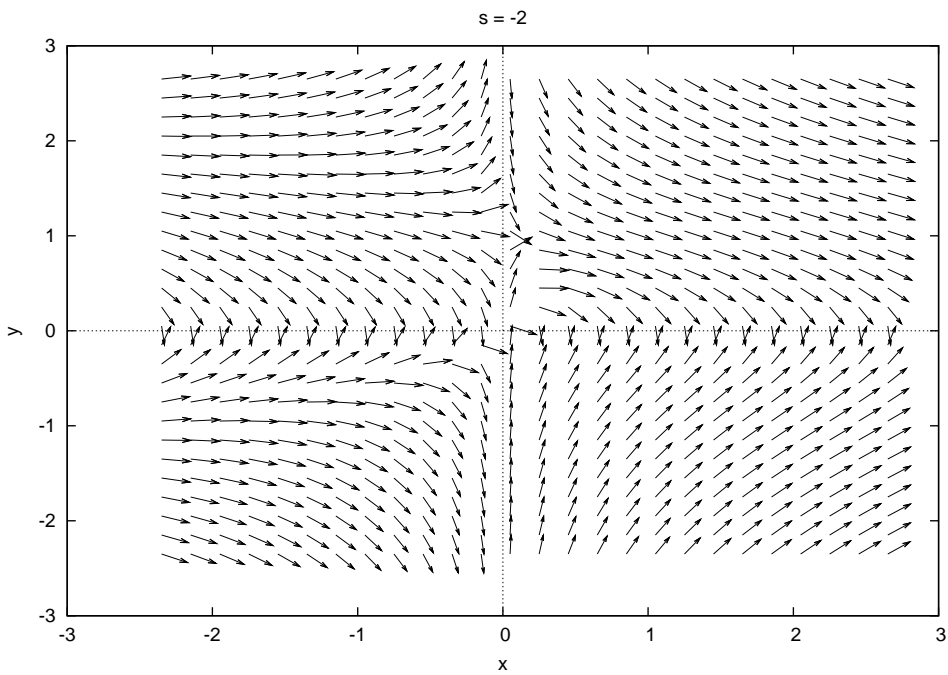, width=13cm}
\caption{The case $P(x)=x$ and $s=-2$. A typical example with
  $s<0$.}\label{s neg}
\end{figure}

Qualitatively the vector fields are rather similar, see Figure
\ref{alltogether}.  The same qualitative picture also remains for values of
$s>0$ where we do not have exact solutions.  For $s<0$ the picture is
somewhat different, see Figure \ref{s neg}, but we still see solutions which die and can still ask whether there are solutions
which exist for all $x>0$

\begin{figure}
\epsfig{file=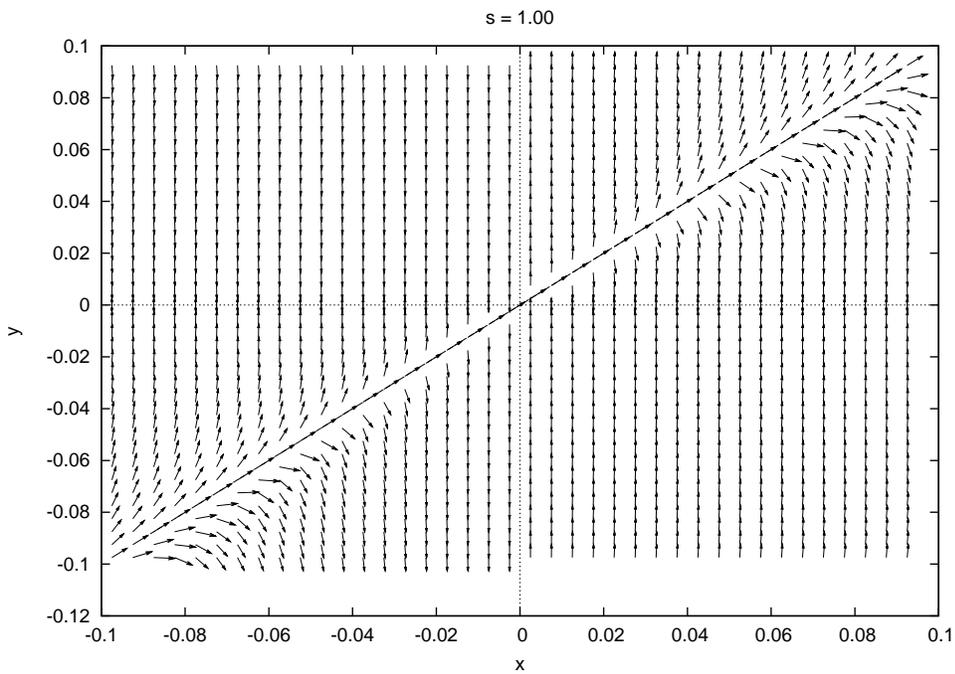, width=13cm}
\caption{The case $s=1$ compared to the curve $\gamma_1(x)=x$}
\label{s1serieszoom}
\end{figure}

In the case $s=1$, $\gamma_1(x) = x$ is manifestly
a solution, so there are solutions which exist for all $x$ for some
values of $s$.  $\gamma_1(x)=x$ is illustrated in Figure
\ref{s1serieszoom}.  
%As it turns out $\gamma_1(x) = x$ is not itself the
%separatrix, since we know such a solution must behave asymptotically as
%$\sqrt{2x}$ for large $x$, see below; though, of course, $\gamma_1(x)=x$ gives the
%behavior of the separatrix for small $x$, see Figure
%\ref{s1serieszoom}.  

Note also that we can, as before, calculate the curve where solutions are
flat for general $s$, and it depends only on the sign of $s$ since
\[
  \gamma_1'(x) = \frac{\gamma_1(x) + \sgn(s)\gamma_1^2(x) - x}{|s|x\gamma_1(x)}
\]
giving the curve
\[
  y = \frac{-1 + \sqrt{1+\sgn(s)4x}}{2}.
\]

%The behavior of the separatrix for large $x$, if it exists, must be asymptotic to
%\[
% \sqrt{\frac{2x}{2-s}}
%\]
%since it must remain above $(-1+\sqrt{1+4x})/2$ which implies that
%for large $x$ $\gamma_1^2 >> \gamma_1$ so we are reduced to
%considering the differential equation
%\[
%  sx y'(x) y(x)  = y(x)^2 - x
%\]
%or with $u(x) = y(x)^2$
%\[
%  s x u'(x)/2 = u(x) - x
%\]
%which has solution
%\[
%  u(x) = \frac{-2x}{s-2} + Cx^{2/s}
%\]
%***********

\section{QED as a single equation}

In this section we are interested in the case where $m=2$, and $s=1$
in \eqref{de}.  In view of the Ward identities and the work of
Johnson, Baker, and Willey \cite{jbw} the QED system can be reduced by
a suitable choice of gauge to
the single equation with those values of $m$ and $s$ describing the
photon propagator.

The first question is how to choose $P(x)$.  To 2 loops 
\[
   P(x) =  \frac{x}{3} + \frac{x^2}{4}
\]
To 4 loops we need to correct the primitives in view of the reductions
of the previous chapters.  Values are from \cite{gkls}.
\[
  P(x) = \frac{x}{3} + \frac{x^2}{4} +
  (-0.0312 + 0.06037)x^3 +(-0.6755 + 0.05074)x^4 
\]

\begin{figure}
\subfloat[$P(x)$ taken to 2 loops.]{\epsfig{file=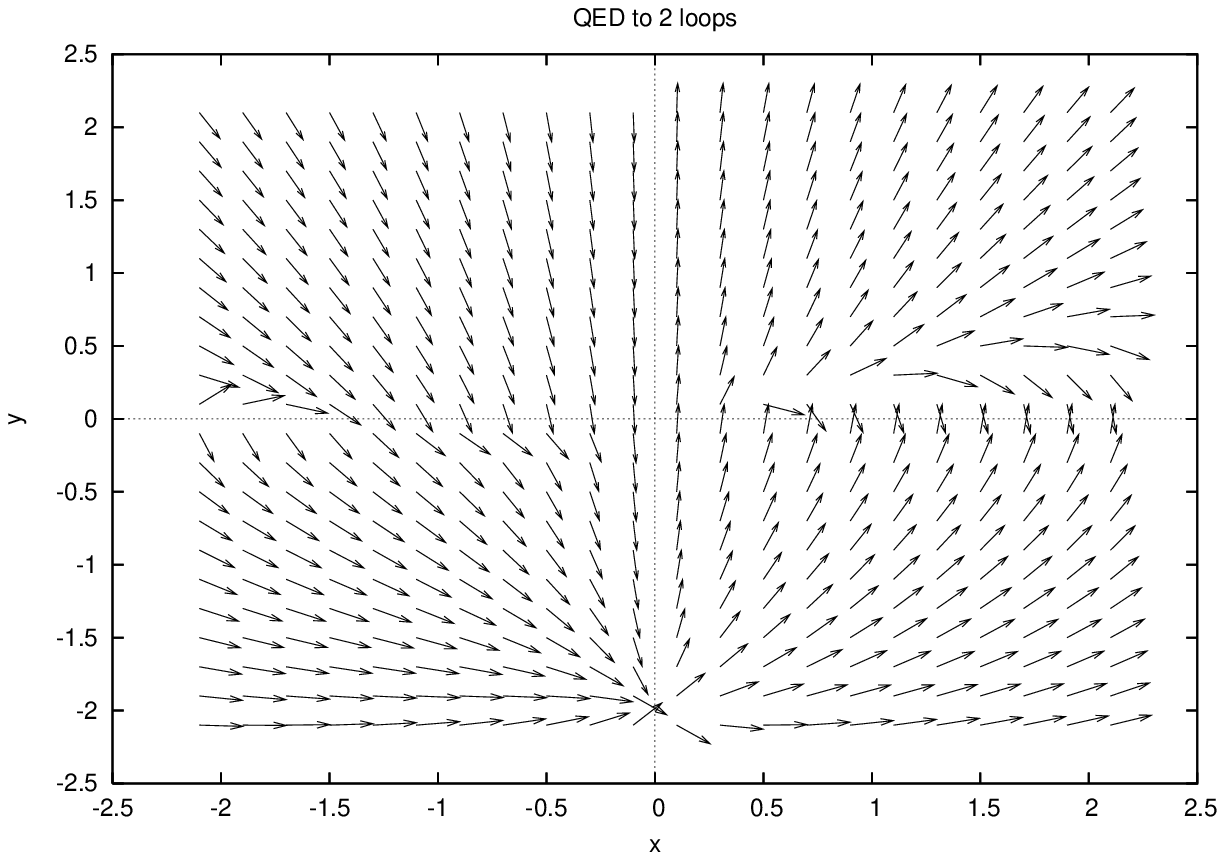,
    width=7.5cm}}
\subfloat[$P(x)$ taken to 4 loops.]{\epsfig{file=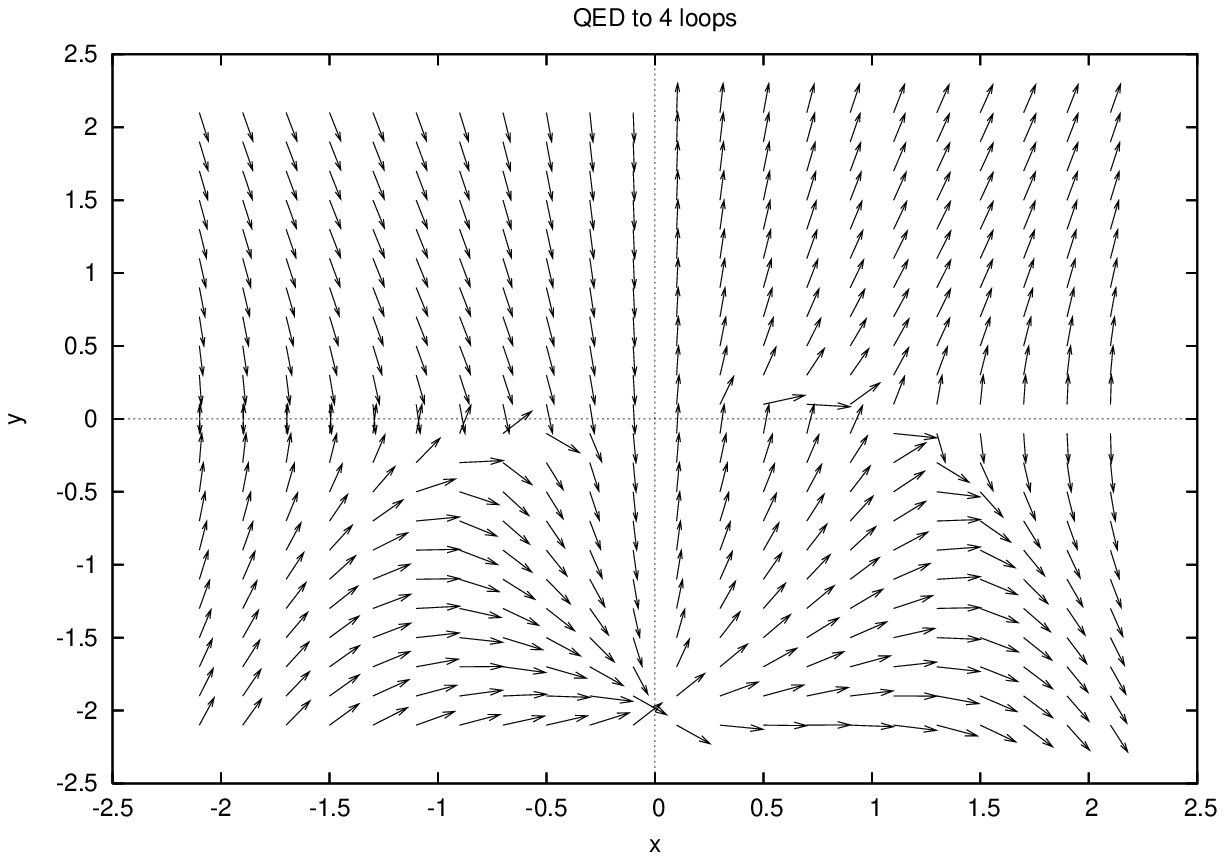,
    width=7.5cm}}
\caption{The vector field of $\gamma_1'(x)$ for QED with different
  choices for $P(x)$.}
\label{QEDloops}
\end{figure}

In the first of these cases little has changed from the simple
examples of the previous sections.
At 4 loops, however, $P(0.992\ldots)=0$ which causes substantial
changes to the overall picture, see Figure \ref{QEDloops}.  

\begin{figure}
\epsfig{file=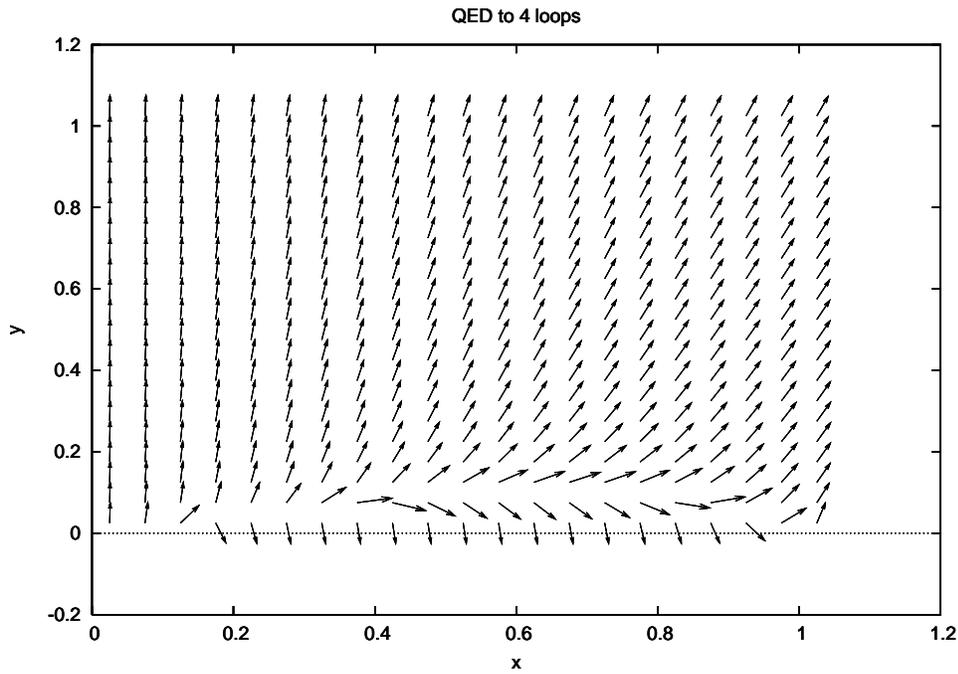, width=13cm}
\caption{The region between $x=0$ and $x=1$ in the vector field of $\gamma_1'(x)$ for QED with $P(x)$ taken
  to 4 loops.}
\label{QED4loopssmall}
\end{figure}

%At 4 loops, however, $P(0.992\ldots)=0$ which causes substantial
%changes to the overall picture, see Figure \ref{QED4loops}.  
%Let $x_0
%= 0.992\ldots$
%be that value.  Then at $x=x_0, y=0$ we see that 
%\[
%  \gamma_1'  =  
%\] 
%***Discuss how the solution doesn't go through that point even though
%it appears to***.  
This zero
in $P(x)$ is expected to be spurious, due only to taking the 4 loop
approximation out beyond where it is valid, and the qualitative
behavior of the solutions looks much more familiar if we restrict our
attention to $0\leq x < 0.992\ldots$, see Figure \ref{QED4loopssmall}.

%\begin{figure}
%\epsfig{file=QED4loops.eps, width=13cm}
%\caption{The vector field of $\gamma_1'(x)$ for QED with $P(x)$ taken
%  to 4 loops.}
%\label{QED4loops}
%\end{figure}

Note that if $P(x) > 0$ for $x>0$ then by the same analysis as in the
$P(x)=x$ case we can determine the curve where the solutions are
flat.  The curve is
\[
  y = \frac{-1 + \sqrt{1+4P(x)}}{2}.
\]

\begin{figure}
\subfloat[Close to the origin, with $x=1/137\ldots$ marked.]{\epsfig{file=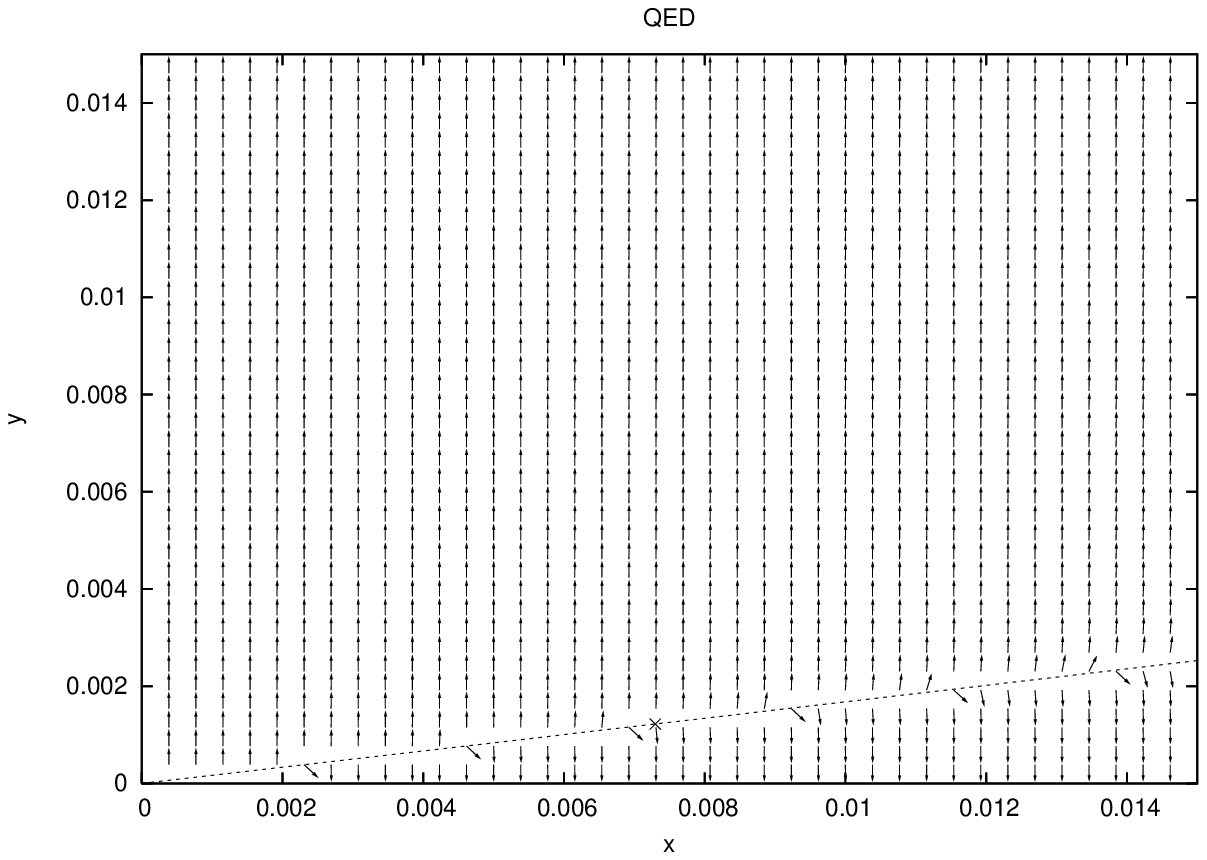, width=7.5cm}}
\subfloat[The limits of the validity of the four loop approximation.]{\epsfig{file=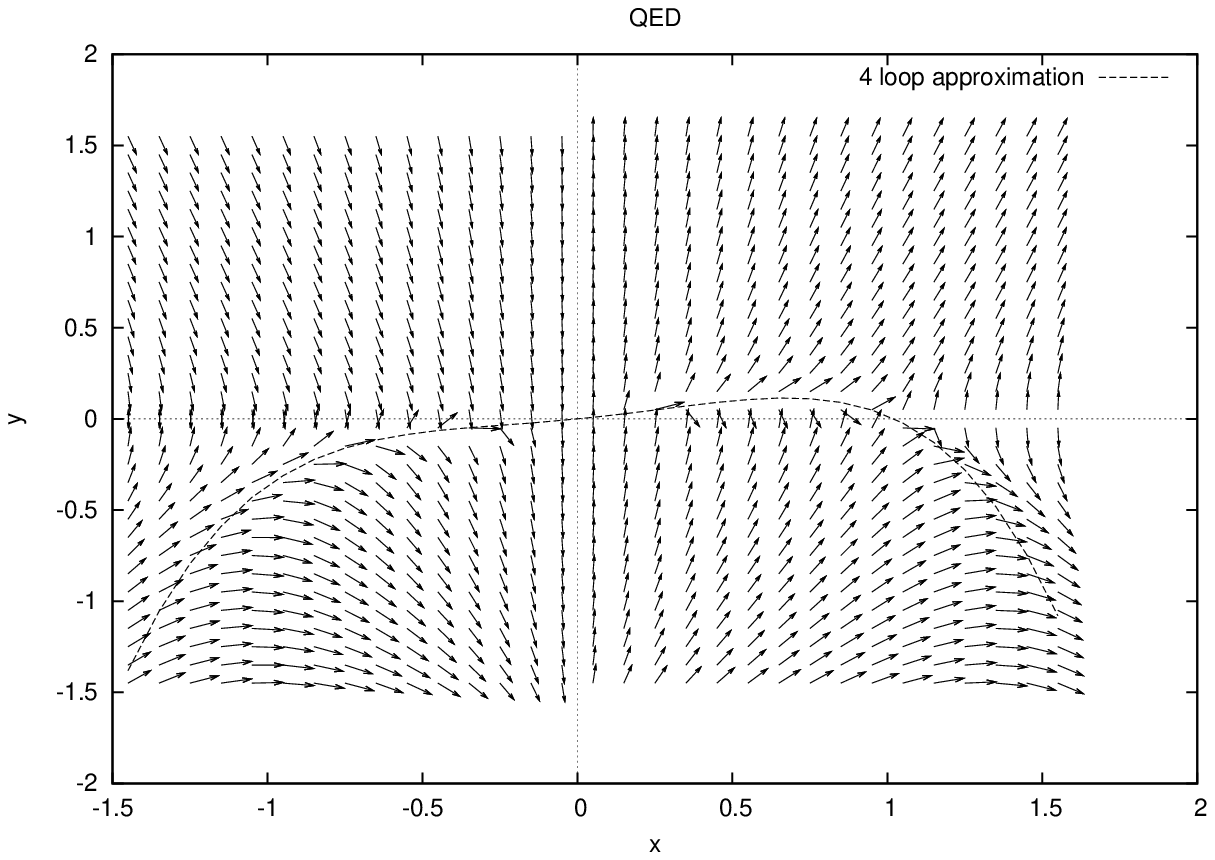, width=7.5cm}}
\caption{The four loop approximation to $\gamma_1(x)$ for QED.}
\label{youarehere}
\end{figure}

The first four loops of perturbation theory
give a good approximation to reality, and also as expected match the
apparent separatrix for small values of $x$, which is illustrated quite
strikingly in Figure \ref{youarehere}.

\section{$\phi^4$}

Let us now consider $\phi^4$ as an example which legitimately leads to a
system of equations, but for which it remains possible to create illustrations, and
perhaps even to analyze.
Taking advantage of the graphical similarity between the vertex and
propagator in $\phi^4$ and the symbols $+$ and $-$ respectively we
will write the 
specialization of \eqref{system} for $\phi^4$ as
the system
\begin{align*}
  \gamma_1^+(x) & = P^+(x) + \gamma_1^+(x)^2 + (\gamma_1^+(x)+ 2 \gamma_1^-(x))x\partial_x\gamma_1^+(x)\\
  \gamma_1^-(x) & = P^-(x) - \gamma_1^-(x)^2 + (\gamma_1^+(x)+ 2 \gamma_1^-(x))x\partial_x\gamma_1^-(x)
\end{align*}

\begin{figure}
\epsfig{file=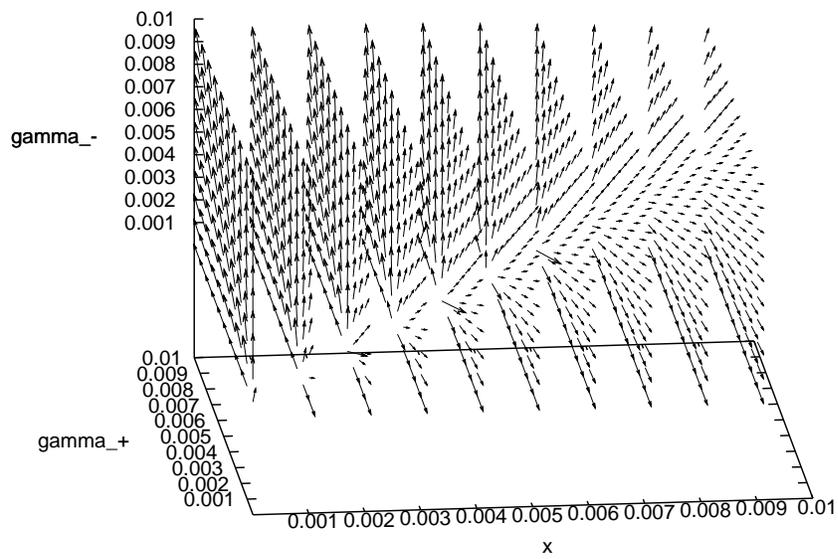, width=14cm}
\caption{$\phi^4$ near the origin.}
\label{phi4mysign}
\end{figure}

The values of $\gamma_1^+$ and $\gamma_1^-$ up to order $x^5$ can be
obtained from \cite{kns-fclphi4} and hence so can those of $P^+$ and $P^-$.
Close to the origin we see a distinguished solution, see Figure \ref{phi4mysign}.  As in subsection \ref{erfc}, this may not indicate
a solution which exists for all $x$, but we hope that this solution is
physical.  

There are many tantalizing features appearing in these examples which
will hopefully be the genesis for future work linking to different
fields.  The equations derived in Chapter \ref{second recursion} seem
considerably more tractable than the original Dyson-Schwinger
equations when viewed either as recursive equations or as differential
equations.  They have already led to physically interesting results as
in Chapter \ref{radii} and hold much promise for the future.

\chapter*{List of Journal Abbreviations}
\begin{center}
\begin{tabular}{lp{0.56\textwidth}}
  Adv. Math. \dotfill & Advances in Mathematics\\
  Annals Phys. \dotfill & Annals of Physics\\
  Commun. Math. Phys.  \dotfill & Communications in Mathematical Physics\\
  IRMA Lect. Math. Theor. Phys.  \dotfill & Institut de Recherche
  Math\'ematique Avanc\'ee Lectures in Mathematics and Theoretical Physics \\
  J. Phys. A \dotfill & Journal of Physics A: Mathematical and Theoretical\\
  Nucl. Phys. B \dotfill & Nuclear Physics B: Particle physics, field
  theory and statistical systems, physical mathematics \\
  Nucl. Phys. B Proc. Suppl. \dotfill & Nuclear Physics B -
  Proceedings Supplements\\
  Phys. Lett. B  \dotfill & Physics Letters B: Nuclear Physics and
  Particle Physics\\
  Phys. Rev. B \dotfill & Physical Review B: Condensed Matter and
  Materials Physics
\end{tabular}
\end{center}

\newpage
%\singlespace
\bibliographystyle{plain}
\bibliography{main}

\end{document}